\PassOptionsToPackage{unicode}{hyperref}
\PassOptionsToPackage{hyphens}{url}
\PassOptionsToPackage{dvipsnames,svgnames,x11names}{xcolor}
\documentclass[
  12pt]{article}

\usepackage{amsmath,amssymb}
\usepackage{iftex}
\ifPDFTeX
  \usepackage[T1]{fontenc}
  \usepackage[utf8]{inputenc}
  \usepackage{textcomp} % provide euro and other symbols
\else % if luatex or xetex
  \usepackage{unicode-math}
  \defaultfontfeatures{Scale=MatchLowercase}
  \defaultfontfeatures[\rmfamily]{Ligatures=TeX,Scale=1}
\fi
\usepackage{lmodern}
\ifPDFTeX\else  
\fi
\IfFileExists{upquote.sty}{\usepackage{upquote}}{}
\IfFileExists{microtype.sty}{% use microtype if available
  \usepackage[]{microtype}
  \UseMicrotypeSet[protrusion]{basicmath} % disable protrusion for tt fonts
}{}
\makeatletter
\@ifundefined{KOMAClassName}{% if non-KOMA class
  \IfFileExists{parskip.sty}{%
    \usepackage{parskip}
  }{% else
    \setlength{\parindent}{0pt}
    \setlength{\parskip}{6pt plus 2pt minus 1pt}}
}{% if KOMA class
  \KOMAoptions{parskip=half}}
\makeatother
\makeatletter
\ifx\paragraph\undefined\else
  \let\oldparagraph\paragraph
  \renewcommand{\paragraph}{
    \@ifstar
      \xxxParagraphStar
      \xxxParagraphNoStar
  }
  \newcommand{\xxxParagraphStar}[1]{\oldparagraph*{#1}\mbox{}}
  \newcommand{\xxxParagraphNoStar}[1]{\oldparagraph{#1}\mbox{}}
\fi
\ifx\subparagraph\undefined\else
  \let\oldsubparagraph\subparagraph
  \renewcommand{\subparagraph}{
    \@ifstar
      \xxxSubParagraphStar
      \xxxSubParagraphNoStar
  }
  \newcommand{\xxxSubParagraphStar}[1]{\oldsubparagraph*{#1}\mbox{}}
  \newcommand{\xxxSubParagraphNoStar}[1]{\oldsubparagraph{#1}\mbox{}}
\fi
\makeatother

\usepackage{longtable,booktabs,array}
\usepackage{calc} % for calculating minipage widths
\usepackage{etoolbox}
\makeatletter
\patchcmd\longtable{\par}{\if@noskipsec\mbox{}\fi\par}{}{}
\makeatother
\IfFileExists{footnotehyper.sty}{\usepackage{footnotehyper}}{\usepackage{footnote}}
\makesavenoteenv{longtable}
\usepackage{graphicx}
\makeatletter
\def\maxwidth{\ifdim\Gin@nat@width>\linewidth\linewidth\else\Gin@nat@width\fi}
\def\maxheight{\ifdim\Gin@nat@height>\textheight\textheight\else\Gin@nat@height\fi}
\makeatother
\setkeys{Gin}{width=\maxwidth,height=\maxheight,keepaspectratio}
\makeatletter
\def\fps@figure{htbp}
\makeatother

\makeatletter
\@ifpackageloaded{caption}{}{\usepackage{caption}}
\AtBeginDocument{%
\ifdefined\contentsname
  \renewcommand*\contentsname{Table of contents}
\else
  \newcommand\contentsname{Table of contents}
\fi
\ifdefined\listfigurename
  \renewcommand*\listfigurename{List of Figures}
\else
  \newcommand\listfigurename{List of Figures}
\fi
\ifdefined\listtablename
  \renewcommand*\listtablename{List of Tables}
\else
  \newcommand\listtablename{List of Tables}
\fi
\ifdefined\figurename
  \renewcommand*\figurename{Figure}
\else
  \newcommand\figurename{Figure}
\fi
\ifdefined\tablename
  \renewcommand*\tablename{Table}
\else
  \newcommand\tablename{Table}
\fi
}
\@ifpackageloaded{float}{}{\usepackage{float}}
\floatstyle{ruled}
\@ifundefined{c@chapter}{\newfloat{codelisting}{h}{lop}}{\newfloat{codelisting}{h}{lop}[chapter]}
\floatname{codelisting}{Listing}

\makeatother
\makeatletter
\@ifpackageloaded{caption}{}{\usepackage{caption}}
\@ifpackageloaded{subcaption}{}{\usepackage{subcaption}}
\makeatother

\ifLuaTeX
  \usepackage{selnolig}  % disable illegal ligatures
\fi
\usepackage[]{natbib}
\usepackage{bookmark}

\IfFileExists{xurl.sty}{\usepackage{xurl}}{} % add URL line breaks if available
\hypersetup{
  pdftitle={Robust estimation and Inference with Categorical Data},
  pdfauthor={Max Welz},
  pdfkeywords={Model misspecification, categorical robustness, asymptotic efficiency, outlier detection, careless responding},
  colorlinks=true,
  linkcolor={blue},
  filecolor={Maroon},
  citecolor={Blue},
  urlcolor={Blue},
  pdfcreator={LaTeX via pandoc}}

\newcommand{\anon}{1}

\usepackage{amsmath, amsthm, bm, amssymb, hyperref, xcolor, caption, subcaption, graphicx, dsfont, multirow, mathtools, booktabs, array,rotating,enumitem,makecell}

\usepackage{grffile}

\renewcommand{\vec}[1]{\boldsymbol{#1}}
\newcommand{\Eoperator}{\mathbb{E}}
\newcommand{\Poperator}{\mathbb{P}}
\newcommand{\E}[2]{\Eoperator_{#1} \left[ #2 \right]}

\newcommand{\var}[2]{\mathbb{V}\mathrm{ar}_{#1} \left[ #2 \right]}
\newcommand{\cov}[2]{\mathbb{C}\mathrm{ov}_{#1} \left[ #2 \right]}

\renewcommand{\Pr}[2]{\Poperator_{#1} \left[ #2 \right]}
\renewcommand{\emph}[1]{\textit{#1}}
\newcommand{\pfun}[2]{p_{#1}\left(#2\right)}
\newcommand{\sfun}[2]{\boldsymbol{s}_{#1}\left(#2\right)}
\newcommand{\pz}[1]{\pfun{#1}{\vec{z}}}

\newcommand{\sz}[1]{\sfun{#1}{\z}}
\newcommand{\sy}[1]{\sfun{#1}{\y}}

\newcommand{\BLambda}{\bm{\Lambda}}
\newcommand{\mat}[1]{\boldsymbol{#1}}
\newcommand{\matfun}[2]{\mat{#1}\left( #2 \right)}
\newcommand{\matfuneps}[2]{\mat{#1}_{\varepsilon}\left( #2 \right)}
\newcommand{\matfunN}[2]{\mat{#1}_{N}\left( #2 \right)}
\newcommand{\vecfun}[2]{\vec{#1}\left( #2 \right)}
\newcommand{\vecfunN}[2]{\vec{#1}_N\left( #2 \right)}
\newcommand{\matinv}[2]{\matfun{#1}{#2}^{-1}}
\newcommand{\matinveps}[2]{\mat{#1}^{-1}_{\varepsilon}\left(#2\right)}
\newcommand{\matinvN}[2]{\mat{#1}^{-1}_{N}\left(#2\right)}

\newcommand{\IF}[1]{\mathrm{IF}\left(#1\right)}
\newcommand{\Btheta}{\bm{\theta}}
\newcommand{\thetahat}{\hat{\Btheta}_N}

\newcommand{\BTheta}{\bm{\Theta}}

\newcommand{\gauss}{\textnormal{N}}
\newcommand{\I}[1]{\mathds{1}\left\{ #1 \right\}}

\renewcommand{\d}{\textnormal{d}}
\renewcommand{\hat}[1]{\widehat{#1}}
\renewcommand{\bar}[1]{\overline{#1}}
\renewcommand{\tilde}[1]{\widetilde{#1}}
\renewcommand{\cal}[1]{\boldsymbol{\mathcal{#1}}}
\newcommand{\z}{\vec{z}}
\newcommand{\y}{\vec{y}}
\newcommand{\Z}{\cal{Z}}
\newcommand{\R}{\mathbb{R}}

\newcommand{\convP}{\stackrel{\Poperator}{\longrightarrow}}
\newcommand{\convweak}{\stackrel{\mathrm{d}}{\longrightarrow}}
\newcommand{\fun}[2]{{#1}\left({#2}\right)}
\newcommand{\feps}{f_{\varepsilon}}
\newcommand{\fepsz}{\feps (\z)}
\newcommand{\fhat}{\hat{f}_{N}}
\newcommand{\fhatz}{\hat{f}_{N}(\vec{z})}

\newcommand{\partialderivative}[2]{\frac{\partial #1 }{\partial #2}}
\newcommand{\partialderivativetwice}[2]{\frac{\partial^2 #1 }{\partial {#2}\partial {#2}^\top}}

\newcommand{\ohp}[1]{o_{\Poperator}\left(#1\right)}
\newcommand{\Ohp}[1]{O_{\Poperator}\left(#1\right)}
\newcommand{\QED}{\hfill$\blacksquare$}

\newcommand{\proglang}[1]{\textsf{#1}}
\newcommand{\pkg}[1]{\texttt{#1}}

\newcommand{\eps}{\varepsilon}
\newcommand{\Dsymbol}{D_{\rho}}
\newcommand{\D}[2]{\Dsymbol\left({#1} \middle\| {#2}\right)}
\newcommand{\Dfp}{\D{\fhat}{\model}}
\newcommand{\Imat}{\mat{\mathbb{I}}_d}
\newcommand{\Imatp}{\mat{\mathbb{I}}_p}

\newcommand{\fx}[2]{#1 \left( #2\right)}
\newcommand{\PRsymbol}{\delta}

\newcommand{\PR}[3]{\PRsymbol_{#1,#2}\left(#3\right)}
\newcommand{\PRzeps}[1]{\PR{#1}{\varepsilon}{\z}}
\newcommand{\PRzn}[1]{\PR{#1}{N}{\z}}

\newcommand{\Bthetahat}{\hat{\Btheta}_N}
\newcommand{\BthetahatMLE}{\Bthetahat^{\mathrm{\ MLE}}}

\newcommand{\gradient}{\nabla_{\Btheta}}
\newcommand{\hessian}{\nabla^2_{\Btheta}}

\newcommand{\Blambda}{\vec{\lambda}}

\newcommand{\BbSigma}{\mat{\Sigma}}
\newcommand{\BF}{\vec{F}}
\newcommand{\BE}{\vec{E}}
\newcommand{\BSigmaE}{\BbSigma_{\BE}}
\newcommand{\BSigmaF}{\BbSigma_{\BF}}
\newcommand{\BZs}{\vec{Z}^*}

\newcommand{\GN}{G_N}
\newcommand{\Ri}[1]{\vec{R}_i \left(#1\right)}
\newcommand{\LN}{L_{N,\rho}}
\newcommand{\Leps}{L_{\varepsilon,\rho}}
\newcommand{\etaNj}[1]{\vec{\eta}_j\left(#1, \fhat\right)}

\newcommand{\Bthetabreve}{\breve{\Btheta}_N}
\newcommand{\remdelta}[1]{\Delta^\delta_{N,\z,#1}}
\newcommand{\cT}[1]{\mathcal{T}_{\rho}\left({#1}\right)}
\newcommand{\Ceps}{\mathcal{C}_\varepsilon(\Btheta_*)}
\newcommand{\Beps}{B_\rho(\eps;\Btheta_*)}
\newcommand{\BRW}{B_{\textnormal{RW}}(\eps;\Btheta_*)}
\newcommand{\Gammarho}[1]{\Gamma_\rho\left( #1 \right)}
\newcommand{\Rfun}[1]{\bar{D}_{#1}(\eps; p_{\Btheta_*})}
\newcommand{\Rrho}{\Rfun{\rho}}
\newcommand{\RRW}{\Rfun{\textnormal{RW}}}
\newcommand{\modelt}{p_{\Btheta_*}}
\newcommand{\epscrit}[1]{\eps_{\textnormal{crit}}(#1,\Btheta_*)}
\newcommand{\gepsdelta}{g_{\eps}^{\Delta_{\z}}}
\newcommand{\fepsdelta}{f_{\eps}^{\Delta_{\z}}}
\newcommand{\heps}{h_{\bar{\Btheta},\eps}}
\newcommand{\pmin}{p_{\textnormal{min}, \Btheta_*}}
\newcommand{\TV}[1]{\textnormal{TV}(#1)}
\newcommand{\kappac}{\kappa_{c_1,c_2}}
\newcommand{\Eeps}[1]{E_{\eps,\Btheta_*}(#1)}
\newcommand{\rRW}[1]{r_{\textnormal{RW},\eps}(#1)}
\newcommand{\rhot}{\tilde{\rho}}
\newcommand{\ueps}{u_\eps}

\newcommand{\beps}{b_{\eps}}
\newcommand{\ct}{\tilde{c}}

\newcommand{\Dphi}[2]{D_{\phi}\left({#1} \middle\| {#2}\right)}
\newcommand{\phisup}[1]{\phi^{\textnormal{#1}}}
\newcommand{\cI}{\boldsymbol{\mathcal{{I}}}}
\newcommand{\pcrit}{p_{\textnormal{crit}}(\eps)}
\newcommand{\Qeps}{Q_{\eps}}
\newcommand{\e}{\textnormal{e}}
\newcommand{\Reps}{R_{\eps,\Btheta_*}}
\newcommand{\Teps}{T_{\eps,\Btheta_*}}
\newcommand{\teps}{t_{\eps,\Btheta_*}}

\newcommand{\Bv}{\vec{v}}

\newcommand{\PRzepsS}[1]{\PR{#1}{\varepsilon}{\z}}
\newcommand{\PRznS}[1]{\PR{#1}{N}{\z}}

\newcommand{\pzS}[1]{\pfun{#1}{\z}}
\newcommand{\model}{p_{\Btheta}}
\newcommand{\szS}[1]{\sfun{#1}{\z}}

\newcommand{\modelk}{\model^{(k)}}
\newcommand{\BZk}{\vec{Z}^{(k)}}
\newcommand{\Zk}{\Z^{(k)}}
\newcommand{\zk}{\z^{(k)}}
\newcommand{\Ak}{\mathcal{A}_k}
\newcommand{\pzk}[1]{p^{(k)}_{#1}\left(\zk\right)}
\newcommand{\fhatk}{\fhat^{(k)}}
\newcommand{\fepsk}{\feps^{(k)}}
\newcommand{\cmp}{\textnormal{cmp}}
\newcommand{\PRzepsk}[1]{\delta_{#1,\eps}^{(k)}\left(\zk\right)}
\newcommand{\PRznk}[1]{\delta_{#1,N}^{(k)}\left(\zk\right)}
\newcommand{\matepscmp}[2]{\mat{#1}_\eps^\cmp\left(#2\right)}
\newcommand{\matNcmp}[2]{\mat{#1}_N^\cmp\left(#2\right)}
\newcommand{\matepsk}[3]{\mat{#1}_\eps^{(#3)}\left(#2\right)}
\newcommand{\skfun}[3]{\vec{s}_{#1}^{(#2)}\left(#3\right)}
\newcommand{\Rik}[2]{\vec{R}_i^{(#1)} \left(#2\right)}
\newcommand{\xik}[2]{\vec{\xi}^{(#1)}_N\left(#2\right)}
\newcommand{\Bthetak}{\Btheta^{(k)}}
\newcommand{\modelkbeta}{p^{(k)}_{\Bthetak}}
\newcommand{\PRk}[3]{\PRsymbol^{(k)}_{#1,#2}\left(#3\right)}

\newcommand{\BX}{\vec{X}}
\newcommand{\x}{\vec{x}}
\newcommand{\X}{\boldsymbol{\mathcal{X}}}
\newcommand{\Y}{\mathcal{Y}}
\newcommand{\yx}{y|\x}
\newcommand{\xy}{\x, y}
\newcommand{\YX}{Y|\BX}
\newcommand{\XY}{\BX, Y}
\newcommand{\pyxfun}[2]{p_{#1}^{\YX}\left(#2\right)}
\newcommand{\syxfun}[2]{\vec{s}_{#1}^{\YX}\left(#2\right)}
\newcommand{\PRyxepsfun}[2]{\delta_{#1,\eps}^{\YX}\left(#2\right)}
\newcommand{\PRyxnfun}[2]{\delta_{#1,N}^{\YX}\left(#2\right)}
\newcommand{\cnd}{\textnormal{cnd}}

\newcommand{\vecfuncndN}[2]{\vec{#1}^{\cnd}_N\left(#2\right)}
\newcommand{\fhatX}{\fhat^{\BX}}
\newcommand{\fhatYX}{\fhat^{\YX}}
\newcommand{\fepsX}{\feps^{\BX}}
\newcommand{\fepsYX}{\feps^{\YX}}
\newcommand{\fepsXY}{\feps^{(\BX,Y)}}

\newcommand{\matepscnd}[2]{\mat{#1}_\eps^\cnd\left(#2\right)}
\newcommand{\modelcnd}{p_{\Btheta}^{\YX}}
\newcommand{\modelcndfun}[2]{p_{#1}^{#2}}
\newcommand{\Dcnd}[2]{\Dsymbol^{\cnd}\left({#1} \middle\| {#2}\right)}
\newcommand{\funsub}[3]{{#1}_{#2}\left(#3\right)}

\newcommand{\ML}{\textnormal{ML}}
\newcommand{\RW}{\textnormal{RW}}

\makeatletter
\newcommand{\pushright}[1]{\ifmeasuring@#1\else\omit\hfill$\displaystyle{#1}$\fi\ignorespaces}
\makeatother

\renewenvironment{proof}{\noindent\textbf{Proof. }}{\QED\bigskip}

\newcommand{\matnorm}[1]{{\left\vert\kern-0.25ex\left\vert\kern-0.25ex\left\vert #1 
    \right\vert\kern-0.25ex\right\vert\kern-0.25ex\right\vert}}

\theoremstyle{plain}
\newtheorem{theorem}{\color{blue}Theorem}
\newtheorem{lemma}{\color{blue}Lemma}

\newtheorem{assumption}{\color{blue}Assumption}

\newtheorem{proposition}{\color{blue}Proposition}
\newtheorem{corollary}{\color{blue}Corollary} 
\theoremstyle{definition}
\newtheorem{definition}{\color{blue}Definition} 
\newtheorem{remark}{\color{blue}Remark} 
\theoremstyle{definition}
\newtheorem{example}{\color{blue}Example} 

\usepackage{mdframed}
\surroundwithmdframed{example}

\usepackage{titletoc}

\begin{document}

\def\spacingset#1{\renewcommand{\baselinestretch}%
{#1}\small\normalsize} \spacingset{1}

%%%%%%%%%%%%%%%%%%%%%%%%%%%%%%%%%%%%%%%%%%%%%%%%%%%%%%%%%%%%%%%%%%%%%%%%%%%%%%

\if1\anon
{
  \title{\bf Robust Estimation and Inference with Categorical Data}
  \author{Max Welz\thanks{
   I thank Andreas Alfons, Isaiah Andrews, Patrick Groenen, Karen Kafadar, Patrick Mair, Anna Mikusheva, Whitney Newey, Mikhail Zhelonkin, Chen Zhou, as well as the participants of various conferences and seminars at Erasmus University, ETH Zurich, MIT, LSE, and the University of Zurich for helpful comments and suggestions. Parts of this work were carried out when I was visiting the economics department at MIT; I thank them for their hospitality. This work was supported by a grant from the Dutch Research Council (NWO), research program Vidi (Project No. VI.Vidi.195.141). I further acknowledge financial support in the form of travel awards by the \emph{Institute of Mathematical Statistics} and \emph{Psychometric Society} to present this work at conferences. This paper is based on Chapter~4 of my PhD thesis.
    }\hspace{.2cm}\\
    University of Zurich\\\href{mailto:max.welz@uzh.ch}{\texttt{max.welz@uzh.ch}}\\
   }
  \maketitle
} \fi

\if0\anon
{
  \bigskip
  \bigskip
  \bigskip
  \begin{center}
    {\LARGE\bf Title}
\end{center}
  \medskip
} \fi

\bigskip
\begin{abstract}
% The text of your abstract. 200 or fewer words.
\noindent 
Categorical data pose a distinctive robustness challenge: contingency-table cells need not have a meaningful magnitude, ordering, or metric, so departures must instead be assessed through discrepancies between observed cell frequencies and model-implied probabilities. We develop $C$-estimation, a unifying framework for robust estimation in structured categorical models. $C$-estimation limits the influence of large frequency discrepancies and can be applied to unconditional, composite, and regression models of categorical data. Building on minimum-disparity estimation, the framework also accommodates clipped nonsmooth loss functions. A common asymptotic theory establishes Fisher consistency, consistency for the population target and asymptotic normality under contamination, and sandwich covariance estimation. At a correctly specified model, regular $C$-estimators retain the first-order efficiency of maximum likelihood. To quantify global robustness, we derive computable lower and upper envelopes for maximum-bias curves and, for a Huber-like loss, connect its clipping constants to a global robustness bound. Simulations support the theory and illustrate the estimators' robustness. An application to questionnaire data illustrates robust estimation of a latent factor model and identifies misfitting response strings that may reflect careless responding. A software implementation is provided.
\end{abstract}

\noindent%  % Keywords: 3 to 6 keywords, that do not appear in the title
\vfill

\newpage
\spacingset{1.0} % changed from default 1.8

%%%%%%%%%%%%% MAIN TEXT STARTS HERE %%%%%%%%%%%%%%%%%%%%%%%
\section{Introduction}
\textit{``Robust approaches to covariance matrices, to time series, to just about any problem yet encountered, \underline{with the possible exception of contingency tables}, have been devised.''}

\hspace*{\fill} ---Stephen M. \citet[][``The changing history of robustness'']{stigler2010}
\vspace{10pt}

The possible exception in Stigler's observation reflects a distinctive challenge. Classical robust methods typically identify unusual observations through their magnitude, order, or distance from a  center---concepts that need not exist for the cells of a contingency table. In categorical data, model departures instead appear as discrepancies between observed relative cell frequencies and model-implied probabilities. Robust estimation must therefore limit the influence of such cellwise discrepancies.

This problem arises throughout the biomedical, demographic, psychological, and social sciences. Departures from a postulated model may result from careless or bot responses in online questionnaires \citep[e.g.,][]{falk2025,arias2020,meade2012}, misspecified latent-variable models for rating data \citep[e.g.,][]{foldnes2022}, or sampling and measurement errors in network analyses \citep[e.g.,][]{henry2024,borsboom2021}. Even small departures can substantially distort maximum-likelihood estimates, especially in structured categorical models whose joint tables are too high-dimensional for direct estimation and are therefore fitted through marginal or conditional models.

Minimum disparity estimation \citep{lindsay1994} provides a natural foundation: it compares each cell's relative frequency with its fitted probability and can attenuate the influence of cells showing substantial disagreement. \citet{markatou1997} connected this principle to weighted-likelihood estimation for discrete models, including logistic regression; \citet{ruckstuhl2001} introduced a Huber-like version for binomial models; and \citet{hooker2016} developed conditional disparity methods and their asymptotic theory. The Huber-like procedure clips a cell's contribution once its probability discrepancy crosses prespecified thresholds, paralleling gradient clipping in robust $M$-estimation.

Building on this principle and line of research, we develop $C$-estimation (``$C$'' for \emph{categorical}), a unified framework for structured categorical models. It covers unconditional models on finite or countably infinite sample spaces as well as composite and conditional models. We establish Fisher consistency, consistency and asymptotic normality under contamination, and sandwich covariance estimation. At a correctly specified model, regular $C$-estimators retain the first-order efficiency of maximum likelihood. Beyond classical smooth minimum-disparity theory, the framework accommodates clipped, nonsmooth discrepancies such as that of \citet{ruckstuhl2001} and makes explicit when standard asymptotic inference remains valid. We further derive computable maximum-bias envelopes and show how the clipping constants determine a global robustness bound. Thus, $C$-estimation brings robust fitting, statistical inference, scalable categorical modeling, and quantitative robustness analysis into a common framework.

Through simulations, we practically demonstrate the robustness of $C$-estimators and corroborate their theoretical properties. We showcase their applied usefulness in an empirical application on a latent factor model for responses to psychological questionnaire items. Compared to conventional maximum likelihood, $C$-estimators provide stronger factor loadings and a fitted model with better explanatory power. In addition, we show how they can be used to identify misfitting response strings, possibly due to careless responding. Finally, we provide a free implementation in the \proglang{R} package \pkg{robcat} at \url{https://CRAN.R-project.org/package=robcat}.

This paper is organized as follows. Section~\ref{sec:Cestimator} defines~$C$-estimators and in Section~\ref{sec:properties} we derive their theoretical properties. Section~\ref{sec:maxbias} quantifies their global robustness through maximum-bias curves. In Sections~\ref{sec:literature} and~\ref{sec:cmp}, we respectively establish their relationship to related methods and extend them to composite and conditional models. Section~\ref{sec:simulations} carries out simulation studies, and Section~\ref{sec:application} provides an empirical application. Section~\ref{sec:conclusion} concludes.

Regarding notation, denote by $\gradient g(\vec{a}) = \partialderivative{g(\Btheta)}{\Btheta}\big|_{\Btheta=\vec{a}}$ the gradient of a function~$g$ with respect to~$\Btheta$ that is evaluated at~$\vec{a}$. Analogously, we denote the Hessian by $\hessian g(\vec{a}) = \partialderivativetwice{g(\Btheta)}{\Btheta}\big|_{\Btheta=\vec{a}}$. The norms $\|\cdot\|$ and $\matnorm{\cdot}$ denote the Euclidean vector norm and Frobenius matrix norm, respectively. %The indicator function $\I{E}$ takes value~1 if an event~$E$ is true, and~0 otherwise. The symbols ``$\convP$'' and ``$\convweak$'' denote convergence in probability and distribution, respectively. 
The convention $0\log 0 \equiv 0$ is used.  
All proofs are deferred to the appendix.

\section{$C$-estimation}\label{sec:Cestimator}
This section defines $C$-estimators for \emph{unconditional} probability models. We will deal in later sections with  composite models and conditional models for regression problems.

\subsection{Unconditional probability models}
Let $\vec{Z} = (Z_1,\dots, Z_p)^\top$ be a~$p$-dimensional categorical random vector taking values in an at most countable sample space~$\Z = \{\z_1,\z_2,\dots\}$. The events in~$\Z$ are mutually exclusive and $|\Z|\geq 2$. 
Suppose the statistician postulates an unconditional model for the vector~$\vec{Z}$ that is subject to a deterministic $d$-variate parameter vector~$\Btheta\in\BTheta\subset\R^d$. This model, denoted $\{\model : \Btheta\in\BTheta\}$ here, is characterized by assigning to each discrete event $\z \in\Z$ a nonnegative probability
$
	\pzS{\Btheta} = \Pr{\Btheta}{\vec{Z} = \z}
$ 
such that $\sum_{\z\in\Z}\pz{\Btheta}=1$.  
The function~$\model$ is thus a \emph{probability mass function} (PMF). We impose the following minimal  conditions.

\begin{assumption}\label{ass:model}
The postulated unconditional model~$\model$ satisfies the following properties.
\begin{enumerate}[label={\textnormal{A.\theassumption.\arabic*.}}, ref={\textnormal{A.\theassumption.\arabic*}}, itemindent=20pt]
	\item $\sum_{\z\in\Z}\pzS{\Btheta} = 1$ for all $\Btheta\in\BTheta$,
	\item $\pzS{\Btheta}$ is twice continuously differentiable with respect to~$\Btheta\in\BTheta$ for all $\z\in\Z$, \label{ass:differentiability}%
	\item For every $\Btheta_1 \neq \Btheta_2$ in $\BTheta$, there exists a $\z\in\Z$ such that $\pzS{\Btheta_1} \neq \pzS{\Btheta_2}$, \label{ass:identification}%
	%\item $\left\| \partialderivative{\pz{\Btheta}}{\Btheta}\right\| < \infty$ for all~$\Btheta\in\BTheta,\z\in\Z$,\label{ass:boundedgradient}%
	\item $\pzS{\Btheta} > 0$ for all~$\Btheta\in\BTheta$, $\z\in\Z$.\label{ass:positiveprobability}%
\end{enumerate}
\end{assumption}
 % Specifically, we require point-identifiability of the parameters
%Assumption~\ref{ass:identification} imposes point-identifiability of the model's parameters. Otherwise, consistent estimation may not be feasible \citep[e.g.,][p.443]{lehmann1998}. 

Assumption~\ref{ass:model} requires~$\model$ to be a non-degenerate PMF that is  sufficiently smooth, its model parameters to be point-identifiable, and the individual probabilities to be strictly positive. %The positivity condition is not strictly required, but it makes the exposition easier. % and avoids having to deal with unconstructive technicalities. 

This setup is very general and encompasses many well-known unconditional models for categorical data, such as the binomial distribution, the multinomial distribution, and the Poisson distribution. It also encompasses more complex latent variable models, such as \emph{structural equation models} \citep[SEMs; e.g.,][]{bartholomew2011,muthen1984} for ordinal data, which are often used for the analysis of questionnaire responses.\medskip

\begin{example}[SEM for ordinal variables]\label{ex:factor}
Let $\vec{Z}$ denote ordinal responses to $p$ questionnaire items where the $j$-th item has~$m_j$ response options, $j=1,\dots,p$. Without loss of generality, encode the responses by consecutive integers so that the support of~$\vec{Z}$ is $\Z = \bigtimes_{j=1}^p\{1,2,\dots,m_j\}$. Suppose the response~$Z_j$ to the $j$-th item is governed by an underlying latent continuous variable~$Z_j^*$ through the unobserved discretization  process
\[
	Z_j = z \quad\textnormal{ if and only if }\quad \tau^{(j)}_{z-1} \leq Z_j^* < \tau^{(j)}_z,\quad z = 1,\dots, m_j,
\]
where $-\infty = \tau^{(j)}_0 < \tau^{(j)}_1 < \dots < \tau^{(j)}_{m_j-1} < \tau^{(j)}_{m_j} = +\infty$ are unknown \emph{threshold} parameters. The vector of latent variables~$\BZs = (Z_j^*)_{j=1}^p$ is assumed to admit a factor decomposition
$
	\BZs = \BLambda \BF + \BE,
$ 
where $\BF\sim \gauss_r(\vec{0}, \BSigmaF)$ is a vector of $r \ll p$ latent \emph{factors}, $\BE\sim\gauss_p(\vec{0}, \BSigmaE)$ is a $p$-vector of latent exogenous \emph{measurement errors}, and~$\BLambda = (\Blambda_1^\top, \dots, \Blambda_p^\top)^\top$ is a $p\times r$ matrix of unknown factor \emph{loading} parameter vectors~$\Blambda_j$. For identification, one imposes that~$\BSigmaF$ has only 1's on its main diagonal and that $\BSigmaE = \Imatp - \text{diag}(\BLambda\BSigmaF\BLambda^\top)$, where~$\Imatp$ is the $p\times p$ identity matrix. Consequently, the vector of model parameters~$\Btheta$ holds all thresholds, all factor loadings, and the unique off-diagonal elements of~$\BSigmaF$, a total of $d = \sum_{j=1}^pm_j - p + r(p + (r-1)/2)$ parameters. Under this model, the probability of an ordinal response vector $\z\in\Z$ at~$\Btheta$ reads
\begin{equation}\label{eq:factorfull}
	\pzS{\Btheta} 
	=
	\Pr{\Btheta}{\vec{Z} = \z} =
	\int_{\tau^{(1)}_{z_1 - 1}}^{\tau^{(1)}_{z_1}}	
	\cdots
	\int_{\tau^{(p)}_{z_p - 1}}^{\tau^{(p)}_{z_p}}
	\phi_p\left(\vec{z}^*; \vec{0}, \BLambda\BSigmaF\BLambda^\top + \BSigmaE\right) \d z_p^*\cdots \d z_1^*,
\end{equation}
where $\vec{z}^* = (z_j^*)_{j=1}^p$, and $\phi_p$ denotes the density of the $p$-variate normal distribution. Orientation restrictions, including a sign convention when $r=1$, must be imposed to ensure point-identifiability (Assumption~\ref{ass:identification}).
\end{example}

%Estimation of the SEM probabilities~\eqref{eq:factorfull} in Example~\ref{ex:factor} becomes computationally intractable when there is more than a handful of observed variables due to the $p$-dimensional integral. Therefore, this model is often estimated in composite manner by breaking up the joint model~\eqref{eq:factorfull} into smaller sub-models. Section~\ref{sec:cmp} is co such composite models.  

\subsection{Contamination}

In the spirit of robust statistics, we consider model misspecification due to data contamination. We therefore assume that~$\vec{Z}$ is distributed according to the sampling PMF
% we adopt the contamination model of \citet{ruckstuhl2001}, which assumes that observed data of~$\vec{Z}$ is distributed according to a perturbed  version of the postulated model. This perturbed model is characterized by the PMF
\begin{equation}\label{eq:hubermodel}
 	\fepsz = (1-\varepsilon) \pzS{\Btheta_*} + \varepsilon h(\z), \qquad \z\in\Z,
\end{equation}
where the unobserved but fixed $\varepsilon\in [0,0.5)$ is called the \emph{contamination fraction/probability}, and~$h(\cdot)$ is an unspecified unknown PMF on~$\Z$, called the \emph{contamination PMF}. Furthermore, $\Btheta_*\in\BTheta$ in~\eqref{eq:hubermodel} is the \emph{true} parameter value under which the postulated model generates uncontaminated observations, in contrast to~$h(\cdot)$, which generates contaminated observations.  Given  $\varepsilon>0$, we say that the postulated model is \emph{misspecified} if for all $\Btheta\in\BTheta$ there exists at least one $\z\in\Z$ such that $\fepsz \neq \pzS{\Btheta}$. If $\eps=0$, contamination is absent. The contamination model in~\eqref{eq:hubermodel} was originally proposed by \cite{ruckstuhl2001} and adapts the well-known contamination model of \citet{huber1964} to categorical data. 

\subsection{Residual adjustment functions}
%\subsection{Empirical risk minimization}
Let $(\vec Z_i)_{i=1}^N$ be independent observations from~$\feps$ and let $\fhatz=N^{-1}\sum_{i=1}^N\I{\vec Z_i=\z}$, which converges in probability to~$\fepsz$ for every $\z\in\Z$. At $\Btheta\in\BTheta$, the cellwise \emph{Pearson residual} (PR) \citep{lindsay1994} is $\PRznS{\Btheta}=\fhatz/\pzS{\Btheta}-1$. It lies in $[-1,\infty)$ and compares the observed relative frequency of a cell with its model probability. If $\eps=0$, then $\PRznS{\Btheta_*}\convP 0$ for every cell; under misspecification, no $\Btheta\in\BTheta$ makes all PRs converge to~0.

A \emph{residual adjustment function} (RAF) $A:[-1,\infty)\to\R$ moderates the contribution of discrepant cells \citep{lindsay1994}. The corresponding estimating equation is
\begin{equation}\label{eq:estimatingeq}
	\sum_{\z\in\Z} A\big( \PRznS{\Btheta}\big)\gradient\pzS{\Btheta} = \vec{0}.
\end{equation}
We define in a moment which roots are $C$-estimators, after imposing conditions on the RAF.

 \begin{assumption}\label{ass:RAF}
An RAF $A:[-1,\infty)\to \R$  satisfies the following properties.
\begin{enumerate}[label={\textnormal{A.\theassumption.\arabic*.}}, ref={\textnormal{A.\theassumption.\arabic*}}, itemindent=20pt]
    \item $A$ is non-constant and continuous, \label{ass:RAF-continuous}%
    \item $A$ is differentiable, with differentiability at $-1$ understood as right differentiability, except on a finite (possibly empty) subset of $[-1,\infty)$; its derivative~$A'$ is continuous and nonnegative on its domain,\label{ass:RAFderivative}
    \item $A(0)=0$ and, if it exists, $A'(0)=1$.\label{ass:RAForigin}%
\end{enumerate}
\end{assumption}
Let~$A_0$ be a function satisfying Conditions~\ref{ass:RAF-continuous}--\ref{ass:RAFderivative}. Then, if $A_0'(0) >0$, the normalized $A(\delta) = (A_0(\delta)-A_0(0))/A_0'(0)$ satisfies Condition~\ref{ass:RAForigin} and preserves the roots of~\eqref{eq:estimatingeq} because  $\sum_{\z\in\Z}\gradient\pzS{\Btheta}=\vec{0}$ whenever differentiation and summation may be interchanged. If~$A_0'(0)$ does not exist, the centered function $A(\delta)=A_0(\delta)-A_0(0)$ preserves the same roots and also satisfies the condition, whose derivative requirement is then vacuous. Thus, Condition~\ref{ass:RAForigin} entails a loss of generality only when~$A'_0(0)$ exists and equals zero.

\subsection{Divergence formulation}\label{sec:divergence}
For a convex function $\rho:[-1,\infty)\to\R$ that is differentiable on $(-1,\infty)$, define
\begin{equation}\label{eq:divergence}
	\Dfp 
	=	\sum_{\z\in\Z}\rho\left(\frac{\fhatz}{\pz{\Btheta}} -1 \right)\pz{\Btheta}
	= \sum_{\z\in\Z}\rho\big(\PRznS{\Btheta}\big)\pzS{\Btheta},
\end{equation}
the discrete $f$-divergence between the empirical PMF~$\fhat$ and the model~$\model$ \citep{csiszar1963}. For $\delta>-1$, define the RAF $A(\delta)=(\delta+1)\rho'(\delta)-\rho(\delta)$ and, when the limit exists, define $A(-1)$ by continuous extension. Whenever differentiation and summation may be interchanged, the left-hand side of~\eqref{eq:estimatingeq} equals $-\gradient\Dfp$. The following assumption imposes conditions on~$\rho$. 
Conversely, one can also construct from a given RAF a~$\rho$ meeting this assumption; see Appendix~\ref{app:RAF-to-rho} for details.

% The negative gradient of the divergence $\Dfp$ with respect to~$\Btheta$ equals 
%\begin{equation}\label{eq:gradient}
%	-\gradient\Dfp
%	=
%	\sum_{\z\in\Z} 
%	\Bigg( 
%			\frac{\fhatz}{\pzS{\Btheta}} \fx{\psi}{\PRznS{\Btheta}} - \fx{\rho}{\PRznS{\Btheta}}
%	\Bigg)
%	\gradient\pzS{\Btheta},
%\end{equation}
%where we write $\psi = \rho'$ for the derivative of~$\rho$. Setting 
%$
%	A(\delta) \equiv (\delta + 1) \psi(\delta) - \rho(\delta),
%$ 
%this gradient is equal to the left-hand-side of estimating equation~\eqref{eq:estimatingeq}. Hence, the RAF emerges from the estimating equation of the divergence~$\Dfp$. , a discrepancy function can also be obtained from a given RAF; see Appendix~\ref{sec:RAFtorho} for details. 
 %In particular, for $\rho(x) = (x+1)\log(x+1)$, one recovers $A(x) = x$ through~\eqref{eq:RAF} and, therefore, the ML estimating equation in~\eqref{eq:estimatingeq}, which is why the \emph{negative} gradient is used in~\eqref{eq:gradient}. For this particular choice, the empirical risk in~\eqref{eq:loss} is equal to the Kullback-Leibler (KL) divergence \citep{kullback1951} between~$\fhat$ and~$\model$. In fact,~\eqref{eq:loss} is a special case of the~$f$-divergence of \citet{csiszar1963}, which generalizes the KL divergence for discrete distributions. \todo{say that we can obtain discrepancy for given RAF}

\begin{assumption}\label{ass:rho}
The function $\rho:[-1,\infty) \to\R$  satisfies the following properties.
\begin{enumerate}[label={\textnormal{A.\theassumption.\arabic*.}}, ref={\textnormal{A.\theassumption.\arabic*}}, itemindent=20pt]
		\item $\rho(0)=0$, \label{ass:rho0} 
	\item $\rho$ is convex and continuous on $[-1,\infty)$ and continuously differentiable on $(-1,\infty)$,\label{ass:rho'}
	\item $\rho'$ is twice continuously differentiable outside a finite (possibly empty) subset of $(-1,\infty)$, \label{ass:rho''}
		\item $\rho'(0)=0$ and, if it exists, $\rho''(0) = 1$.\label{ass:rhonorm}
\end{enumerate}
\end{assumption}
\noindent 
Functions satisfying Assumption~\ref{ass:rho} are called \emph{discrepancy functions}; strict convexity is not required. No derivative of~$\rho$ is required at~$-1$, which does not preclude its induced RAF from having the right derivative permitted by Condition~\ref{ass:RAFderivative}.  %We thus for the rest of this paper view $C$-estimators as divergence minimizers.  
%While discrepancy functions may take negative values, the divergence~$\Dsymbol$ remains nonnegative.
Condition~\ref{ass:rhonorm} is a normalization ensuring that the induced RAF \(A(\delta)=(\delta+1)\rho'(\delta)-\rho(\delta)\) satisfies Condition~\ref{ass:RAForigin} without further normalization. Indeed, let~$\rho_0$ satisfy Conditions~\ref{ass:rho0}--\ref{ass:rho''}. If $\rho_0''(0)>0$, then
$
\rho(\delta)=
\frac{\rho_0(\delta)-\rho_0'(0)\delta}{\rho_0''(0)}
$ satisfies Condition~\ref{ass:rhonorm}. If \(\rho_0''(0)\) does not exist, the centered function $\rho(\delta)=\rho_0(\delta)-\rho_0'(0)\delta$ also satisfies the condition, whose second-derivative requirement is then vacuous. These transformations preserve divergence minimizers; see Remark~\ref{rem:rho-invariance} below. Thus, Condition~\ref{ass:rhonorm} entails a loss of generality only when~$\rho_0''(0)$ exists and equals zero.

\begin{proposition}\label{prop:divergence-nonnegative}
Grant Assumptions~\ref{ass:model} and~\ref{ass:rho}. Let $f$ be an arbitrary PMF on~$\Z$. Then, $\D{f}{\model} \geq 0$ for all $\Btheta$. Moreover, $\D{f}{\model} = 0$ if and only if $f = \model$.
\end{proposition}

%We are now ready to formally define $C$-estimators. 

\begin{definition}[$C$-estimator] \label{def:Cestimator}
Let $\model$ and~$\rho$ satisfy Assumptions~\ref{ass:model} and~\ref{ass:rho}, and suppose that the induced RAF $A(\delta)=(\delta+1)\rho'(\delta)-\rho(\delta)$ satisfies Assumption~\ref{ass:RAF}. A $C$-estimator is any global minimizer $\Bthetahat\in\BTheta$ of the divergence in~\eqref{eq:divergence}, that is,
$
	\Bthetahat  \in \arg\min_{\Btheta\in\BTheta} \Dfp.
$
If~\eqref{eq:divergence} is differentiable at an interior minimizer $\Bthetahat$, then $\Bthetahat$ solves~\eqref{eq:estimatingeq}. A root of~\eqref{eq:estimatingeq} that is not a global minimizer of~\eqref{eq:divergence} is not a $C$-estimator.
\end{definition}
% Note that the computing cost of~$\Dfp$, and therefore~\eqref{eq:thetahat}, is the same for all $\rho(\cdot)$.

\begin{remark}\label{rem:rho-invariance}
$C$-estimators are invariant to certain linear transformations of a discrepancy function~$\rho$. For constants $a >0$ and $(b,c)$, let  $\widetilde\rho(\delta)=a\rho(\delta)+b\delta+c$. Then $D_{\widetilde\rho}=aD_\rho+c$ because model-weighted PRs sum to zero, so the minimization problem in Definition~\ref{def:Cestimator} remains unchanged when~$\widetilde{\rho}$ is used instead of~$\rho$. 
\end{remark}

\subsection{Examples of discrepancy functions}
\begin{figure}[t]
    \centering
    \includegraphics[width = 0.99\textwidth]{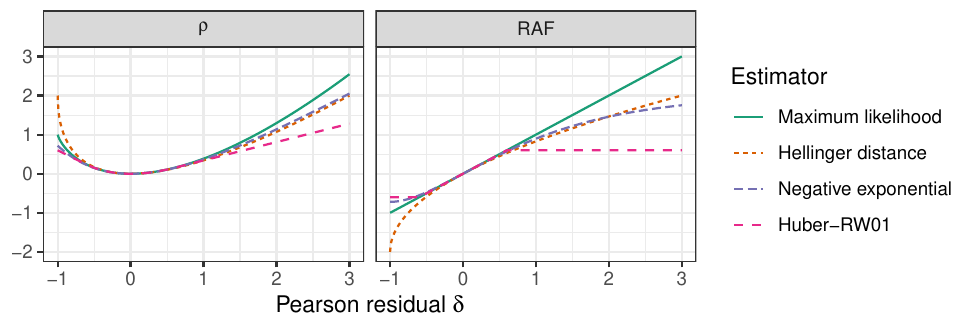}
\caption{Discrepancy functions $\rho$ and  RAFs $A$ for the $C$-estimators in Table~\ref{tab:rhoRAF}. For the Huber-like discrepancy of \citet[][\emph{Huber-RW01}]{ruckstuhl2001}, the tuning constants are set to $c_1 = -0.6$ and $c_2 = 0.6$ here.}
\label{fig:rhoRAF}
\end{figure}

Table~\ref{tab:rhoRAF} presents some discrepancy functions~$\rho$ along their unnormalized versions~$\rho_0$ in which they are often defined \citep[e.g.,][]{lindsay1994}; we normalize them to satisfy Condition~\ref{ass:rhonorm}. By Remark~\ref{rem:rho-invariance}, $C$-estimators are invariant to such normalizations. For maximum likelihood (ML), $\rho_\ML (\delta)=(\delta+1)\log(\delta+1)-\delta$ gives the Kullback--Leibler divergence \citep{csiszar1963}; its unbounded RAF $A(\delta)=\delta$ reflects the sensitivity of ML to contamination \citep[e.g.,][]{hampel1986}. The Hellinger and negative exponential discrepancies \citep{lindsay1994} grow more slowly as $\delta\to\infty$, yielding sublinear or bounded RAFs and attenuating the contribution of cells with large positive PRs. Figure~\ref{fig:rhoRAF} provides visualizations.

\begin{table}[t]
\centering
\small
\resizebox{\textwidth}{!}{%
\begin{tabular}{l |c| c| c | c }
Estimator & Unnormalized $\rho_0(\delta)$ & $\rho(\delta)$ & RAF $A(\delta)$ & $A'(\delta)$ %& $A''(x)$
\\\hline
Maximum likelihood & $(\delta+1)\log(\delta+1)$ & $(\delta+1)\log(\delta+1)-\delta$ & $\delta$ & $1$ %& 0
\\
Hellinger distance & $2(\sqrt{\delta+1}-1)^2$ & $2(\sqrt{\delta+1}-1)^2$ & $2(\sqrt{\delta+1}-1)$ 
& 
$\begin{cases}
(\delta+1)^{-1/2} & \text{if } \delta > -1,\\ \varnothing &\text{if } \delta = -1.
\end{cases}$
\\
Negative exponential & $\textnormal{e}^{-\delta}-1$ & $\textnormal{e}^{-\delta}-1+\delta$ & $2 - (2+\delta)\textnormal{e}^{-\delta}$ & $(\delta+1)\textnormal{e}^{-\delta}$ %& $-x\textnormal{e}^{-x}$
\\
Huber-RW01
&
$\rho_\RW(\delta) + \delta$
&
Eq.~\eqref{eq:rw01}
&
Eq.~\eqref{eq:rw01}
& 
$ \begin{cases} 1 &\text{if } \delta \in (c_1,c_2), \\ \varnothing &\text{if $\delta \in\{c_1,c_2\}$}, \\ 0 &\text{otherwise}. \end{cases}$ 
%&
%$ \begin{cases} \varnothing &\text{if $x \in\{c_1,c_2\}$}, \\ 0 &\text{otherwise}. \end{cases}$ 
\end{tabular}
}
\caption{Selected $C$-estimators, their unnormalized and normalized discrepancy functions, and the corresponding RAFs and derivatives. At $\delta=-1$, $A'$ is the right derivative and $``\varnothing"$ denotes non-existence. The Huber-RW01 cases apply to $\delta>-1$; $A'_+(-1)=\I{c_1=-1}$.}
\label{tab:rhoRAF}
\end{table}

For robust binomial estimation, \citet{ruckstuhl2001} proposed a \cite{huber1964}-like discrepancy. This particular
discrepancy function and its associated RAF are given by
\begin{equation}\label{eq:rw01}
	\resizebox{0.9\linewidth}{!}{$\displaystyle
	\rho_\RW(\delta) = 
	\begin{cases} 
	(\delta+1)(\log(c_1+1) + 1) - c_1-1-\delta &\text{if }{\delta < c_1},
	\\
	(\delta+1)\log(\delta+1) -\delta&\text{if } \delta \in [c_1, c_2], 
	\\ 
	(\delta+1)(\log(c_2+1) + 1) - c_2-1 -\delta &\text{if }{\delta > c_2},
	\end{cases}
	\ 
	A_\RW(\delta) =
	\begin{cases} c_1 &\text{if } \delta < c_1, \\ \delta &\text{if } \delta \in [c_1,c_2], \\ c_2 &\text{if }{\delta > c_2}, \end{cases}
	$}
\end{equation}
where $-1\leq c_1<0\leq c_2\leq\infty$ are prespecified clipping constants; the corresponding outer branch is absent when $c_1=-1$ or $c_2=\infty$. The RAF clips the PR to $[c_1,c_2]$, paralleling Huber's score function \citep{huber1964}. It is bounded when $c_2<\infty$ and nondifferentiable at finite interior clipping points, as permitted by Assumptions~\ref{ass:RAF} and~\ref{ass:rho}. Because it depends only on cell probabilities, the construction applies beyond the binomial setting of \citet{ruckstuhl2001}. Figure~\ref{fig:rhoRAF} provides an illustration. %In particular, the general asymptotic theory for $C$-estimators that we derive in Section~\ref{sec:properties} extends the asymptotics of \cite{ruckstuhl2001} beyond binomial models for this discrepancy function. 

\section{Theoretical properties of $C$-estimation}\label{sec:properties}
% We now derive the properties of unconditional $C$-estimators. Recall that we have a sample~$(\vec{Z}_i)_{i=1}^N$ of independent observations from~$\feps$ and wish to estimate the true parameter~$\Btheta_*$.

\subsection{Estimand}\label{sec:estimand}
The estimand of a $C$-estimator~$\Bthetahat$ is given by the parameter that minimizes the population divergence~$\Dsymbol$ between the sampling density~$\feps$ and the model~$\model$. Formally, with population PRs being defined by
$
	\PRzepsS{\Btheta} = \frac{\fepsz}{\pzS{\Btheta}} - 1
$, 
let~$\Btheta_0$ denote a population minimizer, so that
$
	\Btheta_0 \in \arg\min_{\Btheta\in\BTheta} \D{\feps}{\model}.
$ 
As such,~$\Btheta_0$ depends on the unobserved contamination PMF~$h(\cdot)$,~$\eps$, and~$\rho(\cdot)$. We now establish Fisher-consistency for the true~$\Btheta_*$.

\begin{proposition}\label{prop:fisher-consistency}
A $C$-estimator is Fisher-consistent, that is, $\Btheta_0 = \Btheta_*$ whenever $\varepsilon = 0$.
\end{proposition}

\subsection{Consistency}
Since $C$-estimators are divergence minimizers, define $\LN(\Btheta) \equiv \Dfp$ and $\Leps(\Btheta) \equiv \D{\feps}{\model}$ and refer to them as \emph{empirical loss} and \emph{population loss}, respectively. %Note that this notation omits the dependence on~$\rho$, but it is silently understood that a discrepancy function in the sense of Assumption~\ref{ass:rho} is used.
%For a finite~$\Z$, cellwise convergence $\PRznS{\Btheta}\convP\PRzepsS{\Btheta}$ and continuity of the finitely many summands imply $\LN(\Btheta)\convP \Leps(\Btheta)$. If~$\Z$ is infinite, cellwise PR convergence alone does not suffice for said loss convergence. The following theorem requires uniform convergence, which is sufficient for this purpose.
\begin{theorem}\label{thm:consistency}
Let $\Bthetahat$ be a $C$-estimator as in Definition~\ref{def:Cestimator}. Grant Assumptions \ref{ass:model}--\ref{ass:rho} and
\begin{enumerate}[label={\textnormal{C.\arabic*.}}, ref={\textnormal{C.\arabic*}}, itemindent=10pt]
	\item Uniform convergence of the loss functions: \label{ass:consistency-uniformconvergence}
	$
		\sup_{\Btheta\in\BTheta} \Big| \LN(\Btheta)- \Leps(\Btheta) \Big| \convP 0,
	$
	\item $\Leps$ has a well-separated minimum at $\Btheta_0$, i.e., for all $\alpha > 0$,\label{ass:consistency-cleanminimum}
	$
		\inf_{\Btheta\in\BTheta\setminus\mathcal{B}(\Btheta_0,\alpha)} \Leps(\Btheta) > \Leps(\Btheta_0),
	$
	where $\mathcal{B}(\Btheta_0,\alpha)$ is a closed $d$-ball with center~$\Btheta_0$ and radius~$\alpha$,
	\item $\Bthetahat$ is a near-minimizer of $\LN$: For $\zeta_N = \ohp{1}$,\label{ass:consistency-nearminimizer}
	$
		\LN\left(\Bthetahat\right) \leq \inf_{\Btheta\in\BTheta} \LN(\Btheta) + \zeta_N.
	$
\end{enumerate}
Then, $\Bthetahat$ is consistent for $\Btheta_0$, i.e., $\Bthetahat\convP\Btheta_0$ as $N\to\infty$.
\end{theorem}
Conditions \ref{ass:consistency-uniformconvergence}--\ref{ass:consistency-nearminimizer} are analogous to familiar conditions for establishing consistency of $M$-estimators in \citet[][Theorem~5.7]{vandervaart1998}. As such, they are stronger than necessary and can be weakened. For instance, if~$\BTheta$ is compact, condition~\ref{ass:consistency-cleanminimum} can be replaced by~$\Btheta_0$ being a unique global minimum of~$\Leps$. Condition~\ref{ass:consistency-uniformconvergence} can be replaced by  domination conditions on~$\Leps$ and compactness of~$\BTheta$ to obtain the ``classic'' conditions for a Wald-type consistency result; see \citet[][Chapter~5.2]{vandervaart1998} for a discussion.

\subsection{Asymptotic normality}\label{sec:asy-normality}
To derive asymptotic normality, we require additional notation. Let $\BTheta_0\subsetneq\BTheta$ be an open neighborhood of~$\Btheta_0$. Also, for $\Btheta\in\BTheta$, let 
$
	\matfuneps{U}{\Btheta} \equiv \var{\feps}{\fun{A'}{\PR{\Btheta}{\varepsilon}{\vec{Z}}} \sfun{\Btheta}{\vec{Z}} }
$
%\begin{align*}
%	\matfuneps{U}{\Btheta} 
%	&= \matfuneps{W}{\Btheta} \mat{\Omega}_\eps \matfuneps{W}{\Btheta}^\top
%	\\
%	&=
%	\sum_{\z\in\Z} \fepsz \big(A'(\PRzepsS{\Btheta})\big)^2 \szS{\Btheta}\szS{\Btheta}^\top
%	-
%	\left( \sum_{\z\in\Z} \fepsz A'(\PRzepsS{\Btheta}) \szS{\Btheta}  \right)
%	\left( \sum_{\z\in\Z} \fepsz A'(\PRzepsS{\Btheta}) \szS{\Btheta}  \right)^\top
%\end{align*}
denote the population covariance matrix of the ML score $\sfun{\Btheta}{\vec{Z}} \equiv \gradient\log(\pfun{\Btheta}{\vec{Z}})$ linearized by $\fun{A'}{\PR{\Btheta}{\varepsilon}{\vec{Z}}}$ for $\vec{Z}\sim\feps$. We have that $A'(\delta) = (\delta+1)\rho''(\delta)$. Further, the Hessian matrix of the population loss is given by
$
	\matfuneps{M}{\Btheta}
	\equiv
	\hessian \Leps(\Btheta) =
	\sum_{\z\in\Z} \bigg[ 
		 \fepsz \fun{A'}{\PRzepsS{\Btheta}} \szS{\Btheta} \szS{\Btheta}^\top - \fx{A}{\PRzepsS{\Btheta}}\hessian\pzS{\Btheta}
	\bigg].
$
Denote by $\matfunN{M}{\Btheta} = \hessian \LN(\Btheta)$ its empirical counterpart estimating~$\matfuneps{M}{\Btheta}$. Note that since the RAF may not be differentiable everywhere (Assumption~\ref{ass:RAF}), neither~$\matfuneps{U}{\Btheta}$ nor~$\matfuneps{M}{\Btheta}$ exists if at least one population PR~$\PRzepsS{\Btheta}$ is equal to a non-differentiability point of the RAF. We now state the asymptotic normality theorem, some remarks, and then discuss its assumptions. %This plays an important role in establishing asymptotic normality.

\begin{theorem}\label{thm:asy-normality}
Let $\Bthetahat$ be a $C$-estimator as in Definition~\ref{def:Cestimator}. Grant Assumptions \ref{ass:model}--\ref{ass:rho} and the following conditions:
\begin{enumerate}[label={\textnormal{C.\arabic*.}}, ref={\textnormal{C.\arabic*}}, itemindent=10pt]
\setcounter{enumi}{3}
	\item Interior estimand: $\Btheta_0\in\BTheta_0\subset \textnormal{int}(\BTheta)$. \label{ass:interior}
	\item Consistency of the estimator: $\Bthetahat \convP\Btheta_0$. \label{ass:consistency}
	\item $\Bthetahat$ is a near-zero of the empirical loss gradient: $\gradient\LN\left(\thetahat\right) = \ohp{N^{-1/2}}$.\label{ass:nearzero}
		\item Local properties of the population Hessian: The map $\Btheta\mapsto\matfuneps{M}{\Btheta}$ is well-defined on~$\BTheta_0$ and continuous as well as positive definite at~$\Btheta_0$. \label{ass:Mposdef}
	\item Local uniform convergence of~$\mat{M}_N$:  $\sup_{\Btheta\in\BTheta_0}\matnorm{ \matfunN{M}{\Btheta} - \matfuneps{M}{\Btheta} } \convP 0$.\label{ass:Muniform}
	\item Square-integrability of the linearized score:
	$
	\E{\feps}{\left\|A'\big(\PR{\Btheta_0}{\varepsilon}{\vec Z}\big)
	\sfun{\Btheta_0}{\vec Z}\right\|^2}<\infty.
	$ 
	In particular, $\matfuneps{U}{\Btheta_0}$ exists and has finite norm. \label{ass:MUfinite}
	\item $\fun{A'}{\PRzepsS{\Btheta_0}}$ exists for all $\z\in\Z$, i.e., all PRs are RAF points of differentiability.\label{ass:PRRAF}
	\item There is a neighborhood~$\BTheta_1\subseteq\BTheta_0$ of~$\Btheta_0$ on which the series defining~$\Leps(\Btheta)$ and its first two derivatives converge absolutely and
locally uniformly, and in both derivatives summation over~$\Z$ may be interchanged with differentiation.\label{ass:countable-interchange}
	\item  For 
	$
		\vecfunN{\Delta}{\Btheta} = \sum_{\z\in\Z} \gradient \pz{\Btheta}\big[ \fun{A}{\PRzn{\Btheta}} - \fun{A}{\PRzeps{\Btheta}} - 
		\fun{A'}{\PRzeps{\Btheta}}\left(\PRzn{\Btheta} - \PRzeps{\Btheta} \right) \big]
	$, 
	it holds true that $\|\vecfunN{\Delta}{\Btheta_0}\|<\infty$ almost surely, as well as $\sqrt{N}\|\vecfunN{\Delta}{\Btheta_0}\|\convP 0$. \label{ass:countable-remainder}
\end{enumerate}
Then, as $N\to\infty$, the following asymptotic linearization holds true:
\[
	\sqrt{N}\left(\Bthetahat - \Btheta_0\right) 
	=
	\big(\matfuneps{M}{\Btheta_0}\big)^{-1}\frac{1}{\sqrt{N}}
	\sum_{i=1}^N\left[\sum_{\z\in\Z} \big( I_i(\z) - \fepsz \big) \fun{A'}{\PRzepsS{\Btheta_0}} \szS{\Btheta_0}\right] + \ohp{1},
\]
where $I_i(\z) = \I{\vec{Z}_i = \z}$. In particular,
$
	\sqrt{N}\left(\Bthetahat - \Btheta_0 \right) \convweak \gauss_d\Big(\vec{0}, \matfuneps{\Sigma}{\Btheta_0}\Big),
$
where 
$
	\matfuneps{\Sigma}{\Btheta_0} = \matinveps{M}{\Btheta_0}\matfuneps{U}{\Btheta_0}\matinveps{M}{\Btheta_0}
$ 
is positive definite under the additional assumption that $\matfuneps{U}{\Btheta_0}$ is positive definite. 
\end{theorem}

\begin{remark} \label{rem:Sigmahat}
	An estimator of~$\matfuneps{\Sigma}{\Btheta_0}$ can be constructed as follows. Replace~$\feps$ by~$\fhat$ in~$\matfuneps{U}{\Btheta}$ and denote the ensuing matrix by~$\matfunN{U}{\Btheta}$. Under the additional condition $\sup_{\Btheta\in\BTheta_0}\matnorm{\matfunN{U}{\Btheta}-\matfuneps{U}{\Btheta}}\convP 0$, Conditions~\ref{ass:consistency},~\ref{ass:Mposdef}, and~\ref{ass:Muniform} together with continuous mapping imply 
$
		\matfunN{\Sigma}{\Bthetahat}
		=
		\matinvN{M}{\Bthetahat}\matfunN{U}{\Bthetahat}\matinvN{M}{\Bthetahat}
		\convP
		\matfuneps{\Sigma}{\Btheta_0}.
$
\end{remark}

\begin{remark}
If $A'(0)$ exists, then at the postulated model ($\varepsilon = 0$) the asymptotic covariance matrix~$\matfuneps{\Sigma}{\Btheta_0}$ equals the inverse Fisher information; see Lemma~\ref{lem:IFlimits}. Hence, $C$-estimators are fully first-order efficient at the postulated model under the theorem's assumptions.
\end{remark}

\begin{remark}
	The construction of~$\matfuneps{\Sigma}{\Btheta}$ is analogous to the sandwich covariance of \cite{white1982}. It reduces to the usual misspecified-ML sandwich for the ML discrepancy, whereas other discrepancy functions generally produce different bread and meat matrices if $\eps>0$.
\end{remark}

Conditions~\ref{ass:interior}--\ref{ass:Mposdef} in Theorem~\ref{thm:asy-normality} are analogous to standard conditions for asymptotic normality of $M$-estimators \citep[e.g.,][Theorem~5.23]{vandervaart1998}. Conditions~\ref{ass:MUfinite}--\ref{ass:countable-remainder} control the linearized score and the possibly infinite sums specific to the present discrepancy formulation. Condition~\ref{ass:MUfinite} is sufficient for finiteness of the ``meat'' matrix in the sandwich-type asymptotic covariance matrix. Condition~\ref{ass:PRRAF} ensures that the RAF is differentiable at the population PRs relevant in the asymptotic linearization. Condition~\ref{ass:countable-interchange} allows for interchanging differentiation and summation in the matrices~$\matfuneps{M}{\Btheta_0}$ and~$\matfuneps{U}{\Btheta_0}$. The random vector~$\vecfunN{\Delta}{\Btheta_0}$ adds up remainder terms from Taylor expansions in the proof: All its summands individually are of order~$\ohp{N^{-1/2}}$, and the condition ensures that also a sum over~$\Z$ of such terms vanishes at this rate.  If~$\Z$ is finite, Conditions \ref{ass:countable-interchange} and~\ref{ass:countable-remainder} reduce to the ordinary differentiability and finite-sum Taylor argument already implicit in the other conditions. If~$\Z$ is countably infinite, both conditions must be verified in a concrete model, for example by truncating the summation to increasing finite subsets of~$\Z$ and supplying summable envelopes for the remainder~$\vec{\Delta}_N(\Btheta_0)$, loss, gradient, and Hessian summands that make the summands in the tail uniformly negligible. %\footnote{For instance, let~$\Z_m\uparrow\Z$ be finite, and let~$\mathcal \vec{\Delta}_{N,m}^{\mathrm{tail}}(\Btheta_0)$ be defined as~$\vec{\Delta}_N(\Btheta_0)$ in Condition~\ref{ass:countable-remainder}, but with the sum restricted to~$\Z\setminus\Z_m$. Condition~\ref{ass:countable-remainder} follows if, for every~$\eta>0$,
%$
%\lim_{m\to\infty}\limsup_{N\to\infty}
%\Pr{}{\sqrt N\left\|\mathcal \vec{\Delta}_{N,m}^{\mathrm{tail}}(\Btheta_0)\right\|>\eta}=0.
%$
%Indeed, the assumed differentiability of~$A$ makes the corresponding sum over each fixed~$\Z_m$ equal to~$\ohp{N^{-1/2}}$.}

If $\fun{A'}{\PRzepsS{\Btheta_0}}$ does not exist for at least one $\z\in\Z$, Condition~\ref{ass:PRRAF} fails and Theorem~\ref{thm:asy-normality} is inapplicable because a crucial Taylor expansion in the proof cannot be performed. No converse conclusion follows: the estimator may have a Gaussian, non-Gaussian, or set-valued limit, and this must be determined by a separate nonsmooth analysis. For example, in the saturated Bernoulli model $p_\theta(1)=\theta$, the empirical PMF~$\fhat$ belongs to the postulated model with probability tending to one. Proposition~\ref{prop:divergence-nonnegative} then makes every $C$-estimator equal to the sample proportion, which is asymptotically normal for any discrepancy. For the Huber-like discrepancy in~\eqref{eq:rw01}, Theorem~\ref{thm:asy-normality} applies only when no population PR at the estimand equals either clipping point~$c_1$ or~$c_2$. In particular, this makes the theorem inapplicable for that discrepancy with $c_2=0$ in a correctly specified model $(\varepsilon = 0)$. This property has been described by \cite{ruckstuhl2001} for the binomial model.

\begin{remark}\label{rem:A'notexist}
Potential non-differentiability also affects the plug-in covariance formula for~$\mat{\Sigma}_N$ in Remark~\ref{rem:Sigmahat}. For example, the Hellinger RAF is not differentiable at an empty cell, where the empirical PR equals~$-1$. Thus empty cells in finite samples may make the raw formula undefined or unstable. If~$\Z$ is countably infinite, empty cells occur at every finite sample size, so this Hellinger plug-in covariance as stated is inapplicable.
\end{remark}

\subsection{Influence function}
Recall that the ML estimator (MLE) is obtained by choosing as discrepancy function $\rho_\ML(\delta) = (\delta+1)\log(\delta+1)-\delta$. The influence function (IF) of an MLE~$\BthetahatMLE$ at $\z\in\Z$ is given by
$
	\IF{\z, \BthetahatMLE, p_{\Btheta_*}}
	=
	\matinv{J}{\Btheta_*}\sz{\Btheta_*},
$
where $\matfun{J}{\Btheta} = \sum_{\z\in\Z}\pz{\Btheta}\sz{\Btheta}\sz{\Btheta}^\top$ denotes the Fisher information matrix of model~$\model$. We now establish the IF of $C$-estimators. %The IF describes the bias caused by an infinitesimally small contamination on an estimate \citep{markatou1997}. 
%The following theorem states the IF of $C$-estimators. 

\begin{theorem}\label{thm:IF}
Grant the assumptions of Theorem~\ref{thm:asy-normality}. Then, under additional regularity conditions ensuring local directional differentiability, $\IF{\z, \Bthetahat, p_{\Btheta_*}}= \IF{\z, \BthetahatMLE, p_{\Btheta_*}}$.
\end{theorem}

The additional regularity conditions are stated in the theorem's proof in the appendix.

\begin{remark}
	Theorem~\ref{thm:IF} makes no statement on the IF when $A'(0)$ does not exist (i.e., Condition~\ref{ass:PRRAF} in Theorem~\ref{thm:asy-normality} fails). In that case, the covariance formula in Theorem~\ref{thm:asy-normality} does not apply either, and a separate nonsmooth analysis is required.
\end{remark}

Whenever the conditions of Theorem~\ref{thm:IF} hold, the discrepancy functions in Table~\ref{tab:rhoRAF} yield the same IF as ML, except that the theorem is inapplicable to the Huber-like discrepancy of \cite{ruckstuhl2001} when $c_2=0$, because its RAF is not differentiable at~0. Overall, the fact that the IF of $C$-estimators under the conditions of Theorem~\ref{thm:IF} is equal to the IF of the non-robust MLE suggests that the IF may not be suitable as robustness measure. We therefore consider a different robustness criterion, namely maximum bias.

\section{Maximum bias of $C$-estimators}\label{sec:maxbias}

The IF measures the effect of infinitesimal contamination. We now quantify robustness against a fixed contamination fraction~$\eps$. Let $\mathcal{P}(\Z)$ be the set of all PMFs on~$\Z$ and let 
$
	\Ceps = \big\{f: f = (1-\eps)p_{\Btheta_*} +\eps  h,\ h\in\mathcal{P}(\Z)\big\}
$
denote the contamination neighborhood of the true model~$p_{\Btheta_*}$, which collects all PMFs satisfying the Huber-like contamination model in~\eqref{eq:hubermodel} at a fixed $\eps\in [0, 0.5)$.  Define the set-valued population $C$-functional at a PMF $f\in\Ceps$ as
$
	\cT{f}
	=
	\arg\min_{\Btheta\in\BTheta}\D{f}{\model},
$
which is the population counterpart of a $C$-estimator. Set-valued notation guards against nonunique population minima. Throughout this section, assume that $\cT{f}\neq\emptyset$ for every $\eps\in[0,0.5)$ and every $f\in\Ceps$.

\begin{definition}[Maximum-bias curve (\citeauthor{huber2009}, \citeyear{huber2009},~p.12, adapted)]\label{def:maxbias}
For $\eps\in[0,0.5)$, the maximum-bias (maxbias) curve of the $C$-functional at~$p_{\Btheta_*}$ is
\begin{equation}\label{eq:maxbias-definition}
	\eps\mapsto\Beps
	=
\sup_{f\in\Ceps}
\sup_{\Btheta\in\cT{f}}
\|\Btheta-\Btheta_*\|.
\end{equation}
%While $\|\cdot\|$ is the Euclidean norm throughout, any meaningful metric may be used.
\end{definition}

The maximum-bias curve measures the worst possible bias with respect to the true parameter that a $C$-estimator can incur in contamination model~\eqref{eq:hubermodel} at a given contamination fraction. 
The following result gives an exact optimization representation of lower and upper bounds (i.e., envelopes) for the maximum-bias curve that are directly computable from the postulated model. 
For two PMFs $g_1,g_2$ on $\Z$, define their two-center divergence radius on~$\Ceps$ as
$
	\Gammarho{g_1,g_2}
	=
	\inf_{f\in\Ceps}
	\max\big\{\D{f}{g_1}, \D{f}{g_2}\big\} 
$
and, for fixed $\eps\in[0,0.5)$, denote the maximum divergence with respect to~$\modelt$ on~$\Ceps$ by
$
	\Rrho = \sup_{f\in\Ceps} \D{f}{\modelt}.
$

\begin{theorem}\label{thm:maxbias-general}
Grant Assumptions~\ref{ass:model}--\ref{ass:rho} and let $\eps\in[0, 0.5)$ be fixed. Then, the maximum divergence is characterized by
\begin{equation}\label{eq:maxbias-maxdiv}
	\Rrho = \sup_{\z\in\Z}\left[ \pz{\Btheta_*} \fun{\rho}{\frac{\eps}{\pz{\Btheta_*}}-\eps} + (1-\pz{\Btheta_*})\rho(-\eps) \right].
\end{equation}
If $\Z$ is finite, $\Rrho$ is attained by the particular point-mass contaminated PMF $\feps^{\Delta_{\z}}=(1-\eps)\modelt+\eps\Delta_{\z}$ that maximizes $\D{\feps^{\Delta_{\z}}}{\modelt}$ over $\z\in\Z$, where $\Delta_{\z}(\y)=\I{\y=\z}$. If~$\Z$ is countably infinite, then~$\Rrho$ is a supremum approached by an infinite sequence of such point-mass contamination PMFs.

Moreover, for $\epscrit{\Btheta} = 1 - \inf_{\z\in\Z} \frac{\pz{\Btheta}}{\pz{\Btheta_*}}$, we have the maximum-bias envelope
\begin{equation}\label{eq:maxbias-sandwich}
	\sup_{ \{\Btheta\in\BTheta : \epscrit{\Btheta} \leq \eps \} } \|\Btheta-\Btheta_*\|
	\leq
	\Beps
	\leq
	\sup_{ \left\{\Btheta\in\BTheta : \Gammarho{\modelt,\model} \leq \Rrho \right\} } \|\Btheta-\Btheta_*\|.
\end{equation}
All suprema are understood in the extended-real-line sense.
\end{theorem}

\begin{remark}
	$\epscrit{\Btheta}$ is the smallest contamination fraction for which a given model PMF~$\model$ can be represented as a contaminated PMF satisfying~\eqref{eq:hubermodel}. Specifically, if $\epscrit{\Btheta} \leq \eps$, then there exists a $h\in\mathcal{P}(\Z)$ such that $\model = (1-\eps)\modelt + \eps h \in\Ceps$. The divergence with respect to~$\modelt$ is then minimized at~$\Btheta$, yielding a zero-valued divergence and thereby providing a lower bound for the maximum bias. %Furthermore, the bounds in~\eqref{eq:maxbias-sandwich} remain unchanged if the discrepancy~$\rho$ is multiplied by a positive constant.
\end{remark}

The bounds in the preceding theorem are related to the gauge bounds of \cite{donoho1988} for minimum-distance functionals. When the contamination neighborhood is a ball of radius~$r$ and the same metric for constructing this ball is used for fitting, their Proposition~4.1 bounds the maxbias between the model gauge evaluated at~$r$ and~$2r$. Our setting separates the two geometries: Contamination is described by the Huber-like contamination neighborhood~$\Ceps$, whereas fitting is based on the possibly asymmetric and nonmetric divergence~$\Dsymbol$. The quantity~$\Rrho$ converts the contamination fraction into a divergence~$\Dsymbol$, and~$\Gamma_\rho$ replaces the metric midpoint construction.

We now specialize the upper bound in Theorem~\ref{thm:maxbias-general} to the Huber-like discrepancy in~\eqref{eq:rw01} in a nontrivial corollary. Define the normalized total variation (TV) distance between two PMFs $g_1,g_2$ on~$\Z$ as
$
	\TV{g_1,g_2} = \frac{1}{2}\sum_{\z\in\Z} |g_1(\z) - g_2(\z)|,
$
which takes values in $[0,1]$.

\begin{corollary}\label{cor:maxbias-RW}
Let $\rho_\RW$ be the Huber-like discrepancy function in~\eqref{eq:rw01} with fixed constants $-1\leq c_1 < 0 \leq c_2 \leq \infty$. Define $\pmin = \inf_{\z\in\Z}\pz{\Btheta_*}$ and fix $\eps\in [0, 0.5)$.  
If $\Z$ is finite, then $\pmin >0$ and the maximum divergence in Theorem~\ref{thm:maxbias-general} equals
\[
	\RRW = \pmin \fun{\rho}{\frac{\eps}{\pmin}-\eps} + (1-\pmin)\rho(-\eps). 
\]
If the sample space~$\Z$ is countably infinite, the maximum divergence equals
\[
	\RRW = \eps \log(c_2 + 1) + \rho(-\eps),
\] 
which, for $c_2 = \infty$, is understood as $\infty$ (i.e., unbounded) if $\eps > 0$ and zero if $\eps=0$. 

Moreover, define
$
	\kappa_- = \min\left\{\frac{1}{2}, -c_1\right\},
	%\qquad
	\kappa_+ = \min\left\{\frac{1}{2}, \log(c_2+1)\right\},
	%\qquad
	\kappac = \kappa_-+\kappa_+.
$
For an at most countable~$\Z$, the upper bound in~\eqref{eq:maxbias-sandwich} becomes 
\[
	\BRW \leq \sup_{\left\{ \Btheta\in\BTheta : \TV{\model,\modelt} \leq \Eeps{c_1,c_2} \right\}} \|\Btheta - \Btheta_*\|,
\] 
with the TV radius being defined by
$
	\Eeps{c_1,c_2} = \min\left\{ 1, 2 \sqrt{\RRW /\kappac} \right\}.
$
\end{corollary}

Owing to their generality, the upper bounds in Theorem~\ref{thm:maxbias-general} and Corollary~\ref{cor:maxbias-RW} are typically not sharp. To tighten them, one needs to impose restrictions on the postulated model. In particular, once the TV radius reaches~1 in the corollary, the right-hand side reduces to $\sup_{\Btheta\in\BTheta}\|\Btheta-\Btheta_*\|$, which is uninformative and is infinite when the parameter space~$\BTheta$ is unbounded. Nevertheless, until this point, the bounds represent a useful quantification of the enhanced robustness of $C$-estimators with a Huber-like discrepancy. 

\begin{figure}[t]
    \centering
    \includegraphics[width = 0.99\textwidth]{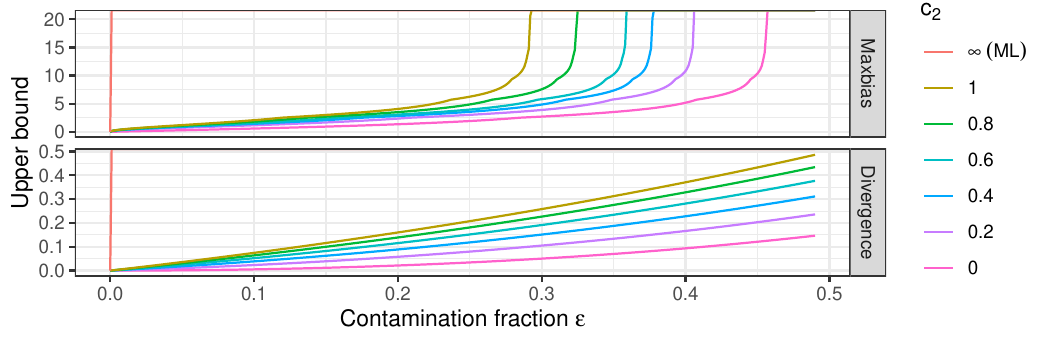}
\caption{Maxbias upper bound and maximum divergence $\RRW$ from Corollary~\ref{cor:maxbias-RW} of the Ruckstuhl-Welsh $C$-estimator with~$c_1=-1$ and varying~$c_2$ at the Poisson model with true $\lambda_* = 1$ and support $\Z=\{0,1,2,\dots\}$. Lines outside the vertical range indicate infinity.} %the trivial TV radius of~1 is reached, at which the bound is $+\infty$.
\label{fig:maxbias}
\end{figure}

Figure~\ref{fig:maxbias} illustrates Corollary~\ref{cor:maxbias-RW} for a Poisson model and tuning constant choice $c_1=-1$, as recommended by \citet{ruckstuhl2001}. The ML choice $c_2=\infty$ has infinite maximum divergence, whereas finite~$c_2$ gives finite maximum divergence and, before the TV radius reaches~1, a nontrivial maximum-bias bound. The bound becomes no larger as~$c_2$ is decreased. The following result formalizes this comparison of tuning constants.

\begin{corollary} \label{cor:maxbias-increases-RW}
Consider the setting of Corollary~\ref{cor:maxbias-RW} where~$\rho_\RW$ is the Huber-like discrepancy. Set $c_1=-1$, fix $\eps\in(0,0.5)$, and write $\Eeps{c_2}=\Eeps{-1,c_2}$. Define the resulting maxbias envelope by
$
	\overline B_\eps(c_2)
	=
	\sup_{\left\{\Btheta\in\BTheta:\,
	\TV{\model,\modelt}\leq\Eeps{c_2}\right\}}
	\|\Btheta-\Btheta_*\|.
$
\begin{enumerate}
	\item If $\Z$ is countably infinite, then both $\Eeps{c_2}$ and $\overline B_\eps(c_2)$ are nondecreasing in~$c_2$.
	\item If $\Z$ is finite with $|\Z|\geq2$, the same conclusion holds whenever
	\begin{equation}\label{eq:pcrit}
		\pmin \leq \pcrit
		\equiv
		\frac{-2(1-\eps)\log(1-\eps)-\eps}
		{(1-\eps)-2(1-\eps)\log(1-\eps)}.
	\end{equation}
	Here $\pcrit\in(0,1)$, and $|\Z|\geq1/\pcrit$ is sufficient for~\eqref{eq:pcrit}.
\end{enumerate} 
\end{corollary}

Corollary~\ref{cor:maxbias-increases-RW} provides a robustness guarantee for tuning: decreasing~$c_2$ cannot enlarge the displayed maximum-bias envelope under its conditions. Proposition~\ref{prop:RW-radius-detailed} in the appendix nests this corollary, gives the exact ranges of strict increase and saturation, and provides a necessary-and-sufficient finite-support characterization.

\section{Related approaches}\label{sec:literature}
Minimum-disparity estimators \citep[MDEs;][]{lindsay1994,markatou1997} provide the divergence-minimization foundation of $C$-estimation. We develop this foundation within a common categorical-data framework by admitting clipped, nonsmooth discrepancies such as that of \citet{ruckstuhl2001} and stating conditions under which standard asymptotic inference remains valid; applying the same cellwise construction to unconditional, composite, and conditional models; and deriving computable maximum-bias envelopes that link the clipping constants to a global robustness guarantee. The latter results adapt the gauge-bound logic of \citet{donoho1988} to categorical Huber-type contamination and possibly asymmetric, nonmetric divergences. Conditional disparity methods \citep{hooker2016} also address regression; our conditional construction specializes the cellwise principle to categorical responses and covariates and embeds it in the same framework as unconditional and composite estimation.

Other robust procedures have been developed for particular classes of categorical models. \citet{simpson1987Mestimator} study $M$-estimation for numerical discrete data. Minimum density power divergence estimation \citep{basu1998} also applies without a numerical ordering of the cells. The MGP estimator of \citet{victoriafeser1997} is designed for grouped data. These procedures generally trade some first-order efficiency for robustness; regular $C$-estimators, by contrast, are first-order equivalent to ML at a correctly specified model. \citet{shane2001} propose least-median-of-squares and least-trimmed-squares estimators for contingency tables but do not derive asymptotic properties, and \citet{rudas1994} develop a goodness-of-fit measure under a mixture model that contains~\eqref{eq:hubermodel} as a special case. We next extend $C$-estimation beyond unconditional joint models.

%Most closely related to $C$-estimation are  \emph{minimum disparity estimators} \citep[MDEs,][]{markatou1997,lindsay1994,simpson1987mde}, which are contained in the class of~$C$-estimators. It is worth mentioning that \emph{minimum power divergence estimators} \citep{cressie1984} can be expressed as MDEs. MDEs are efficient at the postulated model and have the same functional form as $C$-estimators in~\eqref{eq:estimatingeq}. However, there are two key differences. 
%First, MDEs require the $\rho$-function to be thrice differentiable and the RAF to be twice differentiable. As such, MDEs rule out $\rho$-functions such as the one of \citet[][p.~1124]{ruckstuhl2001} in~\eqref{eq:rw01}. As seen in Section~\ref{sec:properties}, allowing for such non-differentiability is more than a mere technicality because it has major consequences for the asymptotics since it may prevent asymptotic normality and covariance estimation. Second, the conditions in the asymptotic analysis are different. In particular, by separating consistency from asymptotic normality, the conditions of Theorem~\ref{thm:asy-normality}  mimic those of $M$-estimation and avoid the \emph{``fairly strong''} assumption \citep[][p.~1107]{lindsay1994} of MDEs that $A'(\delta)$ and $A''(\delta)\delta$ are bounded.

%Also related are \cite{baraud2017}, who use Hellinger distances for regression, and \cite{hooker2016}, who studies MDEs for regression problems using kernel densities.  

\section{Composite $C$-estimation}\label{sec:cmp}
Estimators of models for categorical data are often computationally intractable in higher dimensions. An example is the SEM in Example~\ref{ex:factor}, which requires solving~$p$ integrals. To still be able to fit such models, composite likelihood approaches are popular \citep[e.g.,][]{lindsay1988}, where the intractable joint model is broken up into a sum of smaller submodels, called a \emph{composite} model. $C$-estimation can be used to robustly estimate composite models.

\subsection{Setup}
Consider the setup of Section~\ref{sec:Cestimator}. Suppose the statistician postulates a composite model comprising~$K$ marginal submodels, where a $k$-th submodel~$\modelk$ is a model for a subset~$\Ak$ of the~$p$ variables so that~$\modelkbeta$ is a marginal PMF of~$\model$ with respect to the variables in~$\Ak$. Each submodel~$\modelkbeta$ depends on subvectors~$\Bthetak$ of~$\Btheta\in\BTheta$; note that $\modelkbeta = \modelk$. Denote by $\BZk = (Z_j)_{j\in\Ak}$ a $|\Ak|$-dimensional subvector of~$\vec{Z}$ holding the variables of the $k$-th submodel. The subvector~$\BZk$ takes values in~$\Zk$, which is the $k$-th submodel's sample space. The sets $\Zk$,~$\Ak$, and parameter space for~$\Bthetak$ may overlap across the submodels.
\begin{assumption}\label{ass:submodel}
	For every $k=1,\dots,K$ and $\Btheta\in\BTheta$, $\modelk$ is a strictly positive PMF on~$\Zk$, and each map $\Btheta\mapsto\pzk{\Btheta}$ is twice continuously differentiable. Furthermore, the collection of submodels is globally identifying: for any $\Btheta_1,\Btheta_2\in\BTheta$, if
		$p_{\Btheta_1}^{(k)}=p_{\Btheta_2}^{(k)}$ for every $k=1,\dots,K$, then $\Btheta_1 = \Btheta_2.$
\end{assumption}

\begin{example}\label{ex:cmp}
	Consider the SEM from Example~\ref{ex:factor}. As a computationally leaner alternative, \cite{katsikatsou2012} propose estimating this model through composite ML via pairwise submodels. A $k$-th submodel, $k=1,\dots,K$, models a pair of distinct items  $\Ak = \{i_k,j_k\}$ through bivariate standard normality,
\[
	\pzk{\Btheta} = \Pr{\Btheta}{ Z_{i_k} = a, Z_{j_k} = b }
	=
	\int^{\tau^{(i_k)}_a}_{\tau^{(i_k)}_{a-1}}
	\int^{\tau^{(j_k)}_b}_{\tau^{(j_k)}_{b-1}}
	\tilde{\phi}_2\big(t,s;\varrho_{i_kj_k}(\Btheta)\big)\d s\d t,
\]
where $\zk = \left(a,b\right)\in\Zk = \{ (a,b) : a = 1,\dots,m_{i_k}, b = 1,\dots, m_{j_k} \}$ and $\tilde{\phi}_2(\cdot,\cdot;\varrho_{i_kj_k}(\Btheta))$ denotes the bivariate standard normal density with correlation coefficient $\varrho_{i_kj_k}(\Btheta) = \Blambda_{i_k}^\top\BSigmaF\Blambda_{j_k}$. Each submodel is on a unique item pair so that there are $K = \binom{p}{2}$ submodels. The ensuing composite log-likelihood is a sum over the~$K$ submodel log-likelihoods. 
\end{example}

\subsection{Estimation and inference}
Given a sample $\vec{Z}_1,\dots,\vec{Z}_N$ of independent copies of~$\vec{Z}$, denote by
$
	\fhatk\left(\zk\right) = N^{(k)}_{\zk} / N = \frac{1}{N}\sum_{i=1}^N \I{\vec{Z}_i^{(k)} = \zk},  \zk\in\Zk,
$
the empirical relative frequency function associated with the $k$-th submodel, $k=1,\dots,K$. If the submodel is correctly specified for all observations, then~$\fhatk$ pointwise converges in probability to~$p^{(k)}_{\Btheta_*}$, where $\Btheta_*\in\BTheta$  denotes the true parameter vector. However, if contamination is present, then this convergence may fail. Thus, we can robustify composite estimation by downweighting PRs on the submodel-level. 
Specifically, given a discrepancy function~$\rho$, the composite loss reads
\[
	\LN^\cmp(\Btheta) = \sum_{k=1}^K w_k\cdot \D{\fhatk}{\modelk},
\]
where the $w_k>0$ are known weights, and~$\Dsymbol$ is the divergence measure defined in~\eqref{eq:divergence}. For the choice $\rho_\ML(\delta)=(\delta+1)\log(\delta+1)-\delta$, minimizing $\LN^\cmp(\Btheta)$ is equivalent to maximizing the  composite log-likelihood $\mathcal{L}_N^{\textnormal{cmp-logl}}(\Btheta) = \sum_{k=1}^Kw_k \sum_{\zk\in\Zk}N^{(k)}_{\zk}\log\left(\pzk{\Btheta}\right)$.
\begin{definition}[Composite $C$-estimator]
Suppose the $\rho$-function in~$\LN^\cmp$ satisfies Assumption~\ref{ass:rho} and the submodels satisfy Assumption~\ref{ass:submodel}. A composite $C$-estimator is any parameter~$\Bthetahat^\cmp$ that globally minimizes~$\LN^\cmp(\Btheta)$ with respect to $\Btheta\in\BTheta$. If~$\Bthetahat^\cmp$ is an interior point and~$\LN^\cmp(\Btheta)$ is differentiable at it, then~$\Bthetahat^\cmp$ is necessarily a root of~$\gradient\LN^\cmp(\Btheta)$.
\end{definition}

Like in unconditional $C$-estimation, we assume the Huber-type contamination model from~\eqref{eq:hubermodel} so that~$\vec{Z}$ is distributed according to~$\fepsz = (1-\eps)\pz{\Btheta_*} + \eps h(\z), \z\in\Z$. Consequently, the variables being modeled by the $k$-th submodel are distributed according to $\fepsk\left(\zk\right) = (1-\eps)\pzk{\Btheta_*} + \eps h^{(k)}\left(\zk\right), \zk\in\Zk$, where $h^{(k)}$ is the marginal PMF of~$h$ with respect to the variables in~$\Ak$, $k=1,\dots,K$. 
It follows that the estimand of~$\Bthetahat^\cmp$ is given by the parameter~$\Btheta_0^\cmp$ minimizing the population composite loss $\Leps^\cmp(\Btheta) = \sum_{k=1}^K w_k\cdot \D{\fepsk}{\modelk}$ over $\Btheta\in\BTheta$. For further reference, denote by $\PRzepsk{\Btheta} = \fepsk (\zk)/ \pzk{\Btheta}-1, \zk\in\Zk$ the submodel population PRs of the $k$-th submodel, and let $\skfun{\Btheta}{k}{\zk} = \gradient \log\left(\pzk{\Btheta}\right)$ be the log-likelihood scores. 

\begin{proposition}\label{prop:fisher-consistency-cmp}
Composite $C$-estimators are Fisher-consistent, that is, $\Btheta_0^\cmp = \Btheta_*$ whenever $\eps = 0$, in which case~$\Leps^\cmp(\Btheta)$ attains its global minimum of~$0$ at~$\Btheta_0^\cmp$.
\end{proposition}

%%%%%%%%%%%%%%%%%%%%%%%%%%%%%%%%%%%%%%%
\begin{theorem}\label{thm:cmp-normality-reduced}
Grant Assumptions~\ref{ass:rho} and~\ref{ass:submodel}. Under mild regularity conditions,
$
	\Bthetahat^\cmp\convP\Btheta_0^\cmp,
$
as $N\to\infty$. 
Under additional regularity conditions, in particular that $\fun{A'}{\PRzepsk{\Btheta_0^\cmp}}$ exists for all $\zk\in\Zk$ and all submodels $k=1,\dots,K$, it furthermore holds true that
$
		\sqrt{N}\left(\Bthetahat^\cmp - \Btheta_0^\cmp\right)
		\convweak\gauss_d
		\big(
					\vec{0},
					\mat{\Sigma}_\eps^\cmp\left(\Btheta_0^\cmp\right)
		\big)
$ 
as $N\to\infty$, 
where $\matepscmp{\Sigma}{\Btheta} = \left(\matepscmp{M}{\Btheta}\right)^{-1} \matepscmp{U}{\Btheta} \left(\matepscmp{M}{\Btheta}\right)^{-1}$, with $\matepscmp{M}{\Btheta} = \hessian \Leps^\cmp(\Btheta)$, $\matepscmp{U}{\Btheta} = \sum_{k=1}^K\sum_{l=1}^K w_kw_l\matepsk{U}{\Btheta}{k,l}$, and
\begin{equation*}
\begin{split}
		\matepsk{U}{\Btheta}{k,l}
		=
		\cov{(\BZk, \vec{Z}^{(l)})\sim\feps^{(k,l)}}{ \fun{A'}{\delta^{(k)}_{\Btheta,\eps}\left(\BZk\right)}  \skfun{\Btheta}{k}{\BZk} ; \fun{A'}{\delta^{(l)}_{\Btheta,\eps}\left(\vec{Z}^{(l)}\right)} \skfun{\Btheta}{l}{\vec{Z}^{(l)}} },
\end{split}
\end{equation*}
where $\feps^{(k,l)}$ is the marginal distribution of~$\feps$ with respect to the variables in $\Ak\cup\mathcal{A}_l$, and $\cov{}{\vec{A}; \vec{B}}$ denotes the cross-covariance matrix of two $d$-variate random vectors~$\vec{A}$ and~$\vec{B}$. 
\end{theorem}
The regularity conditions in Theorem~\ref{thm:cmp-normality-reduced} are specified in Appendix~\ref{app:cmp}. They adapt Conditions~\ref{ass:consistency-uniformconvergence}--\ref{ass:countable-remainder} to composite models. We further describe in Appendix~\ref{app:cmp} how to construct a pointwise consistent estimator~$\hat{\mat{\Sigma}}_N^\cmp$ for~$\mat{\Sigma}_\eps^\cmp$. 

\begin{remark}
In the zero-contamination case ($\eps=0$), if $A'(0)$ exists, then~$\matepscmp{\Sigma}{\Btheta_0^\cmp}$ equals the asymptotic covariance matrix of the composite ML estimator. Consequently, smooth composite $C$-estimators suffer no first-order efficiency loss. %This conclusion is not supplied by the present theorem when the RAF is nondifferentiable at~$0$.  
\end{remark}

Moreover, in Appendix~\ref{sec:cnd} we define and analyze $C$-estimators for regression problems, which are characterized by conditional PMFs, thereby further demonstrating their flexibility.

\section{Simulation experiments}\label{sec:simulations}

%\subsection{Composite estimation of latent factor model}
%Responses to questionnaire items may contain careless responses, which may be viewed as contaminated data points. 
Careless responding has been identified as a major threat to the validity of analyses of questionnaire data \citep[e.g.,][]{meade2012}. For instance, in the psychometrics literature \cite{arias2020} find that having a carelessness prevalence rate of as low as~5\% of all respondents can introduce severe biases in parameter estimates of latent factor models when conventional approaches are used. Careless responding may be seen as contamination.

In this simulation, we therefore consider the latent factor model from Example~\ref{ex:factor} with contamination in the observed questionnaire responses. We set the number of questionnaire items to $p=4$, where each item has $m \equiv m_j = 5$ response options. To keep things simple, we use a single latent factor ($r=1$) with the true loadings matrix set to $\mat{\Lambda}_* = (\lambda_{*,j})_{j=1}^4 = (0.9, 0.8, 0.7, 0.6)$ so that~$\BSigmaF=1$ is a scalar. The true discretization thresholds are set to 
$\vec{\tau}^{(j)}_*  = (-1.5, -0.5, 0.5, 1.5)^\top$ 
for all items $j=1,\dots,p$. Overall, each of the~$p$ items takes values in $\{1,\dots,5\}$, and there are $d=20$ parameters in total.
We generate $N=1,000$ uncontaminated $p$-variate vectors of responses in this way. 

We introduce contamination as follows. Given a contamination fraction~$\eps$, we generate~$N\eps$ random draws for~$p$ independent Gumbel-distributed variables with location parameter~0 and scale parameter~3. The ensuing realizations are then discretized separately for each dimension according to the true discretization thresholds~$\vec{\tau}_*^{(j)}$. We then randomly replace~$N\eps$ uncontaminated observations by the discretized contaminated observations. Since the Gumbel distribution scatters in all directions, this type of contamination intends to emulate the erratic behavior of a careless respondent. We consider $\eps\in\{0, 0.01, 0.05, 0.1, 0.15, 0.2, 0.3, 0.4, 0.49\}$. For each~$\eps$,~1,000 contaminated datasets of size~$N$ are generated.

We then fit the latent factor model with a single factor to each dataset with various composite $C$-estimators~$\Bthetahat^\cmp$. Composite estimation of this model is described in Example~\ref{ex:cmp}. All submodels are weighted equally. As discrepancies, we consider the four functions described in Table~\ref{tab:rhoRAF}. For the Huber-like function, we set the tuning constants to $c_1 = -1$ \citep[as recommended by][]{ruckstuhl2001} and $c \equiv c_2 \in\{0, 0.6\}$. If $\eps=0$, $A'(0)$ does not exist for $c=0$, so Theorem~\ref{thm:cmp-normality-reduced} is inapplicable. The other choice, $c=0.6$, has been motivated by \cite{welz2026polycor}. 
As performance measures, we consider the componentwise estimation bias with respect to the true~$\Btheta_*$, being $\hat{\Btheta}^\cmp_{N,j} - \Btheta_{*,j}, j = 1,\dots, d$,  empirical coverage of the estimated 0.95 confidence intervals (CIs) with respect to the true~$\Btheta_{*,j}$, and CI length. The CIs are constructed from the asymptotic theory in Theorem~\ref{thm:cmp-normality-reduced}.

%coverage with respect to the true~$\Btheta_{*,j}$ of the estimated $(1-\nu)$ confidence interval~$\hat{\text{CI}}_{\nu,j}\left(\Bthetahat^\cmp\right)$ associated with~$\hat{\Btheta}^\cmp_{N,j}$, and (3) average length of the confidence interval~$\hat{\text{CI}}_{\nu,j}\left(\Bthetahat^\cmp\right)$. We use the usual significance level $\nu = 0.05$. Owing to Theorem~\ref{thm:cmp-normality-reduced}, the confidence interval is defined as $\hat{\text{CI}}_{\nu,j}\left(\Bthetahat^\cmp\right) = \left\{\hat{\Btheta}^\cmp_{N,j} \pm q_{1-\nu/2} \cdot\hat{\text{SE}}_j\left(\Bthetahat^\cmp\right)\right\}$, where $q_{1-\nu/2}$ is the $(1-\nu/2)$ standard normal quantile, and the estimated standard error~$\hat{\text{SE}}_j\left(\Bthetahat^\cmp\right)$ is the square root of the  $(j,j)$-th element of the estimated asymptotic covariance matrix $\mat{\Sigma}^\cmp_N\left( \Bthetahat^\cmp \right), j = 1,\dots,d$.

\begin{figure}[t]
    \centering
    \includegraphics[width = 0.99\textwidth]{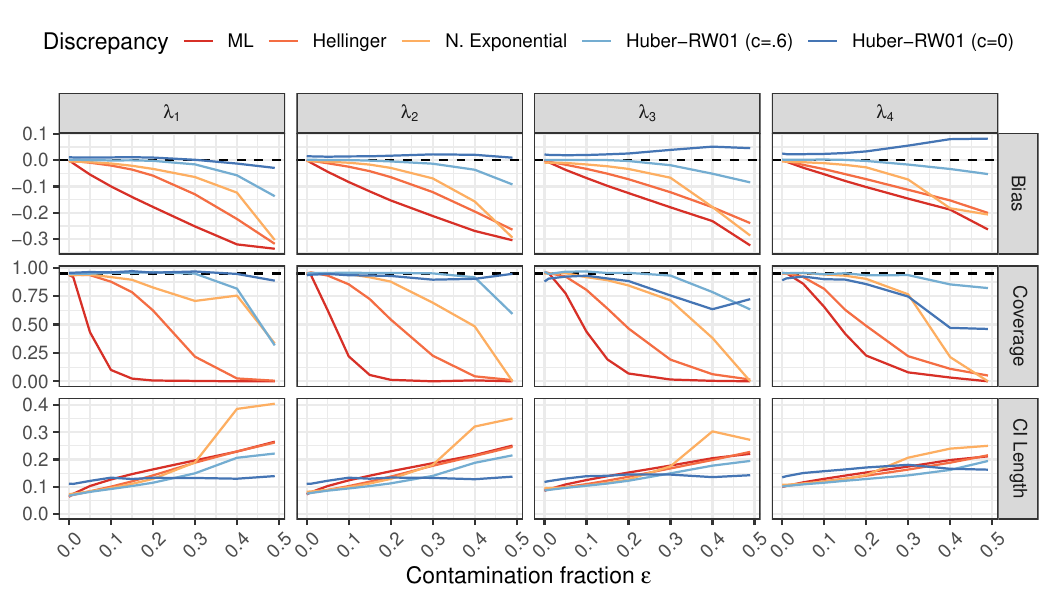}
\caption{Simulation results for composite estimation of the factor loadings parameters $\lambda_j$ using various $C$-estimators: bias with respect to the true value~$\lambda_{*,j}$, coverage of the estimated 0.95~CI with respect to~$\lambda_{*,j}$ (dashed line at nominal level), and length of that CI, $j=1,\dots,4$. Results are averaged over~1,000 repetitions.}
\label{fig:simresults-cmp-loadings}
\end{figure}

Since the threshold parameters are only of secondary interest \citep{katsikatsou2012}, we focus on the loadings parameters here and defer the results of the threshold parameters to Appendix~\ref{app:simulations-additional}.  
Figure~\ref{fig:simresults-cmp-loadings} visualizes the simulation results for the loadings parameters. 

At no contamination, the ML estimator is accurate and achieves coverage at the nominal~0.95 level, as expected. However, once contamination is introduced, its performance rapidly deteriorates: For instance, at $\eps = 0.1$, ML coverage is already below 10\% for~$\lambda_1$. At $\eps=0.2$, the coverage drops further to near~0 for both~$\lambda_1$ and~$\lambda_2$, and is below~0.25\% for $\lambda_3$ and $\lambda_4$. Hence, although each observed variable can only take~5 values, already small levels of categorical contamination can suffice to render ML unreliable.

Focus now on the alternatives to ML in Figure~\ref{fig:simresults-cmp-loadings} in the presence of contamination $(\eps>0)$. First, all alternatives have smaller bias and higher coverage levels than ML across all parameters and positive contamination fractions. Especially the Huber-like discrepancy function succeeds at achieving robustness by incurring only minor biases even in excessive contamination levels. In particular, at the choice $c=0.6$, coverage at the nominal level is approximately retained for all loadings parameters up until $\eps=0.3$. 

Conversely, at $\eps=0$, all discrepancy functions except the Huber discrepancy at $c=0$ yield estimators whose standard errors are as large as those of ML, thereby reflecting the efficiency of $C$-estimators at the postulated model. The Huber discrepancy at $c=0$, though, leads to under-coverage at a correctly specified model $(\eps=0)$. This is due to~$A'(0)$ not existing, so the normality-based standard errors from Theorem~\ref{thm:cmp-normality-reduced} are inapplicable. Also, note that at $\eps=0$, this particular discrepancy leads to a small point-estimation bias, which is seemingly at odds with Fisher consistency. Due to natural sampling variation, sample PRs may exceed the value~0 even when the model is correctly specified. Since this particular discrepancy function downweights PRs exceeding~0, some events may thus be erroneously downweighted, resulting in a finite-sample approximation error. This behavior has been described before by \cite{ruckstuhl2001} and \cite{welz2026polycor}.

Overall, this simulation demonstrates that $C$-estimators can be substantially robust against categorical contamination, unlike ML. It also showcases the importance of differentiability of RAFs for the asymptotic distribution. It also illustrates that Huber discrepancy functions with strictly positive tuning constants may be a preferable choice.

In Appendix~\ref{app:simulations}, we conduct additional simulations on regression problems, with similar findings. For simulations on non-composite models, we refer to \cite{welz2026polycor}, where $C$-estimation is successfully applied to robustly estimate polychoric correlations. % against careless responses. 

\section{Empirical application}\label{sec:application}
\subsection{Study design}
In this section, we robustly fit the latent factor model from Example~\ref{ex:factor} to questionnaire data from empirical psychology. Specifically, we use the data of \citet[][Sample~1]{arias2020}, who collect responses of $N=725$ individuals to an online rating-scale questionnaire measuring the three personality traits of \emph{neuroticism}, \emph{extraversion}, and \emph{conscientiousness}. The data have been made publicly available by \cite{arias2020} at \url{https://osf.io/n6krb}.

%However, \cite{arias2020} suspect that careless responding is a sizable problem in their data, and classify about~10\% as careless. Due to the suspected presence of contaminated observations, this dataset is suitable for a practical demonstration of robust $C$-estimators.

For the ease of demonstration, we restrict ourselves to the $p=12$ items measuring the \emph{neuroticism} personality trait in the data of \cite{arias2020}. %The results remain qualitatively similar for the other two traits; we report them in Appendix \todo{}. 
Each item is a single English adjective (such as ``relaxed'' or ``stable''), and participants are asked to indicate how accurately each such adjective describes their personality on a 5-point Likert-type rating scale (\emph{1 = very inaccurate, 2 = moderately inaccurate, 3 = neither accurate nor inaccurate, 4 = moderately accurate, 5 = very accurate}). Importantly, the~12 items are~6 pairs of polar opposite adjectives, such as ``calm'' and its opposite ``angry''. Table~\ref{tab:neuroticism-items} provides an overview. Because of this structure, one expects that the mutual opposites within an item pair correlate negatively \citep{arias2020}. 

\begin{table}[t]
\centering
\small
\begin{tabular}{l|| l |l |l |l |l |l}
Item pair & N1 & N2 & N3 & N4 & N5 & N6 
\\ \hline %\vspace{-0.4cm}
Positive opposite (P) & calm & relaxed & at ease & not envious & stable & contented
\\
Negative opposite (N) & angry & tense & nervous & envious & unstable & discontented
\end{tabular}
\caption{The $p=12$ single-adjective items measuring the \emph{neuroticism} trait in \cite{arias2020}. The items are~6 pairs of polar opposites with the positive (P) opposite (top), and the negative (N) opposite (bottom). Item identifiers work as follows: ``N2\_P'' refers to the positive opposite in the 2nd neuroticism pair (``relaxed''), while  ``N5\_N'' would refer to the negative opposite in 5th neuroticism pair (``unstable''). Each item has~5 response options.}
\label{tab:neuroticism-items}
\end{table}

However, \cite{arias2020} find that about~10\% of the participants in their data give implausible response patterns, such as strongly agreeing to \emph{both} ``envious'' and ``not envious'', and deduce that such response patterns are indicative of careless responding. Thus, their data are suitable for an application of robust $C$-estimation because (1) they are suspected to be contaminated, and (2) the mutual opposite structure of item pairs logically indicates potential carelessness. We can utilize the latter property to check if downweighted response patterns indeed correspond to such implausible ones. %In other questionnaires, careless responding may not be inferrable from item labels.

Following \cite{arias2020}, we fit a one-factor latent factor model to the $N=725$ observations for the $p=12$ \emph{neuroticism} items in their data. This model is described in Example~\ref{ex:factor} and features $d=60$ parameters here. Estimation is done in composite manner as described in Example~\ref{ex:cmp}, with equal weights for each bivariate submodel. We apply composite ML estimation as well as composite $C$-estimation with the Huber-like discrepancy function of \cite{ruckstuhl2001}, with tuning constants set to $c_1 = -1$ and $c\equiv c_2=0.6$. Adhering to standard practices for fitting factor models to questionnaire data, we reversely code all negatively worded items (that is, the negative opposites). For instance, for a negatively worded item, a given response ``5'' is coded as ``1'', a given response ``4'' is coded as ``2'', etc. Consequently, all loadings parameter estimates are expected to be positive.

\subsection{Results}

\begin{table}[t]
\centering
\small
\begin{tabular}{c c r c c r c}
\small
& & \multicolumn{2}{c}{Robust} & & \multicolumn{2}{c}{ML}
\\
\noalign{\smallskip}\cline{3-4}\cline{6-7}\noalign{\smallskip}
Loading & & Estimate & $\widehat{\text{SE}}$ & & Estimate & $\widehat{\text{SE}}$ 
\\
\noalign{\smallskip}\hline\noalign{\smallskip}
 %\vspace{-0.4cm}
N1\_P & & 0.780 & 0.029 & & 0.727 & 0.028 
\\ 
N1\_N & & 0.674 & 0.031 & & 0.628 & 0.028 
\\ %\vspace{-0.4cm}
N2\_P & & 0.849 & 0.026 & & 0.781 & 0.025 
\\ \smallskip
N2\_N & & 0.776 & 0.024 & & 0.726 & 0.026 
\\ %\vspace{-0.4cm}
N3\_P & & 0.860 & 0.021 & & 0.793 & 0.025 
\\ \smallskip
N3\_N & & 0.749 & 0.024 & & 0.711 & 0.026 
\\  %\vspace{-0.4cm}
N4\_P & & 0.483 & 0.050 & & 0.377 & 0.045 
\\ \smallskip
N4\_N & & 0.567 & 0.039 & & 0.510 & 0.037
\\  %\vspace{-0.4cm}
N5\_P & & 0.776 & 0.026 & & 0.734 & 0.025 
\\ \smallskip
N5\_N & & 0.761 & 0.026 & & 0.726 & 0.024 
\\  %\vspace{-0.4cm}
N6\_P & & 0.654 & 0.032 & & 0.589 & 0.033 
\\ \smallskip
N6\_N & & 0.737 & 0.027 & & 0.676 & 0.027
\\  %\vspace{-0.4cm}
\end{tabular}
 %\vspace{-0.8cm}
\caption{Estimates for parameters and standard errors ($\widehat{\text{SE}}$) for the factor loadings~$(\lambda_j)_{j=1}^{12}$ of the \emph{neuroticism} trait in the data of \citet{arias2020} of the composite $C$-estimator with Huber discrepancy function and tuning constant $c=0.6$,  and composite ML.}
\label{tab:arias2020-loadings}
\end{table}

\begin{figure}[t]
    \centering
    \includegraphics[width = 0.99\textwidth]{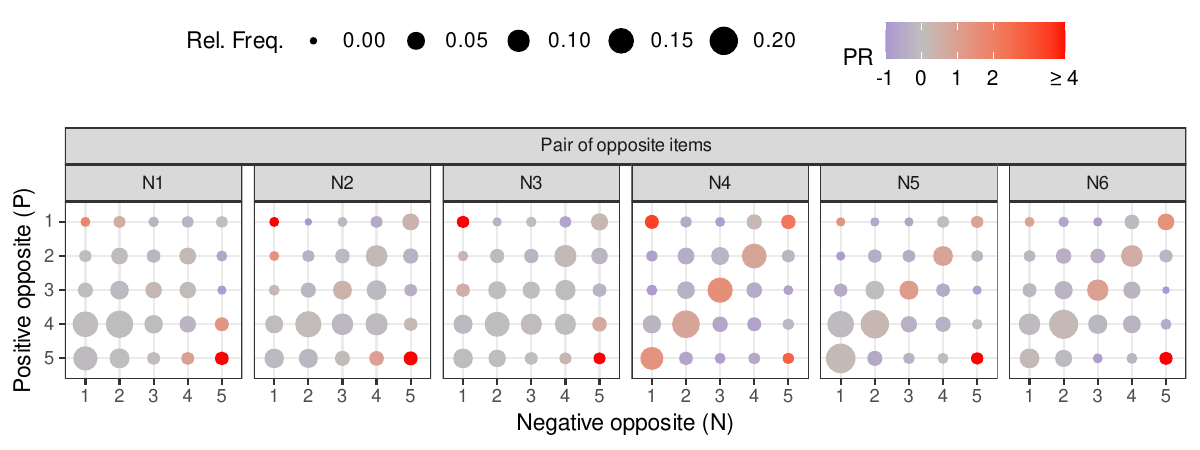}
\caption{ Dot plot of the opposite item pairs in Table~\ref{tab:neuroticism-items} in the data. Each item has five response options ranging from~1 (``very inaccurate'') to~5 (``very accurate''). The size of each dot is proportional to the relative frequency of the corresponding response pattern in the pairwise submodel involving the polar opposite items. A dot's color varies with the value of the submodel PR of that response pattern, computed from a Huber-type composite $C$-estimator at $c=0.6$. Because some PRs have extremely large values (given in Table~\ref{tab:arias2020-PRs} in the appendix),  we truncated the color scale so that a bright red is assigned to all PRs with value $\geq 4$. The reverse-coding of negative items has been undone in this visualization.}
\label{fig:arias2020-antonym}
\end{figure}

Table~\ref{tab:arias2020-loadings} lists the estimates for the $p=12$ factor loadings parameters; the threshold estimates are provided in Table~\ref{tab:arias2020-thresholds} in the appendix. Each robust loadings estimate is stronger than that of ML. On average, the communalities (squared factor loadings) of the robust estimator are about~17\% larger than those of ML, meaning that the latent \emph{neuroticism} factor can on average explain~17\% more variation in item responses when the model is estimated robustly. Particularly the loading of the item N4\_P (``not envious'') stands out with a substantially stronger robust estimate (0.483) than ML estimate (0.377).

The robust estimator providing different estimates than  ML suggests that it has downweighted certain response patterns in the estimation process. To obtain insights into this downweighting behavior, Figure~\ref{fig:arias2020-antonym} visualizes the robustly estimated submodel PRs for the six submodels involving the polar opposite items (without recoding the negatively worded items). Recall that agreeing or disagreeing to \emph{both} polar opposites is indicative of careless responding in this dataset \citep{arias2020}. Figure~\ref{fig:arias2020-antonym} reveals that such contradictory response patterns are present and that their corresponding PRs are large. In fact, their PRs are so large that the figure's color gradient had to be truncated for a meaningful visualization. For instance, the largest PR in the polar opposite pair corresponds to the observed response $(1,1)$ to the pair N3\_P-N3\_N (``at ease''-``nervous'') with a value of~78.4. There are~8 respondents with this pattern, and, following \cite{arias2020}, it seems unlikely that an attentive respondent would strongly disagree to being ``at ease'' while simultaneously strongly disagreeing to being ``nervous''. In similar fashion, the second and third largest PRs have values~67.4 and~54.95 and occur for respondents who strongly disagree to both items in the pair ``relaxed''-``tense'' (N2), and those who strongly agree to both items in ``stable''-``unstable'' (N5), respectively. Although no information on item labels was used in the estimation process, such response patterns were identified as being discordant with the postulated factor model, and were subsequently downweighted. 

Conversely, not downweighted were, as indicated by PRs of mostly near~0 in Figure~\ref{fig:arias2020-antonym}, response patterns reflecting internal consistency by agreeing to the positive opposite while disagreeing to the negative opposite (and vice versa). The only exception occurs in the ``envious''--``not envious'' item pair (N4), whose robustly estimated loadings in Table~\ref{tab:arias2020-loadings} are the weakest among all loadings and those differing most strongly with their corresponding ML estimates, indicating issues in general measurement quality in these two items. In summary, these results suggest that the robust $C$-estimator downweights response patterns that are discordant with the fitted one-factor model while largely retaining internally consistent patterns. We stress that the former may not necessarily be due to carelessness, but other reasons causing them to be poorly represented by this model.

\section{Discussion and conclusion}\label{sec:conclusion}
Categorical robustness requires model departures to be assessed through discrepancies between observed cell frequencies and model-implied probabilities rather than through the numerical extremeness of observations. Building on minimum-disparity estimation \citep{lindsay1994}, we developed $C$-estimation as a unifying framework for structured categorical models. The same cellwise principle applies to unconditional models on finite or countably infinite sample spaces, composite models, and categorical regression models, while admitting clipped Huber-type loss functions. Such loss functions are particularly appealing for robustness. We established Fisher consistency, consistency for the population target and asymptotic normality under contamination, and sandwich covariance estimation. At a correctly specified model, regular $C$-estimators retain the first-order efficiency of maximum likelihood. We also derived computable maximum-bias envelopes and, for a Huber-like loss, linked its clipping constants to a global robustness bound. Simulations support these theoretical results, and the \proglang{R} package \pkg{robcat} provides a software implementation.

We demonstrate the practical usefulness of $C$-estimators in an empirical application on a latent factor model with psychological questionnaire data. $C$-estimators produce stronger factor loadings and downweight response patterns inconsistent with the intended item structure, some of which may reflect careless responding.

This paper suggests a number of extensions. First, for computationally challenging composite models, especially categorical ones, \cite{alfonzetti2025jasa} recently proposed a stochastic approximation of the composite ML estimator. It seems plausible that one could also approximate composite $C$-estimators in the same manner. 
Second, robustness criteria complementary to the maximum-bias curves we considered may be useful, such as  breakdown values. Third, the derived asymptotics may allow for the construction of a statistical test for  contamination on the cell-level, such as outlying response strings in questionnaires.

From a practical perspective, $C$-estimators might be particularly useful for robustifying the analyses of questionnaire response data against careless responding or bot responses, which is an increasing concern in large internet-collected datasets \citep[e.g.,][]{falk2025,arias2020,meade2012}; see also this paper's empirical application. By unifying robust fitting, statistical inference, and structured categorical modeling, $C$-estimation helps make categorical data less of the ``possible exception'' identified by \citet{stigler2010}.

{\small \bibliography{bibliography.bib}}

%%%% APPPENDIX
\newpage
\appendix
\renewcommand\thefigure{\thesection.\arabic{figure}}   
\renewcommand\thetable{\thesection.\arabic{table}}   
\renewcommand{\theequation}{\thesection.\arabic{equation}}
\renewcommand{\thefootnote}{\roman{footnote}}
\setcounter{figure}{0} 
\setcounter{table}{0}   
\setcounter{equation}{0}  
\setcounter{footnote}{0}  

\startcontents[appendices]
\section*{Appendix contents}
{
\small
\renewcommand{\baselinestretch}{1}\selectfont
\printcontents[appendices]{}{1}[2]{}
}

\newpage
\section{$C$-estimation: Technical details}\label{app:lemmas}
This section provides technical details of $C$-estimators. Specifically, in Subsection~\ref{app:RAF-to-rho} we construct a normalized discrepancy function from an RAF, and in Subsections~\ref{app:uncond-lemmas} and~\ref{app:uncond-lemmas-proofs} we list and prove, respectively, a number of technical lemmas that shall be useful in the proofs of the properties of $C$-estimators stated in Section~\ref{sec:properties}. The proofs of these properties are given in the subsequent Appendix~\ref{app:proofs-uncond}. Throughout the entire appendix, we use the notation and definitions introduced in the main text.

\subsection{Constructing a discrepancy function from an RAF}\label{app:RAF-to-rho}
Conversely, let $A$ satisfy Assumption~\ref{ass:RAF}. For $\delta>-1$, solving the linear differential equation $A(\delta)=(\delta+1)\rho'(\delta)-\rho(\delta)$ subject to $\rho(0)=0$ gives\footnote{\citet[][Eq.~(15)]{lindsay1994} provides an equivalent characterization of~\eqref{eq:rho-from-RAF} that depends on the derivative of~$A$, which always exists under the assumptions of that paper but may not under ours.}
\begin{equation}\label{eq:rho-from-RAF}
	\rho_A(\delta) = \rho(\delta) = (\delta + 1) \int_0^{\delta} \frac{A(t)}{(t+1)^2} \d t.
\end{equation}
Continuity of~$A$ at $-1$ gives $\lim_{\delta\downarrow-1}\rho_A(\delta)=-A(-1)$, so define $\rho_A(-1)=-A(-1)$. On $(-1,\infty)$, direct differentiation yields
\begin{equation*}
	\rho_A(0)=\rho_A'(0)=0,\qquad
	(\delta+1)\rho_A'(\delta)-\rho_A(\delta)=A(\delta),\qquad
	\rho_A''(\delta)=\frac{A'(\delta)}{\delta+1},
\end{equation*}
where the last identity holds wherever $A'$ exists in $(-1,\infty)$. Hence, $\rho_A$ is convex and satisfies Conditions~\ref{ass:rho0}, \ref{ass:rho'}, and~\ref{ass:rhonorm}; in particular, if $\rho_A''(0)$ exists, then $\rho_A''(0)=A'(0)=1$.

Assumption~\ref{ass:RAF} alone does not imply Condition~\ref{ass:rho''}. A sufficient additional condition is that $A$ be twice continuously differentiable outside a finite subset of $(-1,\infty)$, because then
\begin{equation*}
	\rho_A'''(\delta)=\frac{A''(\delta)}{\delta+1}-\frac{A'(\delta)}{(\delta+1)^2},
\end{equation*}
wherever $A''$ exists. Thus, $\rho_A$ satisfies Assumption~\ref{ass:rho} under this additional condition and is therefore a discrepancy function. Without this additional condition,~\eqref{eq:rho-from-RAF} still gives a normalized convex function linked exactly to~$A$, but Condition~\ref{ass:rho''} may fail.

We now turn to the technical lemmas.

\subsection{Technical lemmas: statements}\label{app:uncond-lemmas}
We first introduce some additional notation that will be used throughout the lemmas and their proofs. Define the empirical process
\begin{align*}
	\GN(\z) = \sqrt{N}\left(\fhatz - \fepsz\right) \qquad \z\in\Z,
\end{align*}
and  the $d$-variate random vector 
\begin{equation}\label{eq:xi}
\begin{split}
\vecfunN{\xi}{\Btheta}
	&=
	N^{-1/2}\sum_{\z\in\Z}\GN(\z)A'(\PRzepsS{\Btheta})\szS{\Btheta}
	\\&=
	\frac{1}{N} \sum_{i=1}^N\sum_{\z\in\Z} \big(I_i(\z) -\fepsz \big) A'(\PRzepsS{\Btheta})\szS{\Btheta},
\end{split}
\end{equation}
where $I_i(\z) = \I{\vec{Z}_i = \z}$ and $\Btheta\in\BTheta$. Furthermore, for $\Btheta\in\BTheta$, $\rho$ a discrepancy function, and~$f$ a PMF on~$\Z$, let
\begin{equation*}
\begin{split}
		\vecfun{\eta}{\Btheta, f} 
		&\equiv -\gradient\D{f}{\model}
		\\
		&= 
		\sum_{\z\in\Z} \fun{A}{\frac{f(\z)}{\pz{\Btheta}}-1} \gradient \pz{\Btheta},
\end{split}
\end{equation*}
where the second equation is due to $A(\delta) = (\delta + 1)\rho'(\delta)-\rho(\delta)$. Note that for simplicity, the dependence on the discrepancy function~$\rho$ is suppressed in the notation~$\vecfun{\eta}{\Btheta, f}$.

The following lemma establishes that if an estimator~$\Bthetabreve$ is consistent for~$\Btheta_0$, then so is the empirical matrix~$\matfunN{M}{\Bthetabreve}$ for the population matrix~$\matfuneps{M}{\Btheta_0}$ under our assumptions.

\begin{lemma}\label{lem:hessian-convergence}
Grant Assumptions~\ref{ass:model} (on the model),~\ref{ass:RAF} (on the RAF), as well as the regularity conditions~\ref{ass:interior},~\ref{ass:Mposdef}, and~\ref{ass:Muniform}. Let~$\Bthetabreve$ be a consistent estimator of~$\Btheta_0$, i.e., $\Bthetabreve\convP\Btheta_0$ as $N\to\infty$. Then
\[
	\matfunN{M}{\Bthetabreve}\convP\matfuneps{M}{\Btheta_0}
\]
as $N\to\infty$.  
\end{lemma}

The following lemma, Lemma~\ref{lem:gradientdecomposition}, decomposes the (negative) gradient into a deterministic and a stochastic part, where the latter is proportional to~$\vec{\xi}_N$ in~\eqref{eq:xi}. As such, since~$\vec{\xi}_N$ is proportional to a sum over the~$N$ observations, Lemma~\ref{lem:gradientdecomposition} provides an important building block in establishing the asymptotic linearization in Theorem~\ref{thm:asy-normality}.

\begin{lemma}\label{lem:gradientdecomposition}
Grant the assumptions of Theorem~\ref{thm:asy-normality}. Then, at~$\Btheta_0$, the decomposition
\begin{align*}
\sqrt{N}\vecfun{\eta}{\Btheta_0, \fhat} =
\underbrace{\sqrt{N}\vecfun{\eta}{\Btheta_0,\feps}}_{deterministic} +
\underbrace{\sqrt{N}\vecfunN{\xi}{\Btheta_0}}_{stochastic\ \&\ linear} + \ohp{1}
\end{align*}
holds true.
\end{lemma}

Lemma~\ref{lem:xi-limit} shows that the linear term~$\sqrt{N}\vecfunN{\xi}{\Btheta_0}$ from Lemma~\ref{lem:gradientdecomposition} is asymptotically Gaussian, which is useful for proving asymptotic normality of $C$-estimators in Theorem~\ref{thm:asy-normality}.

\begin{lemma}\label{lem:xi-limit}
Under the assumptions of Theorem~\ref{thm:asy-normality}, it holds true that
\[
	\sqrt{N}\vecfunN{\xi}{\Btheta_0}\convweak \gauss_d\left(\vec{0}, \matfuneps{U}{\Btheta_0}\right)
\]
as $N\to\infty$, where $\matfuneps{U}{\Btheta_0}$ exists and has finite norm.
\end{lemma}

Finally, Lemma~\ref{lem:IFlimits} supports the influence-function result in Theorem~\ref{thm:IF}, while the subsequent two lemmas support the maximum-bias results.

\begin{lemma}\label{lem:IFlimits}
Grant the assumptions of Theorem~\ref{thm:IF}. If~$A'(0)$ exists, then
\[
	\fun{A'}{\PRzeps{\Btheta_*}}\Big|_{\eps=0} = A'(0),
\]
and
\[
	 \matfuneps{M}{\Btheta_*}\Big|_{\eps=0} = A'(0)\matfun{J}{\Btheta_*},
\]
and
\[
	 \matfuneps{U}{\Btheta_*}\Big|_{\eps=0} = A'(0)^2\matfun{J}{\Btheta_*},
\]
and
\[
	 \matfuneps{\Sigma}{\Btheta_*} \Big|_{\eps=0} = \matinv{J}{\Btheta_*},
\]
where the final property requires $A'(0)$ to exist; Condition~\ref{ass:RAForigin} then gives $A'(0)=1$.
\end{lemma}

\begin{lemma}\label{lem:phi-bounds}
For fixed constants $-1 \leq c_1 < 0 \leq c_2 \leq \infty$, let $\tilde{c}_1 = c_1 + 1$ and $\tilde{c}_2 = c_2 + 1$ and define, for $x\geq 0$, 
\begin{equation*}
\rhot(x)
=
\begin{cases}
	x(\log(\tilde{c}_1) + 1) - \tilde{c}_1 & \textnormal{ if } x < \tilde{c}_1,
	\\
	x\log x & \textnormal{ if } x \in [\tilde{c}_1, \tilde{c}_2],
	\\
	x(\log(\tilde{c}_2) + 1) - \tilde{c}_2 & \textnormal{ if } x > \tilde{c}_2.
\end{cases}
\end{equation*}
As in~\eqref{eq:rw01}, when $c_1=-1$ or $c_2=\infty$, respectively, the lower or upper branch is empty and its displayed expression is not evaluated.
Further define on $[0, \infty)$ the function
\[
	\phi(x) = \rhot(x) - (x - 1)
\]
as well as  the constants $\kappa_- = \min\{1/2, -c_1\}$ and $\kappa_+ = \min\{1/2, \log(c_2+1)\}$.
Then,
\begin{align*}
	\phi(x) &\geq \kappa_-(1-x)^2 \qquad \textnormal{ if }x \in [0, 1),
	\\
	\phi(x) &\geq \kappa_+\frac{(1-x)^2}{x} \qquad \textnormal{ if }x \geq 1.
\end{align*}
\end{lemma}

\begin{lemma}\label{lem:pcrit-equivalence}
	Consider the setup of Corollary~\ref{cor:maxbias-increases-RW} and suppose that $\Z$ is finite. For any $\eps\in(0,0.5)$, the number~$\pcrit$ defined in~\eqref{eq:pcrit} belongs to~$(0,1)$. Moreover,
	\[
		\pmin\leq\pcrit
		\quad\Longrightarrow\quad
		\log\big(\eps/\pmin-\eps+1\big)>1/2.
	\]
	The reverse implication need not hold.
\end{lemma}

\subsection{Technical lemmas: proofs}\label{app:uncond-lemmas-proofs}
In this sub-subsection we prove Lemmas \ref{lem:hessian-convergence}--\ref{lem:pcrit-equivalence}.

\subsubsection{Proof of Lemma~\ref{lem:hessian-convergence}}
We have that
\begin{equation}\label{eq:Mbound}
\begin{split}
	\matnorm{\matfunN{M}{\Bthetabreve} - \matfuneps{M}{\Btheta_0}}
	&=
	\matnorm{\matfunN{M}{\Bthetabreve} - \matfuneps{M}{\Bthetabreve} + \matfuneps{M}{\Bthetabreve} - \matfuneps{M}{\Btheta_0}}
	\\
	&\leq
	\matnorm{\matfunN{M}{\Bthetabreve} - \matfuneps{M}{\Bthetabreve}} +  \matnorm{\matfuneps{M}{\Bthetabreve} - \matfuneps{M}{\Btheta_0}}
	\\
	&=
	\matnorm{\matfunN{M}{\Bthetabreve} - \matfuneps{M}{\Bthetabreve}}  + \ohp{1},
\end{split}
\end{equation}
where the third line follows by the assumed consistency of~$\Bthetabreve$ for~$\Btheta_0$, the continuity in Condition~\ref{ass:Mposdef}, and the continuous mapping theorem. 
Because $\BTheta_0$ is an open neighborhood of~$\Btheta_0$ (Condition~\ref{ass:interior}) and $\Bthetabreve\convP\Btheta_0$, also~$\Bthetabreve\in\BTheta_0$ with probability tending to~1 as~$N\to\infty$. Thus, with probability tending to~1,
\[
	\matnorm{\matfunN{M}{\Bthetabreve} - \matfuneps{M}{\Bthetabreve}}
	\leq
	\sup_{\Btheta\in\BTheta_0}
	\matnorm{ \matfunN{M}{\Btheta} - \matfuneps{M}{\Btheta} }.
\]
Condition~\ref{ass:Muniform} imposes that the right-hand-side vanishes in probability. Combining this with~\eqref{eq:Mbound}, we  conclude that $\matnorm{\matfunN{M}{\Bthetabreve} - \matfuneps{M}{\Btheta_0}} = \ohp{1}$, which completes the proof. \QED

\subsubsection{Proof of Lemma~\ref{lem:gradientdecomposition}}
Fix $\z\in\Z$ and note that $I_i(\z) = \I{\vec{Z}_i = \z}$ follows a Bernoulli distribution with probability parameter~$\fepsz$ for all $i=1\,\dots,N$. Since $\fhatz = \frac{1}{N}\sum_{i=1}^N I_i(\z)$, it holds true that $\E{}{\fhatz} = \fepsz$ and $\var{}{\fhatz} = \frac{1}{N}\fepsz (1-\fepsz) >0$, which are both finite.  Applying the Lindeberg-Lévy central limit theorem and dividing by $\pz{\Btheta} > 0$ gives
\begin{equation}\label{eq:PRtight}
	\sqrt{N}\remdelta{\Btheta} \equiv \sqrt{N}\big( \PRzn{\Btheta} - \PRzeps{\Btheta} \big) = \sqrt{N}\frac{\fhatz - \fepsz}{\pz{\Btheta}} = \Ohp{1} 
\end{equation}
for all $\Btheta\in\BTheta$, where $\remdelta{\Btheta} = \PRzn{\Btheta}-\PRzeps{\Btheta}$. 

By the assumed differentiability of the RAF at $\PRzeps{\Btheta_0}$, it holds true that
\begin{align*}
	\fun{A}{\PRzn{\Btheta_0}} 
	&= \fun{A}{\PRzeps{\Btheta_0} + \remdelta{\Btheta_0}}
	\\
	&= \fun{A}{\PRzeps{\Btheta_0}} + \fun{A'}{\PRzeps{\Btheta_0}}\remdelta{\Btheta_0} + r_{N,\z},
\end{align*}
where the second line is due to a Taylor expansion with remainder $r_{N,\z} = \ohp{\left| \remdelta{\Btheta_0} \right|}$. Using~\eqref{eq:PRtight}, $r_{N,\z} = \ohp{\left| \remdelta{\Btheta_0} \right|} = \ohp{\Ohp{N^{-1/2}}} = \ohp{N^{-1/2}}$. 

Rearranging the previous display,
\begin{align}
	\fun{A}{\PRzn{\Btheta_0}} - \fun{A}{\PRzeps{\Btheta_0}}
	&=
	\fun{A'}{\PRzeps{\Btheta_0}}\remdelta{\Btheta_0} + r_{N,\z} \label{eq:Adiff}
	\\	
	&\iff \nonumber 
	\\
	r_{N,\z} &= \fun{A}{\PRzn{\Btheta_0}} - \fun{A}{\PRzeps{\Btheta_0}} -
	\fun{A'}{\PRzeps{\Btheta_0}}\remdelta{\Btheta_0}. \label{eq:remainder}
\end{align}
We can therefore write 
\begin{align*}
	\sqrt{N}\big( \vecfun{\eta}{\Btheta_0, \fhat} - \vecfun{\eta}{\Btheta_0, \feps} \big)
	&=
	\sum_{\z\in\Z} \gradient \pz{\Btheta_0} \sqrt{N} \big( \fun{A}{\PRzn{\Btheta_0}} - \fun{A}{\PRzeps{\Btheta_0}} \big)
	\\
	&=
	\sqrt{N}\sum_{\z\in\Z} \gradient \pz{\Btheta_0} \big( \fun{A'}{\PRzeps{\Btheta_0}}\remdelta{\Btheta_0} + r_{N,\z}  \big)
	\\
	&=
	\sqrt{N} \vecfun{\xi}{\Btheta_0} 
	+
	\sqrt{N} \sum_{\z\in\Z} r_{N,\z} \gradient\pz{\Btheta_0}.
\end{align*}
The first equality follows by definition of $\vec{\eta}(\cdot,\cdot)$ in conjunction with Assumption~\ref{ass:countable-interchange} to ensure convergence of the series as well as interchangeability of summation and differentiation, which are relevant when~$\Z$ is infinite. The second and third equalities are due to~\eqref{eq:Adiff} in conjunction with Assumption~\ref{ass:MUfinite} and the definition of~$\vecfun{\xi}{\Btheta_0}$, respectively. We will show in a moment that if~$\Z$ is infinite, the second series on the right-hand-side in the third expression is convergent under our assumptions. 

If $\Z$ is finite, then, by Slutsky's Lemma, $\sqrt{N} \sum_{\z\in\Z} r_{N,\z} \gradient\pz{\Btheta_0} = \ohp{1}$ because it is a finite sum of $\ohp{1}$ terms since $r_{N,\z} = \ohp{N^{-1/2}}$ for each~$\z$. The result follows for finite sample spaces.

Let $\Z$ now be countably infinite. The previous display can be expressed as
\[
	\sqrt{N}\big( \vecfun{\eta}{\Btheta_0, \fhat} - \vecfun{\eta}{\Btheta_0, \feps} \big)
	=
	\sqrt{N} \vecfun{\xi}{\Btheta_0} + \sqrt{N} \vecfunN{\Delta}{\Btheta_0},
\]
where $\vecfunN{\Delta}{\Btheta_0} = \sum_{\z\in\Z} r_{N,\z} \gradient\pz{\Btheta_0}$  is equal to the definition from Assumption~\ref{ass:countable-remainder} after substituting~$r_{N,\z}$ for the expression in~\eqref{eq:remainder}. Because each~$r_{N,\z}$ is of order $\ohp{N^{-1/2}}$, the vector~$\sqrt{N}\vecfunN{\Delta}{\Btheta_0}$ in the previous display is an infinite sum over~$\ohp{1}$ terms. Assumption~\ref{ass:countable-remainder} guarantees that this infinite sum is  convergent and also of order~$\ohp{1}$. The result follows for infinite sample spaces and the proof is complete. \QED

\subsubsection{Proof of Lemma~\ref{lem:xi-limit}}
For all sample indices $i = 1,\dots, N$, and $\z\in\Z$, the random variable $I_i(\z) = \I{\vec{Z}_i = \z}$ follows a Bernoulli distribution with probability parameter~$\fepsz$. Further define the $d$-dimensional random vector
\[
	\Ri{\Btheta_0} = \sum_{\z\in\Z} I_i(\z) \fun{A'}{\PRzeps{\Btheta_0}} \sz{\Btheta_0},
\]
which takes values in the set $\{\fun{A'}{\PRzeps{\Btheta_0}} \sz{\Btheta_0} : \z\in\Z \}$ because the events in~$\Z$ are mutually exclusive. By Condition~\ref{ass:PRRAF},  $\fx{A'}{\PRzeps{\Btheta_0}}$ exists for all $\z\in\Z$, so also the~$\Ri{\Btheta_0}$ exist for all observations $i=1,\dots,N$. 

To prove the assertion, it is useful to study the limit distribution of $N^{-1/2}\sum_{i=1}^N \Ri{\Btheta_0}$. By the multivariate Lindeberg-Lévy central limit theorem, as $N\to\infty$,
\begin{equation}\label{eq:Ki_CLT}
	N^{-1/2}\sum_{i=1}^N \Big(\Ri{\Btheta_0} - \E{}{\Ri{\Btheta_0}} \Big) \convweak \gauss_d \Big(\vec{0}, \var{}{\Ri{\Btheta_0}} \Big),
\end{equation}
 provided that the covariance matrix $\var{}{\Ri{\Btheta_0}}$ is finite.

We proceed by showing that $\var{}{\Ri{\Btheta_0}}$ is indeed finite under our assumptions. Because the~$I_i(\z)$ are the only stochastic objects in vector~$\Ri{\Btheta_0}$, the first moment equals
\begin{equation}\label{eq:Ki_E}
	\E{}{\Ri{\Btheta_0}} = \sum_{\z\in\Z} \fepsz \fun{A'}{\PRzeps{\Btheta_0}} \sz{\Btheta_0} 
	=
	\E{\feps}{\fun{A'}{\PR{\Btheta_0}{\varepsilon}{\vec{Z}}}  \sfun{\Btheta_0}{\vec{Z}}}.
\end{equation}
Denoting by $\vec{v}^{\otimes 2} = \vec{v}\vec{v}^\top$ the outer product of a vector~$\vec{v}$, one obtains for the second moment that
\begin{align*}
	\E{}{\Ri{\Btheta_0}^{\otimes 2}}
	&=\E{}{
		\sum_{\z\in\Z}\sum_{\y\in\Z}	I_i(\z) I_i(\y) \fun{A'}{\PRzeps{\Btheta_0}} \fx{A'}{\PR{\Btheta_0}{\varepsilon}{\y}}\sz{\Btheta_0}\sfun{\Btheta_0}{\y}^\top
	}
	\\
	&=
	\E{}{\sum_{\z\in\Z} \overbrace{I_i^2(\z)}^{=I_i(\z)} \left(\fx{A'}{\PRzeps{\Btheta_0}}\right)^2 \sz{\Btheta_0}\sz{\Btheta_0}^\top }
	\\
	&\quad+
	\E{}{\sum_{\z\in\Z}\sum_{\y\neq\z}	\overbrace{I_i(\z) I_i(\y)}^{=0} \fun{A'}{\PRzeps{\Btheta_0}} \fx{A'}{\PR{\Btheta_0}{\varepsilon}{\y}}\sz{\Btheta_0}\sfun{\Btheta_0}{\y}^\top}
	\\
	&= 
	\sum_{\z\in\Z} \fepsz \big(\fun{A'}{\PRzeps{\Btheta_0}} \sz{\Btheta_0}\big)^{\otimes 2} 
	\\
	&= \E{\feps}{\big(\fun{A'}{\PR{\Btheta_0}{\varepsilon}{\vec{Z}}} \sfun{\Btheta_0}{\vec{Z}}\big)^{\otimes 2} }.
\end{align*}
It thus follows by definition of the matrix $\mat{U}_\eps$ in Theorem~\ref{thm:asy-normality} that
\begin{equation}\label{eq:Ki_Var}
	\var{}{\Ri{\Btheta_0}} 
	=
	\var{\feps}{\fun{A'}{\PR{\Btheta_0}{\varepsilon}{\vec{Z}}}  \sfun{\Btheta_0}{\vec{Z}}} = \matfuneps{U}{\Btheta_0},
\end{equation}
which is finite due to Condition~\ref{ass:MUfinite}, so the asymptotic covariance matrix in~\eqref{eq:Ki_CLT} is indeed finite.

We now use the limit result for $N^{-1/2}\sum_{i=1}^N \Ri{\Btheta_0}$ in~\eqref{eq:Ki_CLT} to derive the limit distribution of~$\sqrt{N}\vecfunN{\xi}{\Btheta_0}$. By the definition $\GN(\z) = \sqrt{N}\left(\fhatz - \fepsz\right)$, we have that
\begin{align*}
	\sqrt{N}\vecfunN{\xi}{\Btheta_0} &= \sum_{\z\in\Z} \sz{\Btheta_0} \fun{A'}{\PRzeps{\Btheta_0}} G_N(\z) \qquad \textnormal{(Def. of $\vecfunN{\xi}{\cdot}$)}
	\\
	&=
	N^{-1/2} \sum_{i=1}^N \Bigg[ 
		\sum_{\z\in\Z} I_i(\z) \fun{A'}{\PRzeps{\Btheta_0}} \sz{\Btheta_0}
		-
		\sum_{\z\in\Z} \fepsz \fun{A'}{\PRzeps{\Btheta_0}} \sz{\Btheta_0}
	\Bigg]
	\\
	&=
	N^{-1/2}\sum_{i=1}^N \Big(\Ri{\Btheta_0} - \E{}{\Ri{\Btheta_0}} \Big)
	\\
	&\convweak
	\gauss_d \Big(\vec{0}, \matfuneps{U}{\Btheta_0}\Big)
\end{align*}
where the third line follows by the definition of~$\Ri{\Btheta_0}$ and the fourth line follows from the convergence result  in~\eqref{eq:Ki_CLT} in conjunction with~\eqref{eq:Ki_Var}. The proof is complete. \QED

\subsubsection{Proof of Lemma~\ref{lem:IFlimits}}
At $\eps=0$, the sampling PMF is $p_{\Btheta_*}$ and hence every population PR evaluated at $\Btheta_*$ is exactly zero:
\[
	\PRzeps{\Btheta_*}\big|_{\eps=0}
	=
	\frac{\pz{\Btheta_*}}{\pz{\Btheta_*}}-1
	=0.
\]
Consequently, $A'(\PRzeps{\Btheta_*})|_{\eps=0}=A'(0)$. Direct evaluation of the population Hessian, using $A(0)=0$, gives
\[
\begin{aligned}
	\matfuneps{M}{\Btheta_*}\big|_{\eps=0}
	&=
	\sum_{\z\in\Z}
	\left[
	\pz{\Btheta_*}A'(0)\sz{\Btheta_*}\sz{\Btheta_*}^{\top}
	-A(0)\hessian\pz{\Btheta_*}
	\right]
	\\
	&=A'(0)\matfun{J}{\Btheta_*}.
\end{aligned}
\]
Likewise, the usual score identity $\E{p_{\Btheta_*}}{\sfun{\Btheta_*}{\vec Z}}=\vec 0$ yields
\[
\begin{aligned}
	\matfuneps{U}{\Btheta_*}\big|_{\eps=0}
	&=
	\var{p_{\Btheta_*}}{A'(0)\sfun{\Btheta_*}{\vec Z}}
	\\
	&=A'(0)^2\var{p_{\Btheta_*}}{\sfun{\Btheta_*}{\vec Z}}
	=A'(0)^2\matfun{J}{\Btheta_*}.
\end{aligned}
\]
Condition~\ref{ass:RAForigin} gives $A'(0)=1$. Substitution into the sandwich covariance therefore gives
\[
	\matfuneps{\Sigma}{\Btheta_*}\big|_{\eps=0}
	=
	\matinv{J}{\Btheta_*}\matfun{J}{\Btheta_*}\matinv{J}{\Btheta_*}
	=
	\matinv{J}{\Btheta_*},
\]
and the proof is complete.
\QED

\subsubsection{Proof of Lemma~\ref{lem:phi-bounds}}
To establish the two desired lower bounds for the function $\phi$, we need to distinguish between cases since this function is piecewise due to~$\rhot$ being piecewise.

\paragraph*{Case 1: Evaluate on unclipped branch for $x \in [0, 1]$.}
By definition of $\rhot$, the unclipped branch is defined by $\phisup{un}(x) = x \log x - x + 1$. For $x\in [0,1]$, the function
\[
	x \mapsto \phisup{un}(x) - \frac{1}{2}(1-x)^2
\]
is nonnegative. It follows that for all $x\in[0,1]$,
\begin{align*}
	\phisup{un}(x) &\geq \frac{1}{2}(1-x)^2
	\\
	&\geq
	\kappa_- (1-x)^2,
\end{align*}
since $\kappa_- = \min\{1/2,-c_1\}\leq 1/2$. 

\paragraph*{Case 2: Evaluate on lower-clipped branch for $x \in [0, \ct_1)$.}
If $c_1=-1$, then $\ct_1=0$ and this case is empty. Suppose $c_1>-1$. By definition of $\rhot$, the lower-clipped branch $\phisup{lo}:[0,\ct_1]\to\R$ is defined by $\phisup{lo}(x) = x(\log(\ct_1) + 1) - \ct_1 - x + 1$. For $x\in[0,\ct_1]$, the function
\[
	g_-(x) = \phisup{lo}(x) - \kappa_-(1-x)^2
\] 
is concave because $g''_-(x) = -2\kappa_- < 0$. Also, it is nonnegative at its endpoints, that is, $g_-(0), g_-(\ct_1)\geq0$. Therefore,~$g_-$ is nonnegative throughout $[0,\ct_1]$. Thus,
\[
	\phisup{lo}(x) \geq \kappa_-(1-x)^2 \qquad \textnormal{ for all }x \in [0, \ct_1).
\]

Combining the inequalities from Cases~1 and~2, we arrive at 
\begin{equation}\label{eq:phiineq-upto1}
	\phi(x) \geq \kappa_-(1-x)^2 \qquad \textnormal{ for all }x \in [0, 1]
\end{equation} 
because $\phi$ on $[0,1]$ is either equal to $\phisup{un}$ or $\phisup{lo}$, depending on the value of~$\ct_1\in [0,1)$. This proves the first asserted inequality.

We now evaluate $\phi$ on $x \geq 1$, for which the lower branch is irrelevant.

\paragraph*{Case 3: Evaluate unclipped branch for $x\geq 1$.} 
By definition of $\rhot$, the unclipped branch is defined by $\phisup{un}(x) = x \log x - x + 1$. For $x\geq 1$, the 
function
\[
	x \mapsto \phisup{un}(x) - \frac{1}{2}\frac{(1-x)^2}{x}
\]
is nonnegative, so
\begin{equation}\label{eq:phiunineq}
\begin{split}
	 \phisup{un}(x) 
	 &\geq \frac{1}{2}\frac{(1-x)^2}{x}
	 \\
	 & \geq \kappa_+\frac{(1-x)^2}{x}
	 \qquad \textnormal{ for all }x \geq 1,
\end{split}
\end{equation}
where the final inequality is due to  $\kappa_+ = \min\{1/2,\log\ct_2\}\leq 1/2$.

\paragraph*{Case 4: Evaluate on upper-clipped branch for $ x > \ct_2 \geq 1 $.}
If $c_2=\infty$, then $\ct_2=\infty$ and this case is empty. Suppose $c_2<\infty$. By definition of $\rhot$, the upper-clipped branch $\phisup{up}:(\ct_2,\infty)\to\R$ is defined by $\phisup{up}(x) = x(\log(\ct_2) + 1) - \ct_2 - x + 1$. Additionally define the function
\[
	g_+(x) = \phisup{up}(x) - \kappa_+ (1-x)^2/x, \qquad x > \ct_2.
\] 
Differentiating,
\[
	g_+'(x) = \log \ct_2 - \kappa_+(1-x^{-2}) \geq \log\ct_2 - \kappa_+ \geq 0,
\]
so $g_+$ is nondecreasing on $(\ct_2,\infty)$. Consequently, 
\begin{align*}
	\inf_{x\in(\ct_2,\infty)} g_+(x)
	=
	\lim_{x\downarrow\ct_2}g_+(x)
	&=
	\phisup{un}(\ct_2) - \kappa_+(1-\ct_2)^2/\ct_2
	\\
	&\geq 0,
\end{align*}
where the inequality is due to~\eqref{eq:phiunineq}. It follows that~$g_+$ is nonnegative on $(\ct_2,\infty)$. Thus,  
\[
	\phisup{up}(x) \geq \kappa_+\frac{(1-x)^2}{x} \qquad \textnormal{ for all }x > \ct_2.
\]

Combining the inequalities from Cases~3 and~4, we arrive at 
\begin{equation}\label{eq:phiineq-beyond1}
	\phi(x) \geq \kappa_+\frac{(1-x)^2}{x} \qquad \textnormal{ for all }x \geq 1.
\end{equation} 
because $\phi$ on $[1,\infty)$ is either equal to $\phisup{un}$ or $\phisup{up}$, depending on the value of~$\ct_2\geq 1$. This proves the second asserted inequality.

To summarize, the inequalities in~\eqref{eq:phiineq-upto1} and~\eqref{eq:phiineq-beyond1} jointly yield the assertion, completing the proof. \QED

\subsubsection{Proof of Lemma~\ref{lem:pcrit-equivalence}}
Fix $\eps\in(0,0.5)$. Set $\ueps=1-\eps$ and $\beps=-\ueps\log\ueps$, so that
\[
	\pcrit=\frac{2\beps-\eps}{2\beps+\ueps}.
\]
Because $-\log\ueps>1-\ueps=\eps$ and $\ueps>1/2$, we have $\beps>\ueps\eps>\eps/2$. Thus the numerator $2\beps-\eps$ is positive. The denominator exceeds the numerator by $\ueps+\eps=1$, and hence $\pcrit\in(0,1)$.

We next show that
\[
	\pcrit<\frac{\eps}{\e^{1/2}-\ueps}.
\]
Both denominators are positive, and cross-multiplication gives
\begin{align*}
	\pcrit < \frac{\eps}{\e^{1/2}-\ueps}
	&\iff
	(2\beps - \eps)(\e^{1/2}-\ueps) < \eps(2\beps + \ueps)
	\\
	&\iff	
	2\beps (\e^{1/2}-1) < \eps\e^{1/2}.
\end{align*}
The last inequality holds because $\beps<\eps$ and $2(\e^{1/2}-1)<\e^{1/2}$, so
\[
	2\beps(\e^{1/2}-1)<2\eps(\e^{1/2}-1)<\eps\e^{1/2}.
\]
Consequently, if $\pmin\leq\pcrit$, then
\[
	\pmin<\frac{\eps}{\e^{1/2}-\ueps},
\]
which is equivalent to
\[
	\eps/\pmin-\eps+1>\e^{1/2}
\]
and hence to $\log(\eps/\pmin-\eps+1)>1/2$. This completes the proof. \QED

\newpage
\section{$C$-estimation: Proofs}\label{app:proofs-uncond}
In this section we prove the mathematical statements on $C$-estimation from Sections \ref{sec:Cestimator} and \ref{sec:properties}. These comprise Propositions~\ref{prop:divergence-nonnegative} and~\ref{prop:fisher-consistency}, Theorems~\ref{thm:consistency}--\ref{thm:IF}, and the maximum-bias results in Section~\ref{sec:maxbias}. The proofs partially rely on the technical lemmas established in the preceding Appendix~\ref{app:lemmas}.

\subsection{Proof of Proposition \ref{prop:divergence-nonnegative}}
Denote the PRs by $\delta_{\Btheta}(\z) = f(\z)/\pzS{\Btheta}-1$. 
First, note that for all $\Btheta\in\BTheta$, we have that
\[
	\sum_{\z\in\Z} \delta_{\Btheta}(\z)\pzS{\Btheta} = 0,
\]
which is due to the fact that $\sum_{\z\in\Z} \pzS{\Btheta} = \sum_{\z\in\Z}f(\z)=1$.
Furthermore, convexity of~$\rho$ gives for all $\delta\in [-1,\infty)$ the supporting-line inequality
\begin{align*}
	\rho(\delta) &\geq \rho(0) + \rho'(0)(\delta - 0)
	\\
	&= \rho'(0)\delta.
\end{align*}
Consequently, 
\begin{align*}
	\D{f}{\model} = \sum_{\z\in\Z} \rho(\delta_{\Btheta}(\z))\pzS{\Btheta}
	\geq \rho'(0)\sum_{\z\in\Z}\delta_{\Btheta}(\z)\pzS{\Btheta}
	=0.
\end{align*}

If $f=\model$, then $\delta_{\Btheta}(\z)=0$ for every $\z\in\Z$, so $\D{f}{\model}=0$ is immediate. For the converse, define $g(\delta)=\rho(\delta)-\rho'(0)\delta$. The supporting-line inequality above gives $g(\delta)\geq0$, and
\[
	\D{f}{\model}=\sum_{\z\in\Z}\pzS{\Btheta}g\big(\delta_{\Btheta}(\z)\big) \geq 0.
\]
Thus, if the divergence is zero, strict positivity of the model probabilities (Assumption~\ref{ass:positiveprobability}) implies $g(\delta_{\Btheta}(\z))=0$ for every~$\z$. If any PR were nonzero, the identity $\sum_{\z\in\Z}\pzS{\Btheta}\delta_{\Btheta}(\z)=0$ would imply the existence of PRs $\delta_-<0<\delta_+$. Since we know that $g(\delta_-)$ and $g(\delta_+)$ are both zero, it follows from the definition of~$g$ and the convexity of~$\rho$ that 
\[
	\rho(\delta) = \rho'(0)\delta \quad \textnormal{ for all } \delta\in [\delta_-, \delta_+].
\]
Because $0\in (\delta_-, \delta_+)$, the preceding display implies that after taking the first derivative, there exists an open neighborhood of~0 on which $\rho' (\delta) = \rho'(0)$. Consequently, by definition of the derivative,
\[
	\rho''(0) = \lim_{\delta\to 0} \frac{\rho'(\delta) - \rho'(0)}{\delta} = 0.
\]
Thus, $\rho''(0)$ necessarily exists and equals zero, contradicting Assumption~\ref{ass:rho}. Hence all PRs are zero and it follows that $f=\model$. \QED

\subsection{Proof of Proposition~\ref{prop:fisher-consistency}}
\begin{proof}
 If $\varepsilon=0$, then $\feps=p_{\Btheta_*}$. Proposition~\ref{prop:divergence-nonnegative} shows that $\Leps(\Btheta)=\D{p_{\Btheta_*}}{p_{\Btheta}}$ is nonnegative and equals zero at~$\Btheta_*$. Conversely, any population minimizer must attain this value zero, so Proposition~\ref{prop:divergence-nonnegative} implies $\pz{\Btheta}=\pz{\Btheta_*}$. The point-identification condition in Assumption~\ref{ass:model} then yields $\Btheta=\Btheta_*$. Thus the population minimizer is unique and equals~$\Btheta_*$.
\end{proof}

\subsection{Proof of Theorem~\ref{thm:consistency}}
\begin{proof}
To obtain the result, simply follow the steps of the proof of Theorem~5.7 in \cite{vandervaart1998}, with the maximization problem therein adapted to a minimization problem.
\end{proof}

\subsection{Proof of Theorem~\ref{thm:asy-normality}}
\begin{proof}
First, note that because $\Bthetahat\convP\Btheta_0$ (Condition~\ref{ass:consistency}) and $\Btheta_0\in\BTheta_0$ (Condition \ref{ass:interior}), we have that $\Bthetahat\in\BTheta_0$ with probability tending to~1 as $N\to\infty$. Indeed, because $\BTheta_0$ is a neighborhood of~$\Btheta_0$, there exists $\alpha>0$ such that $\mathcal B(\Btheta_0,\alpha)\subseteq\BTheta_0$. Hence
\[
	\Pr{}{\Bthetahat\in\BTheta_0}
	\geq
	\Pr{}{\|\Bthetahat-\Btheta_0\|<\alpha}
	\longrightarrow 1.
\]

Now write
\[
	\matfun{\eta}{\Btheta,\fhat} =
	 \left( \vec{\eta}_1\left(\Btheta, \fhat\right), \dots, \vec{\eta}_d\left(\Btheta, \fhat\right) \right)^\top 
	 =
	 - \gradient\Dfp
	 = -\gradient\LN(\Btheta)
\]
for the (negative) gradient of the empirical loss. The proof strategy is to first expand $\matfun{\eta}{\Bthetahat,\fhat}$ about~$\Btheta_0$ and then rearrange in such a way that one obtains an approximate expression for $\sqrt{N}\left(\Bthetahat - \Btheta_0 \right)$, which is a standard proof strategy in $M$-estimation \citep[cf.][Chapter~5.3]{vandervaart1998}. However, unlike in $M$-estimation, the ensuing approximate term is \emph{not} a summation over the~$N$ observations, which prohibits a direct application of the central limit theorem.  We therefore subsequently perform a certain second expansion with respect to a Pearson residual to obtain a linearization that is characterized by a summation over the~$N$ observations. Only then can we apply the central limit theorem. 

Fix a coordinate $j\in\{1,\dots,d\}$. We know from Taylor's theorem with the remainder expressed in mean-value form that there exists a $d$-variate random vector $\bar{\Btheta}_N^{(j)}$ on the line segment between~$\Bthetahat$ and~$\Btheta_0$ such that
\begin{equation}\label{eq:etaNj}
	\etaNj{\Bthetahat}
	=
	\vec{\eta}_j\left(\Btheta_0, \fhat\right)
	+
	\left(\gradient \etaNj{\bar{\Btheta}_N^{(j)}} \right)^\top \left(\Bthetahat - \Btheta_0\right).
\end{equation}
Since $\bar{\Btheta}_N^{(j)}$ lies between~$\Bthetahat$ and~$\Btheta_0$ where the former is consistent for the latter (Condition \ref{ass:consistency}), also $\bar{\Btheta}_N^{(j)}\convP\Btheta_0$. 

By definition of~$\mat{M}_N$ as the Hessian of~$\LN$, the vector~$\gradient\etaNj{\bar{\Btheta}_N^{(j)}}$ is equal to the $j$-th column of the corresponding negative Hessian, being~$\left(-\matfunN{M}{\bar{\Btheta}_N^{(j)}}\right)_{\bullet,j}$. Invoking Lemma~\ref{lem:hessian-convergence}, $\left(\matfunN{M}{\bar{\Btheta}_N^{(j)}}\right)_{\bullet,j}\convP\left(\matfuneps{M}{\Btheta_0}\right)_{\bullet,j}$. Thus, we can overall deduce from~\eqref{eq:etaNj} that
\[
	\etaNj{\Bthetahat}
	=
	\vec{\eta}_j\left(\Btheta_0, \fhat\right)
	+
	\left(\left(-\matfuneps{M}{\Btheta_0}\right)_{\bullet,j} + \ohp{1}\right)^\top \left(\Bthetahat - \Btheta_0\right).
\]
This result holds for all $j=1,\dots,d$, so stacking the coordinates on top of each other yields 
\[
	\vecfun{\eta}{\Bthetahat,\fhat} = \vecfun{\eta}{\Btheta_0,\fhat} + \left( -\matfuneps{M}{\Btheta_0} + \ohp{1} \right) \left(\Bthetahat - \Btheta_0\right),
\]
by Slutsky's lemma. Since $\vecfun{\eta}{\Bthetahat,\fhat} = \ohp{N^{-1/2}}$ (Condition~\ref{ass:nearzero}), we obtain after rearranging and premultiplying by $-\sqrt{N}\matinveps{M}{\Btheta_0}$ (which exists by Condition~\ref{ass:Mposdef}) that
\[
	\left(\Imat + \ohp{1}\right) \sqrt{N}  \left(\Bthetahat - \Btheta_0\right) = \matinveps{M}{\Btheta_0} \sqrt{N} \vecfun{\eta}{\Btheta_0,\fhat} + \ohp{1},
\]
where $\Imat$ denotes the $d\times d$ identity matrix. Premultiplying by $\left(\Imat + \ohp{1}\right)^{-1} = \Imat + \ohp{1}$, we arrive at
\begin{equation}\label{eq:Mestimation-linearization}
	\sqrt{N} \left(\Bthetahat - \Btheta_0\right) = \left(\Imat + \ohp{1}\right) \matinveps{M}{\Btheta_0} \sqrt{N}\vecfun{\eta}{\Btheta_0,\fhat} + \ohp{1}.
\end{equation}

So far, all steps were familiar steps for establishing asymptotic normality of $M$-estimators. However, unlike in $M$-estimation, a $C$-estimator's scaled empirical gradient $\vecfun{\eta}{\Btheta_0,\fhat}$ does \emph{not} have a characterization as a sum over~$N$ observations, so we cannot apply the central limit theorem. Yet, by Lemma~\ref{lem:gradientdecomposition}, we have the approximate decomposition
\begin{equation}\label{eq:graddecomp}
\begin{split}
			\sqrt{N}\vecfun{\eta}{\Btheta_0,\fhat} 
			&=
			 \overbrace{\sqrt{N}\vecfun{\eta}{\Btheta_0, \feps}}^{deterministic} + \overbrace{\sqrt{N}\vec{\xi}_N(\Btheta_0)}^{stochastic\ \&\ sum\ over\ N\ obs} + \ohp{1}
			 \\
			 &=
			 0 + \sqrt{N}\vec{\xi}_N(\Btheta_0) + \ohp{1},
\end{split}
\end{equation}
where the second equality follows because~$\Btheta_0$ is an interior minimizer of the differentiable population loss~$\Leps$ under Condition~\ref{ass:interior} and Assumptions~\ref{ass:model}--\ref{ass:rho}, and hence its gradient vanishes. The random vector~$\vec{\xi}_N$ is defined in~\eqref{eq:xi}. Notably,~$\vec{\xi}_N$ is proportional to a sum over the~$N$ observations. Hence, the central limit theorem can be applied to establish asymptotic normality of~$\sqrt{N}\vec{\xi}_N(\Btheta_0)$. Lemma~\ref{lem:xi-limit} works out the details.

Finally, combining~\eqref{eq:graddecomp} with~\eqref{eq:Mestimation-linearization} and the fact that $\sqrt{N}\vec{\xi}_N(\Btheta_0) = \Ohp{1}$ (Lemma~\ref{lem:xi-limit}) yields the first assertion on the asymptotic linearization after writing out the definition of~$\vec{\xi}_N(\Btheta_0)$ in Equation~\eqref{eq:xi}.

Regarding the second assertion on asymptotic normality, 
\begin{align*}
	\sqrt{N}\vecfun{\eta}{\Btheta_0,\fhat} 
	&= 	\sqrt{N}\vec{\xi}_N(\Btheta_0) + \ohp{1} \qquad(Eq.~\eqref{eq:graddecomp})
	\\
	&\convweak \gauss_d\left(\vec{0}, \matfuneps{U}{\Btheta_0} \right),
\end{align*}
by Lemma~\ref{lem:xi-limit}. The assertion on asymptotic normality now follows from~\eqref{eq:Mestimation-linearization} in conjunction with Slutsky's lemma. This completes the proof.
\end{proof}

\subsection{Proof of Theorem~\ref{thm:IF}}
Theorem~\ref{thm:IF} is re-stated in Theorem~\ref{thm:IF-detailed} below, with the additional regularity conditions spelled out. However, before we can state and subsequently prove Theorem~\ref{thm:IF-detailed}, we require additional notation and some background on directional derivatives.

Fix $\z\in\Z$. Consider a Huber-type sampling PMF with point-mass contamination at~$\z$,
\begin{equation*}
	\fepsdelta(\y) = (1-\eps)p_{\Btheta_*}(\y) + \eps \Delta_{\z}(\y), \qquad \y\in\Z,
\end{equation*}
where $\y\mapsto\Delta_{\z}(\y) = \I{\y = \z}$. For this particular sampling PMF and  under the assumptions of Theorem~\ref{thm:asy-normality}, a $C$-estimation estimand~$\Btheta_0$ is characterized as a solution to the estimating equation~$\vec{0} = \sum_{\y\in\Z}\fun{A}{\fepsdelta(\y)\big/\pfun{\Btheta_0}{\y}-1}\gradient\pfun{\Btheta_0}{\y}$, an identity already used in~\eqref{eq:estimatingeq}. When $\eps >0$, the solution~$\Btheta_0$ implicitly depends on the contamination fraction~$\eps$. To accommodate this implicit dependence, reparameterize the estimating equation by defining 
\begin{equation}\label{eq:Psi}
	\vec{\Psi}_{\z}(\Btheta,\eps)
	=
	\sum_{\y\in\Z}
	\fun{A}{\frac{\fepsdelta(\y)}{\pfun{\Btheta}{\y}} - 1}
	\gradient p_{\Btheta}(\y).
\end{equation}
Correspondingly, define by $\Btheta_{\z}(\eps)$ the solution to $\vec{0} = \vec{\Psi}_{\z}(\Btheta,\eps)$ in the first argument, where we have made explicit the dependence of this solution on the fixed~$\z$ and the contamination fraction~$\eps$.

It will be useful to derive the directional derivative of~$\vec{\Psi}_{\z}$ in~\eqref{eq:Psi} along the path $(\Btheta, \eps) = (\Btheta_* + \eps\Bv,\eps)$ in a fixed direction $\Bv\in\R^d$ and evaluated at $(\Btheta,\eps) = (\Btheta_*,0)$. For this purpose, define the PR associated with~$\fepsdelta$ 
\[
	r_{\z,\y}(\Btheta,\eps) = \frac{\fepsdelta(\y)}{\pfun{\Btheta}{\y}}-1
\]
where we have used the same parametrization as for~$\vec{\Psi}_{\z}$ in~\eqref{eq:Psi} 
The definition of $\vec{\Psi}_{\z}$ then gives
\[
	\vec{\Psi}_{\z}(\Btheta_* + \eps\Bv,\eps)
	=
	\sum_{\y\in\Z} \fun{A}{ r_{\z,\y}(\Btheta_* + \eps\Bv,\eps) }\gradient\pfun{\Btheta_* + \eps\Bv}{\y}.
\]
For each fixed cell $\y\in\Z$, the directional derivative of $r_{\z,\y}$ along the path $(\Btheta_* + \eps\Bv,\eps)$ and evaluated at $(\Btheta_*,0)$ is then given by 
\begin{align*}
	\lim_{\eps\downarrow 0} \frac{ r_{\z,\y}(\Btheta_* + \eps\Bv,\eps) - r_{\z,\y}(\Btheta_*, 0) }{\eps}
	&=
	\partialderivative{}{\eps} r_{\z,\y}(\Btheta_* + \eps\Bv,\eps)\Big|_{\eps=0}
	\\
	&=
	\partialderivative{}{\eps}\frac{\fepsdelta(\y)}{\pfun{\Btheta_* + \eps\Bv}{\y}}\Big|_{\eps=0}
	\\
	&=
	\frac{\Delta_{\z}(\y)}{\pfun{\Btheta_*}{\y}} - 1 - \Bv^\top \sfun{\Btheta_*}{\y},
\end{align*}
Thus, provided that differentiation and summation may be interchanged, directionally differentiating~$\vec{\Psi}_{\z}$ along the path $(\Btheta_* + \eps\Bv,\eps)$ with evaluation at $(\Btheta_*,0)$ yields by the chain rule
\begin{equation}\label{eq:directionalderivative}
\begin{split}
	&\lim_{\eps\downarrow 0} \frac{ \vec{\Psi}_{\z}(\Btheta_* + \eps\Bv,\eps) - \vec{\Psi}_{\z}(\Btheta_*, 0) }{\eps}
	\\
	&\quad=
	\partialderivative{}{\eps} \vec{\Psi}_{\z}(\Btheta_* + \eps\Bv,\eps)\Big|_{\eps=0}
	\\
	&\quad=
	\sum_{\y\in\Z}
	\left[
		\left(\partialderivative{}{\eps} \fun{A}{r_{\z,\y}(\Btheta_* + \eps\Bv,\eps)}\right)\gradient\pfun{\Btheta_*+\eps\Bv}{\y}
		+
		\fun{A}{r_{\z,\y}(\Btheta_* + \eps\Bv,\eps)} \left( \partialderivative{}{\eps}\gradient\pfun{\Btheta_*+\eps\Bv}{\y}\right)
	\right] \Bigg|_{\eps=0}
	\\
	&\quad=
	\sum_{\y\in\Z}
	\left(\partialderivative{}{\eps} \fun{A}{r_{\z,\y}(\Btheta_* + \eps\Bv,\eps)}\right)\gradient\pfun{\Btheta_*+\eps\Bv}{\y} \Bigg|_{\eps=0}
	\\
	&\quad=	
	A'(0) \sum_{\y\in\Z}
	\left[
			\frac{\Delta_{\z}(\y)}{\pfun{\Btheta_*}{\y}} - 1 - \Bv^\top \sfun{\Btheta_*}{\y}
	\right]\gradient\pfun{\Btheta_*}{\y}
	\\
	&\quad=
	A'(0) \big( \sfun{\Btheta_*}{\z} - \matfun{J}{\Btheta_*}\Bv\big),
	\end{split}
\end{equation}
where we have used in the third equality that $A(0)=0$ and in the final equality that $\gradient p_{\Btheta_*}(\y)=p_{\Btheta_*}(\y)\sy{\Btheta_*}$ and $\sum_{\y}\gradient p_{\Btheta_*}(\y)=\vec{0}$. Equation~\eqref{eq:directionalderivative} is useful because it is the estimating-equation analogue of the first-order von Mises calculation underlying the IF \citep{hampel1986}.

Recall that a $C$-estimate $\Btheta_{\z}(\eps)$ is the solution to $\vec{0} = \vec{\Psi}_{\z}(\Btheta,\eps)$. We will see in the proof of the theorem that this solution possesses the path characterization $\Btheta_{\z}(\eps) = \Btheta_* + \eps \Bv_{\eps}$ for~$\eps$ small enough, where the direction~$\Bv_\eps$ is bounded and depends on~$\eps$. Because of this dependence, we cannot directly apply~\eqref{eq:directionalderivative} to evaluate the directional derivative of~$\vec{\Psi}_{\z}$ along the $C$-estimate~$\Btheta_{\z}(\eps)$ because~\eqref{eq:directionalderivative} requires the direction to be fixed. To be able to apply~\eqref{eq:directionalderivative} to~$\Btheta_{\z}(\eps)$, we impose the following condition, which ensures that the derivative can be evaluated at an as-yet unknown, moving direction.

\begin{enumerate}[label={\textnormal{IF.\arabic*.}}, ref={\textnormal{IF.\arabic*}}, itemindent=25pt]
	\item The fixed-direction derivative in~\eqref{eq:directionalderivative} is locally uniform in the direction: for every $B<\infty$,
	\begin{equation*}
		\sup_{\|\vec{v}\|\leq B}
		\frac{1}{\eps}
		\left\|
		\vec{\Psi}_{\z}(\Btheta_*+\eps\vec{v},\eps)
		-\eps A'(0)\big(\sz{\Btheta_*}-\matfun{J}{\Btheta_*}\vec{v}\big)
		\right\|
		\longrightarrow 0
		\quad\text{as }\eps\downarrow0.
	\end{equation*}
	In particular, the expansion remains valid for every bounded moving sequence~$\vec{v}_\eps$. 
	\label{ass:IF-uniform}
\end{enumerate}

Furthermore, we require the estimand $\Btheta_{\z}(\eps)$ to be well-defined and sufficiently regular when~$\eps$ is small.

\begin{enumerate}[label={\textnormal{IF.\arabic*.}}, ref={\textnormal{IF.\arabic*}}, itemindent=25pt, start=2]
	\item  There is a neighborhood $\mathcal U_{\z}\subset\operatorname{int}(\BTheta)$ of~$\Btheta_*$ such that, for all sufficiently small~$\eps$, every global minimizer of the population loss $\D{\fepsdelta}{\model}$ lies in~$\mathcal U_{\z}$ and $\vec{\Psi}_{\z}(\Btheta,\eps)=\vec{0}$ has a unique root $\Btheta_{\z}(\eps)$ there. Moreover, $\Btheta_{\z}(0)=\Btheta_*$ and $\|\Btheta_{\z}(\eps)-\Btheta_*\|=O(\eps)$ for the sufficiently small~$\eps>0$.
	\label{ass:IF-regular}
\end{enumerate}

We are finally ready to state the detailed theorem, which nests Theorem~\ref{thm:IF} from the main text.

\begin{theorem}\label{thm:IF-detailed}
Grant the assumptions of Theorem~\ref{thm:asy-normality} at the correctly specified model, so $\eps=0$ and $\Btheta_0=\Btheta_*$. Grant the local conditions~\ref{ass:IF-uniform} and~\ref{ass:IF-regular} for every fixed~$\z\in\Z$.
Then the influence function of the $C$-estimator~$\Bthetahat$ exists and is equal to
\[
	\IF{\z, \Bthetahat, p_{\Btheta_*}}
	=
	\matinv{J}{\Btheta_*}\sz{\Btheta_*}
	=
	\IF{\z, \BthetahatMLE, p_{\Btheta_*}}.
\]
\end{theorem}

\begin{proof}
Fix $\z\in\Z$ and abbreviate $\Btheta_\eps=\Btheta_{\z}(\eps)$. By Condition~\ref{ass:IF-regular},
\[
	\Bv_\eps=\frac{\Btheta_\eps-\Btheta_*}{\eps}
\]
is bounded for sufficiently small $\eps>0$. Because $\Btheta_\eps$ is an interior root of the population estimating equation,
\[
	\vec{0}=\vec{\Psi}_{\z}(\Btheta_\eps,\eps)
	=\vec{\Psi}_{\z}(\Btheta_*+\eps\Bv_\eps, \eps).
\]
Condition~\ref{ass:IF-uniform} therefore gives
\begin{equation}\label{eq:IF-proof-linearization}
	\vec{0}
	=
	A'(0)\big(\sz{\Btheta_*}-\matfun{J}{\Btheta_*}\Bv_\eps\big) + o(1).
\end{equation}
The calculation in~\eqref{eq:directionalderivative} identifies the contamination and parameter terms in this linearization.

By Lemma~\ref{lem:IFlimits}, $A'(0)\matfun{J}{\Btheta_*}=\matfuneps{M}{\Btheta_*}|_{\eps=0}$ is positive definite and hence nonsingular. It follows from~\eqref{eq:IF-proof-linearization} that every convergent subsequence of~$\Bv_\eps$ has the same limit,
\[
	\matinv{J}{\Btheta_*}\sz{\Btheta_*}.
\]
Boundedness then implies convergence of the full sequence. Thus the directional derivative of the population functional exists and
\[
	\lim_{\eps\downarrow0}\frac{\Btheta_{\z}(\eps)-\Btheta_*}{\eps}
	=
	\matinv{J}{\Btheta_*}\sz{\Btheta_*},
\]
which is the ML influence function.
\end{proof}

\subsection{Proof of Theorem \ref{thm:maxbias-general}}
Fix a contamination fraction $\eps\in [0,0.5)$. Let $\Z\ni \y \mapsto \Delta_{\z}(\y) = \I{\y = \z}$ be the point-mass probability function at~$\z\in\Z$ and define by
\begin{equation*}
	\gepsdelta = (1-\eps)\modelt + \eps \Delta_{\z}
\end{equation*}
the contaminated PMF with point mass contamination at $\z$; note that $\gepsdelta\in\Ceps$. Every $f\in\Ceps$ can be represented as
\begin{equation}\label{eq:frepresentation}
	f = (1-\eps)\modelt + \eps h = \sum_{\z\in\Z} h(\z)\gepsdelta.
\end{equation}
For a fixed second argument $\modelt$, the map $f\mapsto \D{f}{\modelt}$ is convex because $\rho$ is convex. It therefore follows for every $f\in\Ceps$ that
\begin{equation*}
\begin{split}
	\D{f}{\modelt} 
	&=
	\D{ (1-\eps)\modelt + \eps h }{\modelt}
	\\
	&= 
	\D{\sum_{\z\in\Z} h(\z) \gepsdelta}{\modelt}
	\\
	&\leq
	\sum_{\z\in\Z} h(\z)\D{\gepsdelta}{\modelt}
	\\
	&\leq
	\sup_{\z\in\Z} \D{\gepsdelta}{\modelt},
\end{split}
\end{equation*}
where the second line is due to the representation in~\eqref{eq:frepresentation}, the third line to Jensen's inequality, and the final line to the fact that $h$ is a PMF, so the preceding weighted average is bounded by the supremum of its terms. The right-hand side in the final inequality does not depend on~$f$. Conversely, each $\gepsdelta$ belongs to~$\Ceps$. Therefore
\begin{equation}\label{eq:Dbarsup}
	\Rrho \equiv \sup_{f\in\Ceps} \D{f}{\modelt} = \sup_{\z\in\Z} \D{\gepsdelta}{\modelt}.
\end{equation}
If the right-hand supremum is not attained, choose cells $\z_m$ such that $\D{g_\eps^{\Delta_{\z_m}}}{\modelt}\to\Rrho$; thus the equality remains valid without a single maximizing cell.
The PR of the contaminated PMF $\gepsdelta$ evaluated at the true parameter~$\Btheta_*$ is given by
\[
	\y\mapsto
	\frac{\gepsdelta(\y)}{\modelt(\y)} - 1
	=
	\begin{cases}
		 \frac{\eps}{\modelt(\y)} - \eps  &\textnormal{ if } \y = \z,
		\\
		-\eps &\textnormal{ if } \y \neq \z.
	\end{cases}
\]
Substituting this PR into the definition of the divergence gives
\begin{align*}
	\D{\gepsdelta}{\modelt}
	=
	\sum_{\y\in\Z} \fun{\rho}{ \frac{\gepsdelta(\y)}{\modelt(\y)} - 1 }  \modelt(\y)
	=
	\modelt(\z) \fun{\rho}{  \frac{\eps}{\modelt(\z)} - \eps } + (1-\modelt(\z)) \rho(-\eps).
\end{align*}
Taking suprema over $\Z$,~\eqref{eq:Dbarsup} takes the concrete expression
\[
	\Rrho = \sup_{\z\in\Z}\left[ \pz{\Btheta_*} \fun{\rho}{\frac{\eps}{\pz{\Btheta_*}}-\eps} + (1-\pz{\Btheta_*})\rho(-\eps) \right],
\]
proving the first assertion. 

We now prove the asserted upper bound of the maxbias. Keeping $\eps\in [0,0.5)$ fixed, take any contaminated PMF $f\in\Ceps$ and any associated estimand $\Btheta_f\in\cT{f}$, where the latter's subscript reminds us that the estimand depends on~$f$. Since~$\Btheta_f$ globally minimizes the population divergence, comparison with the true parameter~$\Btheta_*$ gives
\[
	\D{f}{p_{\Btheta_f}} \leq \D{f}{\modelt}\leq \Rrho \qquad \textnormal{ for all } f\in\Ceps \textnormal{ and } \Btheta_f\in \cT{f}.
\] 
By definition of the two-center divergence radius, this inequality implies that
\[
	\Gammarho{p_{\Btheta_f},\modelt} \leq \Rrho \qquad \textnormal{ for all } f\in\Ceps \textnormal{ and } \Btheta_f\in \cT{f}.
\]
It follows from the previous two displays that every parameter $\Btheta_f\in\cT{f}$ produced by any $f\in\Ceps$ belongs to the set $\left\{\Btheta\in\BTheta : \Gammarho{\model,\modelt} \leq \Rrho \right\}$. Taking suprema over this set inclusion gives
\begin{align*}
	\Beps &\stackrel{\textnormal{Def.}~\ref{def:maxbias}}{=} 
	\sup_{f\in\Ceps}
\sup_{\Btheta_f\in\cT{f}}
\|\Btheta_f-\Btheta_*\|
\\
	&\leq
	\sup_{ \left\{\Btheta\in\BTheta : \Gammarho{\modelt,\model} \leq \Rrho \right\} } \|\Btheta-\Btheta_*\|,
\end{align*}
proving the upper bound of the maxbias.

It remains to prove the asserted lower bound of the maxbias. 
Suppose first that $\eps > 0$. Choose a $\bar{\Btheta}\in\BTheta$ such that
\begin{equation}\label{eq:epscrit-cond}
	\epscrit{\bar{\Btheta}} \equiv 1 - \inf_{\z\in\Z} \frac{\pz{\bar{\Btheta}}}{\pz{\Btheta_*}} \leq \eps.
\end{equation}
This condition is equivalent to
\[
	\pz{\bar{\Btheta}} \geq (1-\eps) \pz{\Btheta_*} \qquad \textnormal{ for all } \z\in\Z.
\]
Therefore, the function $\heps$ defined by
\[
	\heps(\z) = \frac{\pz{\bar{\Btheta}} - (1-\eps)\pz{\Btheta_*}}{\eps}, \qquad \z\in\Z,
\]
is nonnegative, sums up to~1, and satisfies the identity
\begin{equation}\label{eq:heps-identity}
	\pz{\bar{\Btheta}} = (1-\eps)\modelt + \eps \heps.
\end{equation}
Thus, $\heps$ is a contamination PMF with the property that when it is used in the Huber-like contamination model~\eqref{eq:hubermodel}, the generated contaminated PMF~$p_{\bar{\Btheta}}$ is contained in the postulated model $\{\model: \Btheta\in\BTheta\}$. In addition, $p_{\bar{\Btheta}}\in\Ceps$. Proposition~\ref{prop:divergence-nonnegative} shows that the divergence between~$p_{\bar{\Btheta}}$ and any model PMF $\model\in\{\model: \Btheta\in\BTheta\}$
\[
	\D{p_{\bar{\Btheta}}}{\model} \stackrel{\textnormal{Eq.}~\eqref{eq:heps-identity}}{=} \D{(1-\eps)\modelt + \eps \heps}{\model}
\]
is minimized at value~0 exactly when $p_{\bar{\Btheta}} = \model$. Point identification in Assumption~\ref{ass:model} then gives $\bar{\Btheta} = \Btheta$, so $\cT{p_{\bar{\Btheta}}} = \{\bar{\Btheta}\}$. Thus, the contamination PMF~$\heps$ forces the bias $\|\bar{\Btheta} - \Btheta_*\|$. Consequently, 
\begin{align*} 
	\Beps &\stackrel{\textnormal{Def.}~\ref{def:maxbias}}{=} 
	\sup_{f\in\Ceps}
	\sup_{\Btheta\in\cT{f}}
	\|\Btheta-\Btheta_*\|
	\\
	&\geq
	\sup_{\Btheta\in\cT{(1-\eps)\modelt + \eps \heps}} \|\Btheta-\Btheta_*\|
	\\
	&=
	\sup_{\Btheta\in\cT{p_{\bar{\Btheta}}}} \|\Btheta-\Btheta_*\|
	\\
	&=
	\|\bar{\Btheta} - \Btheta_*\|.
\end{align*}
This bound holds true for all $\bar{\Btheta}$ satisfying~\eqref{eq:epscrit-cond}. Taking the supremum over all such parameters gives
\[
	\Beps
	\geq
	\sup_{ \{\Btheta\in\BTheta : \epscrit{\Btheta} \leq \eps \} } \|\Btheta-\Btheta_*\|,
\]
thereby proving the lower bound for $\eps > 0$. If $\eps = 0$, the constraint $\epscrit{\Btheta}\leq 0$ in~\eqref{eq:epscrit-cond} forces $\model = \modelt$ and hence $\Btheta = \Btheta_*$ for all $\Btheta\in\BTheta$, so the same conclusion holds trivially. The proof is complete. \QED

\subsection{Proof of Corollary \ref{cor:maxbias-RW}}
Let $(c_1,c_2)$ be the tuning constants in the (normalized) Huber-like discrepancy function~$\rho\equiv\rho_\RW$ from~\eqref{eq:rw01}. For $\eps=0$, all asserted maximum divergences and upper bounds are zero. We therefore fix $\eps\in(0,0.5)$ throughout the remainder of the proof. To maximize the expression in~\eqref{eq:maxbias-maxdiv}, write
\begin{equation}\label{eq:suprRW}
	\RRW = \sup_{\z\in\Z}\big\{ \rRW{\pz{\Btheta_*}}\big\},
\end{equation}
where, for $q\in(0,1]$,
\[
	\rRW{q}
	=
	q\rho\left(\frac{\eps}{q}-\eps\right)+(1-q)\rho(-\eps).
\]
Let $\delta_{\eps,q}=\eps/q-\eps$. Since $\mathrm{d}\delta_{\eps,q}/\mathrm{d}q=-\eps/q^2=-(\delta_{\eps,q}+\eps)/q$, the product and chain rules give
\[
\frac{\d \rRW{q}}{\d q}
=
\rho(\delta_{\eps,q})-(\delta_{\eps,q}+\eps)\rho'(\delta_{\eps,q})-\rho(-\eps)
\leq0.
\]
The inequality follows from convexity because the supporting-line inequality at $\delta_{\eps,q}$ gives $\rho(-\eps)\geq\rho(\delta_{\eps,q})-(\delta_{\eps,q}+\eps)\rho'(\delta_{\eps,q})$. Thus $r_{\mathrm{RW}}$ is nonincreasing in~$q$, and the supremum in~\eqref{eq:suprRW} is obtained, or approached, by cells with the smallest model probabilities. If~$|\Z|$ is finite, the smallest probability is $\pmin>0$, and direct substitution gives
\begin{equation}\label{eq:maxdiver_RW}
\begin{split}
	\RRW
	=
	\pmin \rho\left( \frac{\eps}{\pmin} -\eps \right)+ (1-\pmin) \rho(-\eps).
\end{split}
\end{equation}
This proves the finite-support assertion.

Suppose now that $|\Z|=\infty$. Then $\pmin=0$. If $c_2<\infty$, upper clipping is active for all sufficiently small~$q$. The upper branch of the normalized discrepancy in~\eqref{eq:rw01} simplifies to $\rho(\delta)=(\delta+1)\log(1+c_2)-c_2$. Hence,
\begin{align*}
	\lim_{q\downarrow0}q\rho\left(\frac{\eps}{q}-\eps\right)
	&=
	\lim_{q\downarrow0}
	q\left[\left(\frac{\eps}{q}+1-\eps\right)\log(1+c_2)-c_2\right]
	\\
	&=\eps\log(1+c_2).
\end{align*}
Therefore,
\begin{align*}
	\RRW
	&=
	\lim_{q\downarrow0}
	\left\{q\rho\left(\frac{\eps}{q}-\eps\right)+(1-q)\rho(-\eps)\right\}
	\\
	&=\eps\log(1+c_2)+\rho(-\eps),
\end{align*}
which proves the countably infinite-support formula for finite~$c_2$. If $c_2=\infty$, upper clipping is absent and the normalized middle branch gives
\[
	\lim_{q\downarrow0}q\rho\left(\frac{\eps}{q}-\eps\right)
	=
	\lim_{q\downarrow0}
	\left[
	(\eps+q(1-\eps))\log\left(\frac{\eps}{q}+1-\eps\right)-\eps+q\eps
	\right]
	=\infty,
\]
so $\RRW=\infty$, as claimed.

We next establish the upper bound for the maximum bias. Let $\Z$ be at most countable and define, directly from the normalized discrepancy,
\[
	\phi(x)=\rho(x-1),\qquad x\geq0.
\]
For any PMFs $g_1,g_2$ on~$\Z$, this is merely a reparametrization of the same divergence:
\begin{equation}\label{eq:Dphi-identity}
	\Dphi{g_1}{g_2}
	\equiv
	\sum_{\z\in\Z}\fun{\phi}{\frac{g_1(\z)}{g_2(\z)}}g_2(\z)
	=
	\sum_{\z\in\Z}\fun{\rho}{\frac{g_1(\z)}{g_2(\z)}-1}g_2(\z)
	=
	\D{g_1}{g_2}.
\end{equation}
By Lemma~\ref{lem:phi-bounds}, $\phi$ satisfies the following inequalities:
\begin{align}
	\phi(x) &\geq \kappa_-(1-x)^2 \qquad \textnormal{ if }x \in [0, 1),\label{eq:phiineqmain-upto1}
	\\
	\phi(x) &\geq \kappa_+\frac{(1-x)^2}{x} \qquad \textnormal{ if }x \geq 1.\label{eq:phiineqmain-beyond1}
\end{align}

Let $f\in\Ceps$ be arbitrary and put
\[
		V_f = \TV{f,\modelt} , \qquad \cI_- = \{\z\in\Z : f(\z) < \pz{\Btheta_*} \},\qquad \cI_+ = \{\z\in\Z : f(\z) > \pz{\Btheta_*} \}.
\]
Observe that
\begin{equation}\label{eq:Videntity}
	\sum_{\z\in\cI_-}\big( \modelt(\z) - f(\z) \big)
	=
	\sum_{\z\in\cI_+}\big( f(\z) - \modelt(\z) \big)
	=
	V_f.
\end{equation}

For all $\z\in\cI_-$, it holds true that $f(\z)/\pz{\Btheta_*} \in [0,1)$. Therefore,
\begin{align*}
	\sum_{\z\in\cI_-} \modelt(\z) \fun{\phi}{\frac{f(\z)}{\modelt(\z)}}
	&\geq
	\kappa_- \sum_{\z\in\cI_-} \modelt(\z) \left(1-\frac{f(\z)}{\modelt(\z)}\right)^2
	\\
	&=
	\kappa_- \sum_{\z\in\cI_-} \frac{\big(\modelt(\z) - f(\z)\big)^2}{\modelt(\z)}
	\\
	&\geq
	\kappa_- \left(\sum_{\z\in\cI_-} \big(\modelt(\z) - f(\z)\big)\right)^2 \Bigg/ \sum_{\z\in\cI_-} \modelt(\z)
	\\	
	&\geq
	\kappa_- \left(\sum_{\z\in\cI_-} \big(\modelt(\z) - f(\z)\big)\right)^2 
	\\
	&=
	\kappa_- V_f^2,
\end{align*}
where the first line is due to~\eqref{eq:phiineqmain-upto1}, the third one to Bergström's inequality, the fourth one to the fact that $\sum_{\z\in\cI_-}\modelt(\z)\leq 1$, and the fifth one to~\eqref{eq:Videntity}. 
Following analogous steps on~$\cI_+$, 
\[
	\sum_{\z\in\cI_+} \modelt(\z) \fun{\phi}{\frac{f(\z)}{\modelt(\z)}}
	\geq
	\kappa_+ \sum_{\z\in\cI_+} \frac{\big(\modelt(\z) - f(\z)\big)^2}{f(\z)} \geq \kappa_+ V_f^2,
\]
where we have used~\eqref{eq:phiineqmain-beyond1} to obtain the first inequality and $\sum_{\z\in\cI_+}f(\z)\leq 1$ in the last. Adding the two parts gives
\begin{equation}\label{eq:TV2-ineq}
\begin{split}
	\D{f}{\modelt} &= \Dphi{f}{\modelt}
	\\
	&=
	 \sum_{\z\in\cI_-} \modelt(\z) \fun{\phi}{\frac{f(\z)}{\modelt(\z)}} 
	 +
	 \sum_{\z\in\cI_+} \modelt(\z) \fun{\phi}{\frac{f(\z)}{\modelt(\z)}}
	 \\
	 &\geq
	 \kappa_- V_f^2 + \kappa_+ V_f^2
	 \\
	 &=
	 \kappac \TV{f,\modelt}^2,
\end{split}
\end{equation}
where we have used~\eqref{eq:Dphi-identity} in the first line, the fact that $\phi(1) = 0$ in the second line, and the definition $\kappac = \kappa_- + \kappa_+$ in the final line.

Now let $\Btheta_f \in \cT{f}$ be an estimand at the PMF~$f\in\Ceps$. By the definitions of~$\Btheta_f$ as a global population minimizer and~$\RRW$ as largest divergence, 
\[
	\D{f}{p_{\Btheta_f}} \leq \D{f}{\modelt} \leq \RRW.
\] 
Combining this sandwich inequality with~\eqref{eq:TV2-ineq} yields
\[
	\TV{f,p_{\Btheta_f}} \leq \sqrt{\frac{\RRW}{\kappac}} 
	\qquad \textnormal{and}\qquad
	\TV{f,\modelt} \leq \sqrt{\frac{\RRW}{\kappac}} .
\]
Using these inequalities together with the triangle inequality and the fact that the TV distance is bounded from above by~1 gives
\begin{align*}
	\TV{p_{\Btheta_f}, \modelt}
	&=
	\frac{1}{2}\sum_{\z\in\Z} \Big| \pz{\Btheta_f} - f(\z) + \big(f(\z) - \pz{\Btheta_*}\big) \Big|
	\\
	&\leq \min\left\{ 1, \TV{f,p_{\Btheta_f}} + \TV{f,\modelt} \right\} 
	\\
	&\leq
	\min\left\{ 1, 2\sqrt{\RRW / \kappac}\right\}.
\end{align*}
Since $f$ is arbitrary, this inequality holds for all $f\in\Ceps$. Taking the supremum over all admissible population minimizers yields 
\begin{align*}
	\BRW &= \sup_{f\in\Ceps}\sup_{\Btheta\in\cT{f}}\|\Btheta - \Btheta_*\|
	\\
	&\leq
	 \sup_{\left\{ \Btheta\in\BTheta : \TV{\model,\modelt} \leq  \min\left\{ 1, 2\sqrt{\RRW / \kappac}\right\}\right\}} \|\Btheta - \Btheta_*\|.
\end{align*}
This proves the assertion on the maxbias upper bound, completing the proof. \QED

\subsection{Proof of Corollary~\ref{cor:maxbias-increases-RW}}
Corollary \ref{cor:maxbias-increases-RW} is nested in the following proposition, which we first state and then prove. Table~\ref{tab:pcrit} illustrates the sufficient condition therein.

\begin{table}[H]
\small
\begin{tabular}{l | r r r r r r r r r r}
$\eps$              &0.01    &0.05    &0.10    &0.15   &0.20   &0.25   &0.30   &0.35   &0.40   &0.45  \\\hline
$\pcrit$      &.0098   &.0453   &.0823   &.1121  &.1357  &.1536  &.1662  &.1736  &.1756  &.1719\\
Sufficient $|\Z|$   &103     &23      &13      &9      &8      &7      &7      &6      &6      &6     \\
\end{tabular}
\caption{Values of~$\pcrit$ and the smallest cardinality sufficient for condition~\eqref{eq:pcrit}.}
\label{tab:pcrit}
\end{table}

\begin{proposition}\label{prop:RW-radius-detailed}
Consider the setting of Corollary~\ref{cor:maxbias-increases-RW}, fix $\eps\in(0,0.5)$, and let $L=\log(1+c_2)$. For $r\in[0,1]$, define the radial model envelope
\[
	H_{\Btheta_*}(r)
	=
	\sup_{\{\Btheta\in\BTheta:\,\TV{\model,\modelt}\leq r\}}
	\|\Btheta-\Btheta_*\|.
\]
Thus $\overline B_\eps(c_2)=H_{\Btheta_*}\{\Eeps{c_2}\}$, and $H_{\Btheta_*}$ is nondecreasing.
\begin{enumerate}
	\item Suppose that $\Z$ is countably infinite. Set $C_\eps=\eps+(1-\eps)\log(1-\eps)$, define $Q_\eps(L)=\RRW/\kappa_{-1,c_2}$, and let
	\[
	L_{\mathrm{sat},\eps}=
	\begin{cases}
	0,
	& C_\eps\geq 1/8,
	\\[2pt]
	\displaystyle\frac{1/8-C_\eps}{\eps-1/4},
	& C_\eps<1/8\ \textnormal{and}\ C_\eps+\eps/2\geq1/4,
	\\[8pt]
	\displaystyle\frac{1/4-C_\eps}{\eps},
	& C_\eps+\eps/2<1/4.
	\end{cases}
	\]
	\begin{enumerate}[label={\textnormal{(\alph*)}}, leftmargin=24pt]
		\item The maximum divergence~$\RRW$ is strictly increasing in~$c_2$.
		\item The ratio $Q_\eps(L)$ is strictly increasing in~$L$ and thus also in~$c_2$.
		\item The radius~$\Eeps{c_2}$ is strictly increasing for $0\leq c_2<\exp(L_{\mathrm{sat},\eps})-1$ and equals~$1$ thereafter; the first range is empty when $L_{\mathrm{sat},\eps}=0$.
		\item The envelope~$\overline B_\eps(c_2)$ is nondecreasing in~$c_2$ and is constant after saturation. Before saturation it is strictly increasing if and only if $H_{\Btheta_*}$ is strictly increasing over the corresponding attained radii.
	\end{enumerate}

	\item Suppose that $\Z$ is finite with $|\Z|\geq2$. Retain $\pmin$, and define
	\[
	\teps=\frac{\eps}{\pmin}-\eps+1,
	\qquad
	\Teps=\log(\teps),
	\qquad
	Q_{\eps,\Btheta_*}(L)=\frac{\RRW}{\kappa_{-1,c_2}}.
	\]
	\begin{enumerate}[label={\textnormal{(\alph*)}}, leftmargin=24pt]
		\item The maximum divergence is strictly increasing for $0\leq L<\Teps$ and constant for $L\geq\Teps$.
		\item The ratio $Q_{\eps,\Btheta_*}$ is nondecreasing if and only if $\pmin\leq\pcrit$.
		\item The radius~$\Eeps{c_2}$ is nondecreasing if and only if
		\begin{equation}\label{eq:finite-radius-monotonicity}
		\pmin\leq\pcrit
		\qquad\textnormal{or}\qquad
		Q_{\eps,\Btheta_*}(1/2)\geq1/4.
		\end{equation}
		When~\eqref{eq:finite-radius-monotonicity} holds, set
		\[
		L_{\mathrm{sat},\eps,\Btheta_*}
		=
		\inf\{L\geq0:Q_{\eps,\Btheta_*}(L)\geq1/4\},
		\qquad
		L_{\mathrm{end},\eps,\Btheta_*}
		=
		\min\{\Teps,L_{\mathrm{sat},\eps,\Btheta_*}\},
		\]
		where $\inf\varnothing=\infty$. The radius is strictly increasing for $0\leq c_2<\exp(L_{\mathrm{end},\eps,\Btheta_*})-1$ and constant thereafter. Under $\pmin\leq\pcrit$, its terminal value equals~$1$ when $Q_{\eps,\Btheta_*}(\Teps)\geq1/4$ and equals $2\sqrt{Q_{\eps,\Btheta_*}(\Teps)}<1$ when $Q_{\eps,\Btheta_*}(\Teps)<1/4$.
		\item The number~$\pcrit$ belongs to~$(0,1)$, and $|\Z|\geq1/\pcrit$ is sufficient for $\pmin\leq\pcrit$. Whenever the radius is nondecreasing, so is $\overline B_\eps(c_2)$. On the strict-increase range in part~(c), the envelope is strictly increasing if and only if $H_{\Btheta_*}$ is strictly increasing over the corresponding attained radii; it is constant thereafter.
	\end{enumerate}
\end{enumerate}
\end{proposition}

\begin{proof}
Fix $\eps\in (0,0.5)$. 

First, suppose that $\Z$ is infinite, so $\pmin=0$. We now prove the four assertions made in part~1. 

\paragraph*{Part 1(a): Maximum divergence.}
The maximum divergence~$\RRW$ in Corollary~\ref{cor:maxbias-RW} can be expressed as
\[
	\RRW = \eps L + C_\eps,
\]
where $L=\log(c_2+1)$ and $C_\eps = \eps + (1-\eps)\log(1-\eps)$. Because $\eps>0$ and $L=\log(1+c_2)$ is strictly increasing in~$c_2$, this proves part~1(a).

\paragraph*{Part 1(b): Untruncated ratio.}
For $L\geq0$, define
\begin{equation}\label{eq:Qeps}
	\Qeps(L) = \frac{\eps L + C_\eps}{K(L)}, \qquad K(L) = \frac{1}{2} + \min\left\{\frac{1}{2}, L\right\} = \kappa_{-1,c_2}.
\end{equation}

\paragraph*{For $0\leq L\leq1/2$.}
Here $K(L)=1/2+L$, and therefore
\[
	\Qeps'(L) = \frac{\eps/2 -C_\eps}{(1/2+L)^2}. 
\]
With $\ueps = 1-\eps > 1/2$, the strict inequality $-\log\ueps > 1-\ueps = \eps$ implies
\[
	\eps / 2 - C_\eps = -\eps/2 -\ueps\log\ueps > \eps (\ueps - 1/2) > 0
\]
on $L\in [0, 1/2]$, so $\Qeps'(L) >0$ and thus $\Qeps(L)$ is strictly increasing on this interval.

\paragraph*{For $L>1/2$.}
Here $K(L)=1$ and $\Qeps'(L)=\eps>0$, so $\Qeps(L)$ strictly increases. This proves part~1(b).

\paragraph*{Part 1(c): Truncated radius and saturation.}
The identity $\Eeps{c_2}=\min\{1,2\sqrt{\Qeps(L)}\}$ and part~1(b) show that the radius is strictly increasing until it reaches~1 and is constant thereafter. The factor~2 outside the square root implies that saturation occurs exactly when $\Qeps(L)\geq1/4$. At the endpoints of the first branch,

\[
	\Qeps(0)=2C_\eps,
	\qquad
	\Qeps(1/2)=C_\eps+\eps/2.
\]
If $C_\eps\geq1/8$, the radius is therefore saturated already at~$L=0$. Suppose next that $C_\eps<1/8$ but $C_\eps+\eps/2\geq1/4$. The unique crossing is then in $[0,1/2]$ and solves
\[
	\frac{\eps L+C_\eps}{1/2+L}=\frac14,
	\qquad\textnormal{hence}\qquad
	L=\frac{1/8-C_\eps}{\eps-1/4}.
\]
The conditions of this case imply $\eps>1/4$, so the denominator is positive. Finally, if $C_\eps+\eps/2<1/4$, the crossing occurs for $L>1/2$, where $K(L)=1$, and hence
\[
	L=\frac{1/4-C_\eps}{\eps}.
\]
These three cases give $L_{\mathrm{sat},\eps}$ in the statement. Since $L=\log(1+c_2)$ is strictly increasing, they give the strict-increase endpoint $c_{2,\mathrm{sat},\eps}=\exp(L_{\mathrm{sat},\eps})-1$ and prove part~1(c).

\paragraph*{Part 1(d): Maximum-bias envelope.}
The TV balls defining $H_{\Btheta_*}$ are nested, so $H_{\Btheta_*}$ is nondecreasing. Part~1(c) and the identity $\overline B_\eps=H_{\Btheta_*}\circ E_{\eps,\Btheta_*}$ therefore imply that the envelope is nondecreasing and constant after saturation. Before saturation, it is strictly increasing if and only if $H_{\Btheta_*}$ is strictly increasing over the attained radii. This proves part~1(d) and therefore part~1 altogether.

Suppose now that $\Z$ is finite. Then $0<\pmin<1$ because $|\Z|\geq2$. Reparametrize the maximum divergence~$\RRW$ from Corollary~\ref{cor:maxbias-increases-RW} as
\[
	\Reps(L) = 
	\begin{cases}
		\pmin \big\{\teps (L+1) - \exp(L) \big\} + (1-\pmin)\ueps\log\ueps \quad &\textnormal{ if } L\in [0, \Teps),
		\\
		\pmin\teps\log\teps + (1-\pmin)\ueps\log\ueps \quad &\textnormal{ if } L\geq\Teps,
	\end{cases}	
\]
where $L = \log(c_2+1) \geq 0$, $\teps = \eps/\pmin -\eps+1$, $\Teps = \log(\teps)$, and $\ueps = 1-\eps$. Note that the two branches agree at $L=\Teps$. We proceed by proving the four assertions made in part~2.

\paragraph*{Part 2(a): Maximum divergence.}
Differentiation of the preceding display gives
\[
	\Reps'(L)=\pmin\{\teps-\exp(L)\}>0
	\quad\textnormal{for }0\leq L<\Teps,
\]
whereas $\Reps(L)$ is constant for $L\geq\Teps$. Thus the claimed strict increase and subsequent constancy of the maximum divergence follow. Define
\[
	Q_{\eps,\Btheta_*}(L)=\frac{\Reps(L)}{K(L)},
	\qquad
	K(L)=\frac12+\min\left\{\frac12,L\right\}.
\]

\paragraph*{Part 2(b): Untruncated ratio.}
We determine when $Q_{\eps,\Btheta_*}$ is nondecreasing. Let $L\in[0, \min\{\Teps, 1/2\}]$, so that upper clipping is active and $K(L) = 1/2 + L$. Then,
\[
	\Reps'(L) = \pmin (\teps - \exp(L)) \quad \textnormal{and}\quad K'(L) = 1,
\]
so that
\[
	Q_{\eps,\Btheta_*}'(L) = \frac{N_{\eps,\Btheta_*}(L)}{(1/2+L)^2},
\]
where the numerator is defined by
\[
	N_{\eps,\Btheta_*}(L) = \pmin\big\{ \exp(L) (1/2-L) - \teps/2 \big\} - (1-\pmin) \ueps\log\ueps.
\]
This numerator is strictly decreasing because
\[
	N_{\eps,\Btheta_*}'(L) = -\pmin \exp(L) (L+1/2) <0.
\]
At the left endpoint, with $\beps=-\ueps\log\ueps$,
\[
	N_{\eps,\Btheta_*}(0)=(1-\pmin)(\beps-\eps/2)>0,
\]
because $\beps>\ueps\eps>\eps/2$. At $L=1/2$,
\begin{equation}\label{eq:Neps12}
\begin{split}
	N_{\eps,\Btheta_*}(1/2) &= -\teps \pmin / 2 -(1-\pmin)\ueps\log\ueps
	\\
	&=\beps(1-\pmin)-\frac{\pmin\ueps+\eps}{2}.
\end{split}
\end{equation}
Therefore,
\begin{equation*}
	N_{\eps,\Btheta_*}(1/2)\geq0
	\quad\Longleftrightarrow\quad
	\pmin \leq \frac{2\beps - \eps}{2\beps+\ueps} = \pcrit.
\end{equation*}
Both the numerator and denominator defining~$\pcrit$ are positive by Lemma~\ref{lem:pcrit-equivalence}. We are now ready to prove the equivalence asserted in part~2(b).

\paragraph*{Sufficiency.}
Suppose that $\pmin\leq\pcrit$. Lemma~\ref{lem:pcrit-equivalence} gives $\Teps>1/2$. Because $N_{\eps,\Btheta_*}$ is strictly decreasing and $N_{\eps,\Btheta_*}(1/2)\geq0$, it follows that $N_{\eps,\Btheta_*}(L)>0$ for every $L<1/2$. Hence $Q_{\eps,\Btheta_*}$ is strictly increasing on $[0,1/2]$.

For $L\in(1/2,\Teps)$, $K(L)=1$, and therefore
\begin{align*}
	Q_{\eps,\Btheta_*}'(L) &= \pmin(\teps -\exp(L))
	\\
	&>\pmin(\teps -\exp(\Teps))
	\\
	&=0.
\end{align*}
Thus $Q_{\eps,\Btheta_*}$ is strictly increasing on this interval as well. For $L\geq\Teps$, both $\Reps(L)$ and $K(L)$ are constant, so $Q_{\eps,\Btheta_*}$ is constant. This proves that $\pmin\leq\pcrit$ is sufficient for monotonicity of~$Q_{\eps,\Btheta_*}$.

\paragraph*{Necessity.}
Conversely, suppose that $\pmin>\pcrit$. If $\Teps\geq1/2$, Equation~\eqref{eq:Neps12} gives $N_{\eps,\Btheta_*}(1/2)<0$. By continuity, $Q_{\eps,\Btheta_*}'(L)<0$ on a nonempty interval immediately to the left of~$1/2$, so $Q_{\eps,\Btheta_*}$ is not nondecreasing. If $\Teps<1/2$, then $\Reps$ is constant on $[\Teps,1/2]$ while $K(L)=1/2+L$ is strictly increasing. Hence $Q_{\eps,\Btheta_*}=\Reps/K$ is strictly decreasing on that interval. In either case, $\pmin>\pcrit$ precludes monotonicity. Thus $Q_{\eps,\Btheta_*}$ is nondecreasing if and only if~$\pmin\leq\pcrit$.

\paragraph*{Part 2(c): Truncated radius and strict-increase range.}
Because $\Eeps{c_2}=\min\{1,2\sqrt{Q_{\eps,\Btheta_*}(L)}\}$, part~2(b) immediately gives a nondecreasing radius when $\pmin\leq\pcrit$. Suppose instead that $\pmin>\pcrit$. The preceding argument shows that $Q_{\eps,\Btheta_*}$ has a strictly decreasing portion ending at $L=1/2$. All values on a decreasing portion are at least $Q_{\eps,\Btheta_*}(1/2)$: when $\Teps\geq1/2$, the ratio decreases to $Q_{\eps,\Btheta_*}(1/2)$ and then increases on $(1/2,\Teps)$; when $\Teps<1/2$, it decreases on $[\Teps,1/2]$ and is constant thereafter. Therefore, if $Q_{\eps,\Btheta_*}(1/2)\geq1/4$, truncation fixes the radius at~1 throughout every decreasing portion, and the radius is nondecreasing. Conversely, if $Q_{\eps,\Btheta_*}(1/2)<1/4$, continuity gives a nonempty interval immediately to the left of~$1/2$ on which the ratio is below~$1/4$ and strictly decreases; the radius then strictly decreases there. Hence
\[
	\Eeps{c_2}\ \textnormal{is nondecreasing in }c_2
	\quad\Longleftrightarrow\quad
	\pmin\leq\pcrit\ \textnormal{ or }\ Q_{\eps,\Btheta_*}(1/2)\geq1/4.
\]

Assume this condition and define $L_{\mathrm{sat},\eps,\Btheta_*}$ and $L_{\mathrm{end},\eps,\Btheta_*}$ as in the statement. If $\pmin\leq\pcrit$, part~2(b) shows that the ratio is strictly increasing on $[0,\Teps)$ and constant thereafter. The radius is therefore strictly increasing until $L=\Teps$ or its first saturation at~1, whichever occurs first, and is constant thereafter. Its terminal value follows by evaluating the constant ratio at~$\Teps$.

It remains to consider $\pmin>\pcrit$ and $Q_{\eps,\Btheta_*}(1/2)\geq1/4$. Since $N_{\eps,\Btheta_*}(0)>0$ and $N_{\eps,\Btheta_*}$ is strictly decreasing, the ratio is initially strictly increasing and has at most one turning point before~$1/2$. Its first crossing of~$1/4$ occurs on this increasing portion because every value on a subsequent decreasing portion is at least $Q_{\eps,\Btheta_*}(1/2)\geq1/4$. Thus the radius is strictly increasing before $L_{\mathrm{sat},\eps,\Btheta_*}$ and equals~1 thereafter. Moreover, $L_{\mathrm{sat},\eps,\Btheta_*}\leq\Teps$, so this range also ends at $L_{\mathrm{end},\eps,\Btheta_*}$. This proves part~2(c).

\paragraph*{Part 2(d): Cardinality and maximum-bias envelope.}
Lemma~\ref{lem:pcrit-equivalence} gives $\pcrit\in(0,1)$. Furthermore, $|\Z|\geq1/\pcrit$ is sufficient for~$\pmin\leq\pcrit$ because
\[
	1 = \sum_{\z\in\Z}\pz{\Btheta_*} \geq \sum_{\z\in\Z}\pmin = |\Z|\pmin
	\quad\Longrightarrow\quad
	\pmin \leq 1/|\Z|\leq\pcrit.
\]
To transfer the radius results to the maximum-bias envelope, let $c_2\leq c_2'$ and suppose that $\Eeps{c_2}\leq\Eeps{c_2'}$. Then
\[
	\left\{\Btheta:\TV{\model,\modelt}\leq\Eeps{c_2}\right\}
	\subseteq
	\left\{\Btheta:\TV{\model,\modelt}\leq\Eeps{c_2'}\right\},
\]
so $\overline B_\eps(c_2)\leq\overline B_\eps(c_2')$. On the strict-increase range in part~2(c), the identity $\overline B_\eps=H_{\Btheta_*}\circ E_{\eps,\Btheta_*}$ shows that the envelope is strictly increasing if and only if $H_{\Btheta_*}$ is strictly increasing over the attained radii. The envelope is constant thereafter. This proves part~2(d) and completes the proof. 
\end{proof}

%%%%%%%%%%%%%%%%%%%%%%%%%%%%%%%%%%%%%%%%%%%%%%%%%%%%%%%
\newpage
\section{Composite estimation: Details}\label{app:cmp}

Subsection~\ref{app:cmp-theory} contains three theoretical results on composite $C$-estimation: Proposition~\ref{prop:fisher-consistency-cmp}, Theorem~\ref{thm:cmp-normality-detailed}, and Lemma~\ref{lem:xi-cmp}. The lemma is used in the proof of the theorem, and the proofs of all three statements are given in Subsection~\ref{app:cmp-proofs}. Finally, Subsection~\ref{app:cmp-covmat-comp} describes how to estimate the population asymptotic covariance matrix in Theorem~\ref{thm:cmp-normality-detailed}.

\subsection{Composite estimation: statements}\label{app:cmp-theory}

Theorem~\ref{thm:cmp-normality-detailed} re-states Theorem~\ref{thm:cmp-normality-reduced} from the main text with the regularity conditions added. Define $\Leps^{(k)}(\Btheta) = \D{\fepsk}{\modelk},k = 1,\dots,K$.

\begin{theorem}\label{thm:cmp-normality-detailed}
Grant Assumptions~\ref{ass:rho} and~\ref{ass:submodel}. Consider the following assumptions.
\begin{enumerate}[label={\textnormal{CMP.\arabic*.}}, ref={\textnormal{CMP.\arabic*}}, itemindent=25pt]
		\item   \label{ass-cmp:consistency-uniformconvergence}
		$\sup_{\Btheta\in\BTheta} \big| \LN^\cmp(\Btheta)- \Leps^\cmp (\Btheta) \big| \convP 0$,
	\item For all $\alpha > 0$, 
	$
		\inf_{\Btheta\in\BTheta\setminus\mathcal{B}(\Btheta_0^\cmp,\alpha)} \Leps^\cmp (\Btheta) >  \Leps^\cmp (\Btheta_0^\cmp),
	$
	\item For some~$\zeta_N = \ohp{1}$, \label{ass-cmp:consistency-nearminimizer}
	$
		\LN^\cmp\left(\Bthetahat^\cmp\right) \leq \inf_{\Btheta\in\BTheta} \LN^\cmp (\Btheta) + \zeta_N,
	$
	\item  $\Btheta_0^\cmp\in\BTheta_0^\cmp\subset \textnormal{int}(\BTheta)$, where $\BTheta_0^\cmp$ is a neighborhood of~$\Btheta_0^\cmp$, \label{ass-cmp:interior}
	\item $\Bthetahat^\cmp \convP\Btheta_0^\cmp$ as $N\to\infty$,
	\item $\gradient \LN^\cmp\left(\Bthetahat^\cmp\right) = \ohp{N^{-1/2}}$,
		\item The map $\Btheta\mapsto\matepscmp{M}{\Btheta}=\hessian \Leps^\cmp(\Btheta)$ is well-defined on~$\BTheta_0^\cmp$ and continuous as well as positive definite at~$\Btheta_0^\cmp$.	
		\item $\sup_{\Btheta\in\BTheta_0^\cmp}\matnorm{ \matfunN{M^\cmp}{\Btheta} - \matfuneps{M^\cmp}{\Btheta} } \convP 0$, where $\mat{M}^\cmp_N(\Btheta) = \hessian \LN^\cmp(\Btheta)$,
	\item For all $k=1,\dots,K$, $\E{\vec{Z}^{(k)}\sim\fepsk}{
		\left\|A'\big(\PRk{\Btheta_0^\cmp}{\varepsilon}{\vec{Z}^{(k)}}\big)	\skfun{\Btheta_0^\cmp}{k}{\vec{Z}^{(k)}}\right\|^2
	}<\infty.$ 
	In particular,~$\matepsk{U}{\Btheta_0^\cmp}{k,l}$ defined in Theorem~\ref{thm:cmp-normality-reduced} exists and has finite norm for all $k,l=1,\dots,K$.\label{ass-cmp:MUfinite}
	\item $\fun{A'}{\PRzepsk{\Btheta_0^\cmp}}$ exists for all $\zk\in\Zk$ and all submodels $k=1,\dots,K$. \label{ass-cmp:PRRAF}
	\item  There is a neighborhood~$\BTheta^\cmp_1\subseteq\BTheta_0^\cmp$ of~$\Btheta_0^\cmp$ on which, for every $k=1,\dots,K$, the series defining~$\Leps^{(k)}$ and its first two derivatives converge absolutely and
locally uniformly, and in both derivatives summation over~$\Z^{(k)}$ may be interchanged with differentiation.\label{ass-cmp:countable-interchange}
\item For 
	\begin{align*}	
		\vec{\Delta}^{(k)}_N (\Btheta) &= \sum_{\zk\in\Zk} \gradient \pzk{\Btheta}\bigg[ \fun{A}{\PRznk{\Btheta}} - \fun{A}{\PRzepsk{\Btheta}} - 
		\\		
		&\qquad\fun{A'}{\PRzepsk{\Btheta}}\left(\PRznk{\Btheta} - \PRzepsk{\Btheta} \right) \bigg],
	\end{align*}
	it holds true that $\left\|\vec{\Delta}^{(k)}_N (\Btheta_0^\cmp)\right\|<\infty$ almost surely as well as $\sqrt{N}\left\|\vec{\Delta}^{(k)}_N (\Btheta_0^\cmp)\right\|\convP 0$ for every $k=1,\dots,K$. \label{ass-cmp:countable-remainder}
\end{enumerate}

Under Assumptions~\ref{ass-cmp:consistency-uniformconvergence}--\ref{ass-cmp:consistency-nearminimizer}, composite $C$-estimators are consistent for~$\Btheta_0^\cmp$, i.e.,
\[
	\Bthetahat^\cmp\convP\Btheta_0^\cmp \quad as\ N\to\infty. 
\] 

Under Assumptions~\ref{ass-cmp:interior}--\ref{ass-cmp:countable-remainder}, composite $C$-estimators are asymptotically normal:
\[
	\sqrt{N}\left(\Bthetahat^\cmp - \Btheta_0^\cmp\right)
		\convweak\gauss_d
		\Big(
					\vec{0},
					\mat{\Sigma}_\eps^\cmp\left(\Btheta_0^\cmp\right)
		\Big)
		\quad as\ N\to\infty,
\]
where $\mat{\Sigma}_\eps^\cmp\left(\Btheta_0^\cmp\right)$ is defined in Theorem~\ref{thm:cmp-normality-reduced} and is positive definite under the additional assumption that $\matepscmp{U}{\Btheta^\cmp_0} = \sum_{k=1}^K\sum_{l=1}^K w_kw_l\matepsk{U}{\Btheta_0^\cmp}{k,l}$ is positive definite.
\end{theorem}

Per the same reasoning as in Section~\ref{sec:asy-normality}, Assumptions \ref{ass-cmp:countable-interchange}--\ref{ass-cmp:countable-remainder} are automatically satisfied when~$\Z$ is finite.

The following lemma is an important building block in the theorem's proof.

\begin{lemma}\label{lem:xi-cmp}
Define, for a $k$-th submodel, $k=1,\dots,K$, the random vector
\[
	\xik{k}{\Btheta}
	=
			\sum_{\zk\in\Zk}\left(\fhatk\left(\zk\right)-\fepsk\left(\zk\right) \right)  \fun{A'}{\PRzepsk{\Btheta}}\skfun{\Btheta}{k}{\zk}, \quad \Btheta\in\BTheta.
\]
Under the Assumptions of Theorem~\ref{thm:cmp-normality-detailed}, it holds true that
\[
	\sqrt{N}\sum_{k=1}^K w_k\xik{k}{\Btheta_0^\cmp}
	\convweak 
	\gauss_d\left(\vec{0}, \sum_{k=1}^K\sum_{l=1}^K w_kw_l\matepsk{U}{\Btheta_0^\cmp}{k,l} \right),
\]
as $N\to\infty$, where the $d\times d$ matrices $\matepsk{U}{\Btheta_0^\cmp}{k,l}$ are defined in Theorem~\ref{thm:cmp-normality-reduced}.
\end{lemma}

\subsection{Composite estimation: proofs} \label{app:cmp-proofs}
This subsection contains the proofs of Proposition~\ref{prop:fisher-consistency-cmp}, Theorem~\ref{thm:cmp-normality-detailed}, and Lemma~\ref{lem:xi-cmp}.

\subsubsection{Proof of Proposition~\ref{prop:fisher-consistency-cmp}}
If $\eps = 0$, then $\feps = \modelt$ as well as $\fepsk = p_{\Btheta_*}^{(k)}$ for all $k$. Since $\Leps^\cmp(\Btheta_*) = \sum_{k=1}^K w_k\cdot \D{\fepsk}{p_{\Btheta_*}^{(k)}}$, each of the~$K$ submodel divergences equals zero in this case, so it follows that $\Leps^\cmp(\Btheta_*)=0$. Conversely, $\D{p_{\Btheta_*}^{(k)}}{\modelk}$ is nonnegative and equals zero if and only if $p_{\Btheta_*}^{(k)} = \modelk$ by Proposition~\ref{prop:divergence-nonnegative}. Thus, if we have that $\Leps^\cmp(\Btheta)=0$, then, because the weights are all strictly positive, $p_{\Btheta_*}^{(k)} = \modelk$ for all~$k$. The point-identification condition in Assumption~\ref{ass:submodel} then yields $\Btheta=\Btheta_*$. Thus the composite population minimizer is unique and equals~$\Btheta_*$.
\QED

\subsubsection{Proof of Theorem~\ref{thm:cmp-normality-detailed}}
The first assertion (on consistency) follows from performing the same steps as the proof of Theorem~\ref{thm:consistency}.

We proceed by proving the second assertion on the limit distribution of composite $C$-estimators. 
We repeat the steps in the proof of Theorem~\ref{thm:asy-normality} leading to~\eqref{eq:Mestimation-linearization}, which is feasible under our assumptions. We ultimately arrive at the asymptotic approximation
\begin{equation}\label{eq:linearization-cmp}
	\sqrt{N}\left( \Bthetahat^\cmp - \Btheta_0^\cmp \right)
	=
	- \left( \Imat + \ohp{1} \right)\left( \matfuneps{M^\cmp}{\Btheta_0^\cmp} \right)^{-1} \sqrt{N}\gradient \LN^\cmp\left(\Btheta_0^\cmp\right) + \ohp{1},
\end{equation}
where $\Imat$ denotes the $d\times d$ identity matrix. In the following, like in the proof of Theorem~\ref{thm:asy-normality}, we derive the limit distribution of $-\gradient \LN^\cmp\left(\Btheta_0^\cmp\right)$.

For every submodel $k=1,\dots, K$, applying Lemma~\ref{lem:gradientdecomposition}---whose conditions are satisfied under our assumptions---yields the approximate decomposition
\begin{equation}\label{eq:cmp-gradient-decomposition}
	\sqrt{N}\vec{\eta}^{(k)}\left(\Btheta,\fhatk\right)
	=
	\sqrt{N}\vec{\eta}^{(k)}\left(\Btheta,\fepsk\right)
	+
	\sqrt{N} \vec{\xi}^{(k)}_N (\Btheta) + \ohp{1}
\end{equation}
at any parameter $\Btheta\in\BTheta$ at which $\fun{A'}{\PRzepsk{\Btheta}}$ exists for all $\zk\in\Zk$. In this decomposition, we used the definitions
\begin{align*}
	\vec{\eta}^{(k)}\left(\Btheta,\fhatk\right) 
	&= -\gradient\D{\fhatk}{\modelk},
	\\
	\vec{\eta}^{(k)}\left(\Btheta,\fepsk\right) 
	&= -\gradient\D{\fepsk}{\modelk},
	\\
	\vec{\xi}^{(k)}_N (\Btheta)
	&= 
	\sum_{\zk\in\Zk}\left(\fhatk\left(\zk\right)-\fepsk\left(\zk\right) \right)  \fun{A'}{\PRzepsk{\Btheta}}\skfun{\Btheta}{k}{\zk}.
\end{align*}

Note that by definition of the composite loss, the linearity of the differentiation operator, and~$K$ being finite, we have that $\gradient \LN^\cmp (\Btheta) = \sum_{k=1}^K w_k \gradient\D{\fhatk}{\modelk}$. Therefore, since~$K$ is finite,
\begin{align*}
	-\sqrt{N}\gradient \LN^\cmp (\Btheta)
	&=
	\sqrt{N}\sum_{k=1}^K w_k \vec{\eta}^{(k)}\left(\Btheta,\fhatk\right)
	\\
	&=
	\sqrt{N}\sum_{k=1}^K w_k \vec{\eta}^{(k)}\left(\Btheta,\fepsk\right)
	+
	\sqrt{N} \sum_{k=1}^K w_k \vec{\xi}^{(k)}_N (\Btheta) + \ohp{1},
	\\
	&=
	-\sqrt{N} \gradient \Leps^\cmp (\Btheta) 
	+
	\sqrt{N} \sum_{k=1}^K w_k \vec{\xi}^{(k)}_N (\Btheta) + \ohp{1},
\end{align*}
where we have used in the first and third equation the definition of the~$\vec{\eta}^{(k)}$, and in the second equation the approximate decomposition in~\eqref{eq:cmp-gradient-decomposition} in conjunction with Slutsky's lemma. 

By Assumption~\ref{ass-cmp:interior} and the stated smoothness assumptions,~$\Btheta_0^\cmp$ is an interior minimizer of the differentiable loss~$\Leps^\cmp$, so the necessary first-order condition gives $\gradient \Leps^\cmp \left(\Btheta_0^\cmp\right) = \vec{0}$. Therefore, when evaluating at parameter~$\Btheta_0^\cmp$, the preceding display reduces to
\begin{equation}\label{eq:neggrad-cmp}
	-\sqrt{N}\gradient \LN^\cmp \left(\Btheta_0^\cmp\right)
	=
	\sqrt{N} \sum_{k=1}^K w_k \vec{\xi}^{(k)}_N \left(\Btheta_0^\cmp\right) + \ohp{1},
\end{equation}
where all~$\vec{\xi}^{(k)}_N \left(\Btheta_0^\cmp\right)$ exist under our assumptions.  Lemma~\ref{lem:xi-cmp} implies that
\begin{equation}\label{eq:xiksum-convergence}
\sqrt{N}\sum_{k=1}^K w_k \xik{k}{\Btheta_0^\cmp}
	\convweak
	\gauss_d\big(\vec{0}, \matfuneps{U^\cmp}{\Btheta_0^\cmp} \big) \qquad \textnormal{ as } N\to\infty,
\end{equation}
where $\matfuneps{U^\cmp}{\Btheta_0^\cmp} =  \sum_{k=1}^K\sum_{l=1}^K w_kw_l\matepsk{U}{\Btheta_0^\cmp}{k,l}$ has finite norm (Lemma~\ref{lem:xi-cmp}). 

Combining~\eqref{eq:xiksum-convergence} with~\eqref{eq:neggrad-cmp}, we conclude that $-\sqrt{N}\gradient \LN^\cmp \left(\Btheta_0^\cmp\right) = \Ohp{1}$. It follows that the initial approximation in~\eqref{eq:linearization-cmp} reduces to the asymptotic linearization
\[
	\sqrt{N}\left( \Bthetahat^\cmp - \Btheta_0^\cmp \right)
	=
	\left( \matfuneps{M^\cmp}{\Btheta_0^\cmp} \right)^{-1} 
	\sqrt{N} \sum_{k=1}^K w_k \vec{\xi}^{(k)}_N \left(\Btheta_0^\cmp\right)
	 + \ohp{1}.
\]
The desired result now follows from the limit distribution of~$\sqrt{N} \sum_{k=1}^K w_k \vec{\xi}^{(k)}_N \left(\Btheta_0^\cmp\right)$ in~\eqref{eq:xiksum-convergence}, completing the proof. \QED

\subsubsection{Proof of Lemma~\ref{lem:xi-cmp}}
Fix a submodel index $k=1,\dots,K$. Define the $d$-variate random vector
\[
	\Rik{k}{\Btheta} =
	\sum_{\zk\in\Zk} I_i^{(k)}\left(\zk\right) \fun{A'}{\PRzepsk{\Btheta}}\skfun{\Btheta}{k}{\zk},
\]
where $ I_i^{(k)}\left(\zk\right) = \I{\vec{Z}_i^{(k)} = \zk}$ is the only stochastic object, which follows a Bernoulli distribution with probability parameter~$\fepsk\left(\zk\right)$. Note that $\sqrt{N}\xik{k}{\Btheta_0^\cmp} = N^{-1/2}\sum_{i=1}^N \left( \Rik{k}{\Btheta_0^\cmp} - \E{}{\Rik{k}{\Btheta_0^\cmp}} \right)$, which we know to converge in distribution to $\gauss_d\left(\vec{0}, \matepsk{U}{\Btheta_0^\cmp}{k,k} \right)$ by Lemma~\ref{lem:xi-limit}, whose conditions are satisfied under our assumptions. 
Thus, to obtain the desired result it suffices to show that the $Kd$-vector
\[
\sqrt{N}\vec{\tilde{\xi}}_N(\Btheta_0^\cmp) =
\left(
	\left(\sqrt{N}\vec{\xi}^{(1)}_N(\Btheta_0^\cmp)\right)^\top, 
	\left(\sqrt{N}\vec{\xi}^{(2)}_N(\Btheta_0^\cmp)\right)^\top,
	\dots,
	\left(\sqrt{N}\vec{\xi}^{(K)}_N(\Btheta_0^\cmp)\right)^\top
	\right)^\top
\]
is jointly asymptotically normally distributed with finite covariance matrix. 

For a fixed sample index $i=1,\dots,N$, stack the~$K$ submodel vectors~$\Rik{k}{\Btheta_0^\cmp}$ on top of each other to obtain the $Kd$-vector
\[
	\vec{\tilde{R}}_i(\Btheta_0^\cmp) =
	\left(
		\left(\Rik{1}{\Btheta_0^\cmp}\right)^\top,
		\left(\Rik{2}{\Btheta_0^\cmp}\right)^\top,
		\dots,
		\left(\Rik{K}{\Btheta_0^\cmp}\right)^\top
	\right)^\top.
\]
The covariance matrix of~$\vec{\tilde{R}}_i(\Btheta_0^\cmp)$ has a block structure where each block comprises a $d\times d$ cross-covariance matrix between $\Rik{k}{\Btheta_0^\cmp}$ and $\Rik{l}{\Btheta_0^\cmp}$ for submodels $k,l=1,\dots,K$. For those cross-covariance matrices, we have by construction that
\begin{align*}
	&\cov{}{\Rik{k}{\Btheta_0^\cmp} ; \Rik{l}{\Btheta_0^\cmp}}
	=
	\\
	&\quad\cov{(\BZk, \vec{Z}^{(l)})\sim\feps^{(k,l)}}{ \skfun{\Btheta_0^\cmp}{k}{\BZk}\fun{A'}{\delta^{(k)}_{\Btheta_0^\cmp,\eps}\left(\BZk\right)} ; \skfun{\Btheta_0^\cmp}{l}{\vec{Z}^{(l)}}\fun{A'}{\delta^{(l)}_{\Btheta_0^\cmp,\eps}\left(\vec{Z}^{(l)}\right)} },
\end{align*}
which is finite for all $k,l=1,\dots,K$ by Assumption~\ref{ass-cmp:MUfinite} and equal to~$\matepsk{U}{\Btheta_0^\cmp}{k,l}$. We conclude that all elements in the covariance matrix of the stacked vector~$\vec{\tilde{R}}_i(\Btheta_0^\cmp)$ are finite, so the norm of that matrix is finite, too.

By definition, we have the equality
\[
	\sqrt{N}\vec{\tilde{\xi}}_N(\Btheta_0^\cmp)
	=
	N^{-1/2}\sum_{i=1}^N \left(\vec{\tilde{R}}_i(\Btheta_0^\cmp) - \E{}{\vec{\tilde{R}}_i(\Btheta_0^\cmp)} \right).
\]
Since the sum on the right-hand-side is taken over~$N$ i.i.d. random vectors, we can invoke the Lindeberg-Lévy central limit theorem to conclude that, as~$N$ grows to infinity,~$\sqrt{N}\vec{\tilde{\xi}}_N(\Btheta_0^\cmp)$ is asymptotically $Kd$-variate normally distributed with mean zero and covariance matrix~$\var{}{\vec{\tilde{R}}_i(\Btheta_0^\cmp)}$. This covariance matrix has finite norm, as demonstrated above.

Having shown that the stacked vector $\sqrt{N}\vec{\tilde{\xi}}_N(\Btheta_0^\cmp)$ is asymptotically normal, apply the fixed linear map $(\vec{v}_1^\top,\ldots,\vec{v}_K^\top)^\top\mapsto\sum_{k=1}^K w_k\vec{v}_k$ with $\vec{v}_k = \sqrt{N}\xik{k}{\Btheta_0^\cmp}, k = 1,\dots, K$. We conclude that $\sqrt{N}\sum_{k=1}^K w_k\xik{k}{\Btheta_0^\cmp}$ is asymptotically normal with mean zero and covariance $\sum_{k=1}^K\sum_{l=1}^K w_kw_l\matepsk{U}{\Btheta_0^\cmp}{k,l}$, as claimed. \QED

\subsection{Composite estimation of covariance matrix}\label{app:cmp-covmat-comp}
A pointwise consistent estimator of $\matepscmp{\Sigma}{\Btheta} = \left(\matepscmp{M}{\Btheta}\right)^{-1} \matepscmp{U}{\Btheta} \left(\matepscmp{M}{\Btheta}\right)^{-1}$  in Theorem~\ref{thm:cmp-normality-reduced} can be constructed as follows. In~$\matepscmp{M}{\Btheta}$, replace all $\fepsk$ by their corresponding consistent estimators $\fhatk$ for all submodels $k=1,\dots,K$. Regarding $\matepscmp{U}{\Btheta}=\sum_{k=1}^K\sum_{l=1}^K w_kw_l \matepsk{U}{\Btheta}{k,l}$, we need to consistently estimate the covariances
\[
	\matepsk{U}{\Btheta}{k,l}
		=
		\cov{(\BZk, \vec{Z}^{(l)})\sim\feps^{(k,l)}}{ \fun{A'}{\delta^{(k)}_{\Btheta,\eps}\left(\BZk\right)}  \skfun{\Btheta}{k}{\BZk} ; \fun{A'}{\delta^{(l)}_{\Btheta,\eps}\left(\vec{Z}^{(l)}\right)} \skfun{\Btheta}{l}{\vec{Z}^{(l)}} }.
\]
The submodel-level marginal expectations $\E{\vec{Z}^{(k)}\sim\fepsk}{\fun{A'}{ \delta^{(k)}_{\Btheta,\eps}\left(\BZk\right) }\skfun{\Btheta}{k}{\BZk}}$ can be consistently estimated by replacing~$\fepsk$ by~$\fhatk$ for all submodels $k=1,\dots,K$.  The product moment 
\[
	\E{(\BZk, \vec{Z}^{(l)})\sim\feps^{(k,l)}}{ \fun{A'}{\delta^{(k)}_{\Btheta,\eps}\left(\BZk\right)}  \skfun{\Btheta}{k}{\BZk} \Big( \fun{A'}{\delta^{(l)}_{\Btheta,\eps}\left(\vec{Z}^{(l)}\right)} \skfun{\Btheta}{l}{\vec{Z}^{(l)}}\Big)^\top }.
\]
is estimable as follows. Replace $\feps^{(k,l)}$ by its consistent estimator
\[
	\fhat^{(k,l)}\left( \zk, \z^{(l)} \right)
	=
	\frac{1}{N}\sum_{i=1}^N\I{ \vec{Z}_i^{(k)} = \zk, \vec{Z}_i^{(l)} = \z^{(l)} },
	\qquad \zk\in\Z^{(k)}, \z^{(l)}\in\Z^{(l)},
\]
for all model pairs $k,l=1,\dots,K$. To ensure that 
\[
	\matNcmp{\Sigma}{\Bthetahat^\cmp} = \left(\matNcmp{M}{\Bthetahat^\cmp}\right)^{-1} \matNcmp{U}{\Bthetahat^\cmp} \left(\matNcmp{M}{\Bthetahat^\cmp}\right)^{-1}
\]
is consistent for~$\matepscmp{\Sigma}{\Btheta_0^\cmp}$, impose a local uniform convergence assumption on~$\mat{U}_{N}^\cmp$ in analogous fashion as in Remark~\ref{rem:Sigmahat}.

%%%%%%%%%%%%%%%%%%%%%%%%%%%%%%%%%%%%%%%%%%%%%%%%%%%%%%%%%%%%%%
\newpage

\section{Conditional $C$-estimation}\label{sec:cnd}
\subsection{Definition and asymptotics}
This section introduces $C$-estimation for \emph{conditional} probability models. 
Consider a regression problem where a scalar response variable~$Y$ is modeled as a function of a possibly multivariate covariate vector~$\vec{X}$. Suppose that both~$Y$ and~$\vec{X}$ are categorical and take values in countable sets~$\Y$ and~$\X$, respectively. To formalize the regression problem, the statistician postulates a model for the joint PMF~$p_{\Btheta}^{(\BX,Y)}$ of the joint random vector $(\BX,Y)$, which is subject to a $d$-variate parameter vector $\Btheta\in\BTheta$. The primary interest lies in the conditional distribution of $\YX$, which is modeled through the conditional PMF associated with~$p_{\Btheta}^{(\BX,Y)}$, that is,
$
	\pyxfun{\Btheta}{\yx} = \Pr{\Btheta}{Y = y | \BX = \x}.
$
Being a conditional PMF, for a given $\x\in\X$, the conditional model $\pyxfun{\Btheta}{\cdot|\x}$ has the unit-sum property $\sum_{y\in\Y} \pyxfun{\Btheta}{\yx} = 1$. Let~$\feps^{\BX}$ denote the marginal PMF of~$\BX$ under the sampling distribution---which will be defined below---and define its support as
\[
	\X_0 = \left\{ \x\in\X : \feps^{\BX}(\x) > 0 \right\}.
\]

\begin{assumption}\label{ass:model-cnd}
 For every $\x\in\X_0$ and $\Btheta\in\BTheta$, $\pyxfun{\Btheta}{\cdot\mid\x}$ is a strictly positive PMF on~$\Y$, and each map $\Btheta\mapsto\pyxfun{\Btheta}{\yx}$ is twice continuously differentiable. Furthermore, the conditional family is globally identifying on~$\X_0$: for any $\Btheta_1,\Btheta_2\in\BTheta$,
\[
	\pyxfun{\Btheta_1}{\cdot\mid\x}=\pyxfun{\Btheta_2}{\cdot\mid\x}
	\ \textnormal{for every }\x\in\X_0
	\quad\Longrightarrow\quad
	\Btheta_1=\Btheta_2.
\]
\end{assumption}

\begin{example}[Cumulative logit model]\label{ex:cumulative-logit}
The \emph{cumulative logit model} \citep[][also known as \emph{proportional odds model}]{mccullagh1980} postulates the following relationship between a $p$-variate numerical covariate vector~$\BX$ and an ordinal response~$Y$ with~$K$ response options taking values in $\Y = \{1,2,\dots, K\}$:
$
	\Pr{\Btheta}{Y \leq y | \BX=\x} = \exp\big( \alpha_y + \x^\top\vec{\beta} \big) \big/ \left(1 + \exp\big( \alpha_y + \x^\top\vec{\beta} \big) \right),
$ 
for $y = 1,2,\dots,K-1$.\footnote{\cite{mccullagh1980} uses a sign-flipped parametrization, namely $\alpha_y - \x^\top\vec{\beta}$, which alters parameter interpretation; see \citet[][Section~3.2.2]{agresti2010} for details.}
The model's parameters are $\Btheta = (\alpha_1,\dots,\alpha_{K-1}, \vec{\beta}^\top)^\top$, where $\vec{\beta}\in\R^p$. The parameter space imposes $\alpha_1<\cdots<\alpha_{K-1}$, and global identification in Assumption~\ref{ass:model-cnd} holds when the vectors $(1,\x^\top)^\top$, $\x\in\X_0$, span~$\R^{p+1}$. If $K=2$, the model reduces to the linear logistic regression model. % Cumulative logit models are usually estimated by means of regression ML, which is well-known to be sensitive to contamination \citep[e.g.,][]{cantoni2001glm}.
It can be shown that the individual response probabilities are
\begin{equation*}
	\pyxfun{\Btheta}{\yx} = \Pr{\Btheta}{Y = y | \BX=\x}
	=
	\frac{\exp\big( \alpha_y + \x^\top\vec{\beta} \big)}{1+\exp\big(\alpha_y + \x^\top\vec{\beta} \big)} -
	\frac{\exp\big( \alpha_{y-1} + \x^\top\vec{\beta} \big)}{1+\exp\big(\alpha_{y-1} + \x^\top\vec{\beta} \big)},
\end{equation*}
where $y\in\Y$, with the conventions $\alpha_0 = -\infty$ and $\alpha_K = +\infty$. Here, like in \cite{mccullagh1980}, we focus on categorical covariates~$\BX$ that take values in a finite set~$\X$.
\end{example}

In the following, we propose $C$-estimators for conditional probability models. Suppose we observe a sample $((\BX_i, Y_i))_{i=1}^N$ of~$N$ i.i.d. copies of $(\BX, Y)$. Denote by
$
	\fhatX(\x) = N_{\x,\bullet} / N = \frac{1}{N} \sum_{i=1}^N \I{\BX_i = \x}, 
$ 
$\x\in\X,$  
the empirical marginal PMF of~$\BX$. Similarly, for $\x\in\X$, let
$
		\fhatYX(\yx) = N_{\xy} / N_{\x,\bullet} = \frac{1}{N_{\x,\bullet}} \sum_{i=1}^N \I{\BX_i = \x, Y_i = y},
$
$y\in\Y$, 
be the empirical conditional PMF of $\YX = \x$.\footnote{If $N_{\x,\bullet}=0$, define $\fhat^{Y\mid\BX=\x}$ to be any fixed PMF on~$\Y$. Its choice is immaterial because $\fhatX(\x)=0$.} 
If the postulated model is correctly specified, then, as $N\to\infty$, $\fhatYX(\yx)\convP\pyxfun{\Btheta_*}{\yx}$ for every $\x\in\X_0$ and $y\in\Y$, where $\Btheta_*\in\BTheta$ is the \emph{true} parameter value at which the (uncontaminated) data were generated.  Similarly,~$\fhatX$ is a pointwise consistent estimator for the marginal distribution of~$p_{\Btheta_*}^{(\BX,Y)}$ with respect to~$\BX$. 
On the other hand, if contamination is present in the sample, then said consistency properties of~$\fhatYX$ and~$\fhatX$ may fail. It is therefore indicative of model misspecification if an empirical \emph{conditional} PR $\PRyxnfun{\Btheta}{\yx} = \fhatYX(\yx)\big/ \pyxfun{\Btheta}{\yx}-1$ is away from~0 for all $\Btheta\in\BTheta$. Analogously to unconditional $C$-estimation (Section~\ref{sec:Cestimator}), we construct contamination-robust estimators for conditional models based on the idea of downweighting \emph{conditional} PRs.

The \emph{conditional relative entropy} between~$\fhatYX$ and~$\modelcnd$ equals 
\[
	\sum_{\x\in\X}\fhatX(\x)\sum_{y\in\Y}\fhatYX(\yx)\log\left(\frac{\fhatYX(\yx)}{\pyxfun{\Btheta}{\yx}}\right).
\] 
Regression ML estimators of~$\modelcnd$ are defined as minimizers of this divergence. Motivated by the construction of conditional relative entropy, we propose the following divergence measure~$\Dsymbol^\cnd$ between conditional PMFs. For a given discrepancy function~$\rho$, let
\begin{equation}\label{eq:loss-cnd}
\begin{split}
		\Dcnd{\fhatYX}{\modelcnd}
		&=
		\sum_{\x\in\X}\fhatX(\x) \sum_{y\in\Y} \fun{\rho}{\PRyxnfun{\Btheta}{\yx}}\pyxfun{\Btheta}{\yx}
		\\&=
		\sum_{\x\in\X}\fhatX(\x)\D{\fhat^{\YX = \x}}{p_{\Btheta}^{\YX = \x}},
\end{split}
\end{equation}
where~$\Dsymbol$ is the divergence in~\eqref{eq:divergence} and~$\fhatX$ is always understood as the marginal PMF associated with the conditional PMF~$\fhatYX$. Observe that for the choice $\rho_\ML(\delta)=(\delta+1)\log(\delta+1)-\delta$, the divergence measure~$\Dsymbol^\cnd$ is equal to the conditional relative entropy. For further reference, define the \emph{empirical conditional loss function} as $\LN^\cnd(\Btheta) = \Dcnd{\fhatYX}{\modelcnd}$.

\begin{definition}[Conditional $C$-estimator]
Suppose the $\rho$-function in $\LN^\cnd$ satisfies Assumption~\ref{ass:rho}, i.e., it is a discrepancy function, and let the postulated joint model~$p_{\Btheta}^{(\BX, Y)}$ satisfy Assumption~\ref{ass:model-cnd}. A \emph{conditional $C$-estimator} is any parameter~$\Bthetahat^\cnd$ that globally minimizes the empirical conditional loss~$\LN^\cnd(\Btheta)$ over~$\BTheta$. If~$\Bthetahat^\cnd$ is an interior point and~$\LN^\cnd(\Btheta)$ is differentiable at it, then~$\Bthetahat^\cnd$ is necessarily a root of~$\gradient\LN^\cnd(\Btheta)$.
\end{definition}

To model contamination, we assume the Huber-type contamination model in~\eqref{eq:hubermodel} for the vector~$(\XY)$, which is then distributed according to~$\fepsXY = (1-\eps)p_{\Btheta_*}^{(\XY)} + \eps h^{(\XY)}$, where both the contamination fraction $\eps\in [0,0.5)$ and the joint contamination PMF~$h^{(\XY)}$ are unspecified. Denote by~$\fepsX$ and~$\fepsYX$ the marginal and conditional PMF, respectively, of the sampling distribution~$\fepsXY$. The estimand of a conditional $C$-estimator~$\Bthetahat^\cnd$ is then given by the minimizer~$\Btheta_0^\cnd$ of the population loss $\Leps^\cnd(\Btheta) = \Dcnd{\fepsYX}{\modelcnd} = \sum_{\x\in\X}\fepsX(\x) \sum_{y\in\Y} \fun{\rho}{\PRyxepsfun{\Btheta}{\yx}}\pyxfun{\Btheta}{\yx}$ over $\Btheta\in\BTheta$, where~$\PRyxepsfun{\Btheta}{\yx} = \fepsYX (\yx)\big/ \pyxfun{\Btheta}{\yx}-1$ denotes the population-level conditional PR. 

In similar fashion as unconditional estimation, conditional $C$-estimators are Fisher-consistent and, under mild regularity conditions, asymptotically normal.

\begin{proposition}\label{prop:fisher-consistency-cnd}
Grant Assumption~\ref{ass:model-cnd}. Then, conditional $C$-estimators are Fisher-consistent, that is, $\Btheta_0^\cnd = \Btheta_*$ whenever $\eps = 0$, in which case~$\Leps^\cnd(\Btheta)$ attains its global minimum of~$0$ at~$\Btheta_0^\cnd$. 
\end{proposition}

To show asymptotic normality, we require additional notation. For $(\x,y)\in\X\times\Y$, write
\[
	\vec{b}_{\Btheta,\eps}(\x,y)
	=
	\fun{A'}{\PRyxepsfun{\Btheta}{\yx}}
	\syxfun{\Btheta}{\yx},
	\qquad
	\bar{\vec{b}}_{\Btheta,\eps}(\x)
	=
	\E{\feps^{Y\mid\BX=\x}}{b_{\Btheta,\eps}(\x,Y)},
	\]
where $\syxfun{\Btheta}{\yx}=\gradient\log\big(\pyxfun{\Btheta}{\yx}\big)$ is the conditional log-likelihood score, and define further
\[
	\vec{g}_{\Btheta,\eps}(\x)
	=
	-\gradient
	\D{
	\feps^{Y\mid\BX=\x}}{
	p_{\Btheta}^{Y\mid\BX=\x}
	}
\]
as well as the $d\times d$ matrix
\begin{equation}\label{eq:Ueps-cnd}
	\mat{U}_\eps^\cnd(\Btheta)
	=
	\sum_{\x\in\X}\fepsX(\x)
	\var{Y\mid\BX=\x\sim\feps^{Y\mid\BX=\x}}{
	b_{\Btheta,\eps}(\x,Y)}
	+
	\var{\fepsX}{\vec{g}_{\Btheta,\eps}(\BX)}.
\end{equation}

\begin{theorem}\label{thm:cnd-normality}
Grant Assumptions~\ref{ass:rho} and~\ref{ass:model-cnd}, suppose that~$\X$ and~$\Y$ are finite, and assume $\fepsX(\x)>0$ for every~$\x\in\X$, so that $\X_0=\X$. Consider the following assumptions.
\begin{enumerate}[label={\textnormal{CND.\arabic*.}}, ref={\textnormal{CND.\arabic*}}, itemindent=25pt]
	\item   \label{ass-cnd:consistency-uniformconvergence}
		$\sup_{\Btheta\in\BTheta} \big| \LN^\cnd(\Btheta)- \Leps^\cnd (\Btheta) \big| \convP 0$,
	\item For all $\alpha > 0$, 
	$
		\inf_{\Btheta\in\BTheta\setminus\mathcal{B}(\Btheta_0^\cnd,\alpha)} \Leps^\cnd (\Btheta) >  \Leps^\cnd (\Btheta_0^\cnd),
	$
	\item For some~$\zeta_N = \ohp{1}$, \label{ass-cnd:consistency-nearminimizer}
	$
		\LN^\cnd\left(\Bthetahat^\cnd\right) \leq \inf_{\Btheta\in\BTheta} \LN^\cnd (\Btheta) + \zeta_N,
	$
	\item  $\Btheta_0^\cnd\in\BTheta_0^\cnd$ $\subset \textnormal{int}(\BTheta)$, where $\BTheta_0^\cnd$ is a neighborhood of~$\Btheta_0^\cnd$, \label{ass-cnd:interior}
	\item $\Bthetahat^\cnd \convP\Btheta_0^\cnd$ as $N\to\infty$,
	\item $\gradient \LN^\cnd\left(\Bthetahat^\cnd\right) = \ohp{N^{-1/2}}$,
	\item The map $\Btheta\mapsto\matepscnd{M}{\Btheta}=\hessian \Leps^\cnd(\Btheta)$ is well-defined on~$\BTheta_0^\cnd$ and continuous as well as positive definite at~$\Btheta_0^\cnd$.	
	\item $\sup_{\Btheta\in\BTheta_0^\cnd}\matnorm{ \matfunN{M^\cnd}{\Btheta} - \matfuneps{M^\cnd}{\Btheta} } \convP 0$, where $\mat{M}^\cnd_N(\Btheta) = \hessian \LN^\cnd(\Btheta)$,
	\item $\matnorm{\mat{U}^\cnd_\eps (\Btheta_0^\cnd)} < \infty$, with~$\mat{U}^\cnd_\eps (\Btheta)$ being defined in~\eqref{eq:Ueps-cnd}, \label{ass-cnd:MUfinite}
	\item $\fun{A'}{\PRyxepsfun{\Btheta_0^\cnd}{\yx}}$ exists for all $y\in\Y$ and $\x\in\X$. \label{ass-cnd:PRRAF}
\end{enumerate}

Under Assumptions~\ref{ass-cnd:consistency-uniformconvergence}--\ref{ass-cnd:consistency-nearminimizer}, conditional $C$-estimators are consistent for~$\Btheta_0^\cnd$, i.e.,
\[
	\Bthetahat^\cnd \convP \Btheta_0^\cnd \quad \text{ as } N\to\infty.
\]
Under Assumptions~\ref{ass-cnd:interior}--\ref{ass-cnd:PRRAF}, conditional $C$-estimators are asymptotically normal:
\[
	\sqrt{N}\Big( \Bthetahat^\cnd - \Btheta_0^\cnd \Big)
	\convweak
	\gauss_d\Big( \vec{0}, \mat{\Sigma}_\eps^\cnd(\Btheta_0^\cnd) \Big)
	\quad \text{ as }N\to\infty,
\]
where 
\[
	\matepscnd{\Sigma}{\Btheta} = \left(\matepscnd{M}{\Btheta}\right)^{-1} \matepscnd{U}{\Btheta} \left(\matepscnd{M}{\Btheta}\right)^{-1}
\]
is positive definite at~$\Btheta_0^\cnd$ under the additional assumption that $\mat{U}^\cnd_\eps (\Btheta_0^\cnd)$ is positive definite.
\end{theorem}

\begin{remark}
Assumptions~\ref{ass-cnd:consistency-uniformconvergence}--\ref{ass-cnd:PRRAF} adapt Conditions~\ref{ass:consistency-uniformconvergence}--\ref{ass:PRRAF} from unconditional estimation to regression problems. The assumption that both~$\X$ and~$\Y$ are finite is for simplicity; countably infinite sample spaces can be accommodated by incorporating additional assumptions ensuring that all infinite series appropriately converge locally in similar fashion as Conditions~\ref{ass:countable-interchange} and~\ref{ass:countable-remainder} in unconditional estimation.
\end{remark}

\begin{remark}
If $\varepsilon=0$ and $A'(0)$ exists, then $\vec{g}_{\Btheta_*,\eps}(\x)=\vec{0}$ for every~$\x$, and~$\matepscnd{\Sigma}{\Btheta_*}$ equals the asymptotic covariance matrix of the conditional ML estimator. Thus smooth conditional $C$-estimators maintain first-order efficiency at the postulated model. 
\end{remark}

\begin{remark}
A plug-in estimator of $\matepscnd{\Sigma}{\Btheta}$ is obtained by replacing the population PMFs~$\fepsX$ and~$\fepsYX$ in~$\matepscnd{M}{\Btheta}$, $b_{\Btheta,\eps}$, $\bar b_{\Btheta,\eps}$, and~$g_{\Btheta,\eps}$ by~$\fhatX$ and~$\fhatYX$. Under the finite-support and regularity conditions of Theorem~\ref{thm:cnd-normality}, this estimator is consistent.
\end{remark}

\begin{remark}
Conditional $C$-estimation is intended for regression problems where \emph{both}~$Y$ and~$\BX$ are categorical. For situations where~$\BX$ is continuously distributed, $M$-estimators for regression can be used for robust estimation \citep{cantoni2001glm}. However, unlike $C$-estimators, robust $M$-estimators generally trade some model efficiency for robustness. Alternatively, one could consider the regression MDE method  of \citet{hooker2016}, which is based on kernel densities.
\end{remark}

Simulation studies on conditional $C$-estimators are conducted in Appendix \ref{app:simulation-logistic regression}. We now state additional technical lemmas and then prove Proposition~\ref{prop:fisher-consistency-cnd} and Theorem~\ref{thm:cnd-normality} along said lemmas.

\subsection{Conditional estimation: technical lemmas} 

For a parameter~$\Btheta\in\BTheta$ at which every required RAF derivative exists, define
\[
	\vec{q}_{\Btheta,\eps}(\x,y)
	=
	\vec{b}_{\Btheta,\eps}(\x,y)-\bar{\vec{b}}_{\Btheta,\eps}(\x)+\vec{g}_{\Btheta,\eps}(\x),
\]
and further define
\begin{equation}\label{eq:xicnd}
	\vecfuncndN{\xi}{\Btheta}
	=
	\frac{1}{N}\sum_{i=1}^N
	\vec{q}_{\Btheta,\eps}(\BX_i,Y_i).
\end{equation}

We are now ready to state the lemmas.

\begin{lemma}\label{lem:T-moments}
Let $\BX$ and $Y$ be a random vector and a random variable that are supported on the finite sets~$\X$ and~$\Y$, respectively. Further suppose that~$\BX$ and~$Y$ are jointly distributed according to the PMF~$\feps^{(\BX, Y)}$. Define the random variable
\[
	T(\xy) = \I{\BX = \x, Y = y} - \I{\BX = \x} \fepsYX(\yx), \qquad (\x,y)\in\X\times\Y,
\]
where $\fepsYX$ is the conditional PMF of~$\YX$. Then, for fixed $(\x,y)\in\X\times\Y$,
\[
	\var{}{T(\x,y)}
	=
	\fepsX(\x)\fepsYX(\yx)\left(1-\fepsYX(\yx)\right)
	<
	\infty
\]

Moreover, for a sample of $N$ i.i.d. copies $((\BX_i, Y_i))_{i=1}^N$ of $(\BX,Y)$, one has for all $(\xy)\in\X\times\Y$ that
\[
	 N^{-1/2}\sum_{i=1}^N T_i(\xy) \convweak \gauss_1\big(0, \var{}{T(\xy)} \big)
	\quad \text{ as }N\to\infty,
\]
where $T_i(\xy) = \I{\BX_i = \x, Y_i = y} - \I{\BX_i = \x} \fepsYX(\yx)$.
\end{lemma}

\begin{lemma}\label{lem:graddecomp-cnd}
Under the assumptions of Theorem~\ref{thm:cnd-normality}, it holds true that
\[
	-\sqrt{N}\left(\gradient \LN^\cnd(\Btheta_0^\cnd)\right)	
	=	
	\sqrt{N}\vec{\xi}_N^\cnd(\Btheta_0^\cnd) + \ohp{1},
\]
as $N\to\infty$.
\end{lemma}

\begin{lemma}\label{lem:xilimit-cnd}
Under the assumptions of Theorem~\ref{thm:cnd-normality}, it holds true that
\[
	\sqrt{N}\vecfuncndN{\xi}{\Btheta_0^\cnd} 
	\convweak \gauss_d\big(\vec{0}, \mat{U}_\eps^\cnd(\Btheta_0^\cnd) \big)
	\qquad \text{ as } N\to\infty,
\]
where $\mat{U}_\eps^\cnd$ is defined in~\eqref{eq:Ueps-cnd}. 
\end{lemma}

\subsection{Conditional estimation: proofs} \label{app:cnd-proofs}

The following subsubsections prove Proposition~\ref{prop:fisher-consistency-cnd}, Theorem~\ref{thm:cnd-normality}, Lemma~\ref{lem:T-moments}, Lemma~\ref{lem:graddecomp-cnd}, and Lemma~\ref{lem:xilimit-cnd}, in this order.

\subsubsection{Proof of Proposition~\ref{prop:fisher-consistency-cnd}}
When $\varepsilon=0$, we have
\[
	\Leps^\cnd(\Btheta)
	=\sum_{\x\in\X}p_{\Btheta_*}^{\BX}(\x)
	\D{p_{\Btheta_*}^{\YX=\x}}{p_{\Btheta}^{\YX=\x}}.
\]
Every summand is nonnegative by  Proposition~\ref{prop:divergence-nonnegative}, and all are zero at~$\Btheta_*$. If another parameter attained the same minimum, each summand with $p_{\Btheta_*}^{\BX}(\x)>0$ would have to be zero. Proposition~\ref{prop:divergence-nonnegative} would then give $p_{\Btheta}^{\YX=\x}=p_{\Btheta_*}^{\YX=\x}$ for such~$\x$, and the point-identification condition in Assumption~\ref{ass:model-cnd} yields $\Btheta=\Btheta_*$. \QED

\subsubsection{Proof of Theorem~\ref{thm:cnd-normality}}
The first assertion (on consistency) follows from performing the same steps as the proof of Theorem~\ref{thm:consistency}.

We proceed by proving the second assertion on the limit distribution of conditional $C$-estimators. 
We repeat the steps in the proof of Theorem~\ref{thm:asy-normality} leading to~\eqref{eq:Mestimation-linearization}, which is feasible under our assumptions. We ultimately arrive at the asymptotic approximation
\begin{equation}\label{eq:cnd-proof-linearization}
	\sqrt{N}\left( \Bthetahat^\cnd - \Btheta_0^\cnd \right)
	=
	-\left( \Imat + \ohp{1} \right)\left( \matfuneps{M^\cnd}{\Btheta_0^\cnd} \right)^{-1} \sqrt{N}\gradient \LN^\cnd\left(\Btheta_0^\cnd\right)  + \ohp{1},
\end{equation}
where $\Imat$ denotes the $d\times d$ identity matrix. By Lemma~\ref{lem:graddecomp-cnd}, the negative gradient can be approximated through
\begin{align*}
		-\sqrt{N}\left(\gradient \LN^\cnd(\Btheta_0^\cnd)\right)	
	&=	
	\sqrt{N}\vec{\xi}_N^\cnd(\Btheta_0^\cnd) + \ohp{1}
	\\
	&
	\convweak \gauss_d\big(\vec{0}, \mat{U}_\eps^\cnd(\Btheta_0^\cnd) \big) \quad \textnormal{ as } N\to\infty,
\end{align*}
where the second line is due to Lemma~\ref{lem:xilimit-cnd}. Combining this convergence result with~\eqref{eq:cnd-proof-linearization} yields the assertion. \QED

\subsubsection{Proof of Lemma \ref{lem:T-moments}}
Let $(\x,y)\in\X\times\Y$ be arbitrary. Observe that $\I{\BX = \x, Y = y}$ is Bernoulli distributed with probability parameter~$\fepsXY(\xy)$, and that $\I{\BX = \x}$ is Bernoulli distributed with probability parameter~$\fepsX(\x)$. For the first moment of $T(\xy)$, we therefore have
\[
	\E{}{T(\xy)}
	=
	\fepsXY(\xy) - \fepsX(\x)\fepsYX(\yx)
	=0,
\]
which follows from Bayes' theorem.
Regarding the random variable's second moment,
\begin{align*}
	\E{}{T^2(\xy)}
	&=
	\E{}{\big( \I{\BX=\x,Y=y} - \I{\BX=\x}\fepsYX(\yx) \big)^2}
	\\
	&=
	(1-2\fepsYX(\yx))\fepsXY(\xy) + \fepsX(\x)\fepsYX(\yx)^2
	\\
	&=	
	\fepsXY(\xy)\big( (1-2\fepsYX(\yx)) + \fepsYX(\yx)  \big)
	\\
	&=
	\fepsX(\x)\fepsYX(\yx) (1-\fepsYX(\yx)),
\end{align*}
where we have repeatedly used Bayes' theorem. It follows that $\var{}{T(\xy)} = \fepsX(\x)\fepsYX(\yx) (1-\fepsYX(\yx))$, which is finite.

The final assertion on the limit distribution of $N^{-1/2}\sum_{i=1}^N T_i(\xy)$ follows by a straightforward application of the Lindeberg-Lévy central limit theorem. The proof is complete. \QED

\subsubsection{Proof of Lemma \ref{lem:graddecomp-cnd}}
For fixed $\x\in\X$ and an arbitrary conditional PMF $q=q(\cdot|\x)$ on~$\Y$, define the negative gradient of the divergence at~$\Btheta_0^\cnd$ by
\[
	\vec{v}_{\x}(q) = -\gradient\D{q}{\modelcndfun{\Btheta_0^\cnd}{\YX = \x}}.
\]
The relationship between~$\rho$ and its RAF, already used in~\eqref{eq:estimatingeq}, gives the representation
\begin{equation}\label{eq:cnd-lemma-gradient-derivative}
	\vec{v}_{\x}(q) = \sum_{y\in\Y}
	\fun{A}{\frac{q(y|\x)}{\pyxfun{\Btheta_0^\cnd}{\yx}}-1}
	\gradient \pyxfun{\Btheta_0^\cnd}{\yx}.
\end{equation}
In particular, evaluation at the population conditional PMF gives
\[
	\vec{v}_{\x}(\feps^{\YX=\x})
	=
	\funsub{\vec{g}}{\Btheta_0^\cnd,\eps}{\x} 
\]
by the definition of $\vec{g}_{\Btheta,\eps}$.

We next calculate the derivative of~\eqref{eq:cnd-lemma-gradient-derivative} evaluated at $\feps^{\YX=\x}$. For every $y\in\Y$,
\begin{equation}\label{eq:cnd-b}
	\partialderivative{\vec{v}_{x}(q)}{q}\Bigg|_{q=\feps^{\YX=\x}(\yx)}
	=
	\fun{A'}{\PRyxepsfun{\Btheta_0^\cnd}{\yx}}\syxfun{\Btheta_0^\cnd}{\yx}
	=\funsub{\vec{b}}{\Btheta_0^\cnd,\eps}{\x,y}
\end{equation}
This derivative exists by Assumption~\ref{ass-cnd:PRRAF}. 

Our aim is to expand $\vec{v}_{\x}$ about $\feps^{\YX = \x}$. For this purpose, it is useful to study the limit properties of the following empirical process. On the event $\fhatX(\x) > 0$, for every $y\in\Y$,
\begin{equation}\label{eq:cnd-lemma-conditional-rate}
\begin{split}
	&\sqrt{N}\fhatX(\x)
	\big(\fhatYX(\yx)-\fepsYX(\yx)\big)
	\\
	&\qquad=
	\frac{1}{\sqrt{N}}\sum_{i=1}^N
	\left[
	\I{\BX_i=\x,Y_i=y}
	-
	\I{\BX_i=\x}\fepsYX(\yx)
	\right]
	\\
	&\qquad\equiv
	\frac{1}{\sqrt{N}}\sum_{i=1}^N T_i(\x,y).
\end{split}
\end{equation}
The same equality holds with both sides being equal to zero when $\fhatX(\x) = 0$. In the previous display, Lemma~\ref{lem:T-moments} makes the final equality~$\Ohp{1}$. Furthermore, by the law of large numbers, $\fhatX(\x)\convP \fepsX(\x)>0$. Dividing the previous display by~$\sqrt{N}$ therefore shows that for every~$(\x,y)\in\X\times\Y$,
\begin{equation}\label{eq:cnd-lemma-conditional-op}
	\fhatYX(\yx) - \fepsYX(\yx) = \Ohp{N^{-1/2}}.
\end{equation}

Because~$\Y$ is finite,~\eqref{eq:cnd-lemma-gradient-derivative} permits the desired first-order expansion of~$\vec{v}_{\x}$ about~$\feps^{\YX=\x}$:
\begin{equation}\label{eq:cnd-lemma-stratum-expansion}
\begin{split}
	-\gradient\D{\fhat^{\YX=\x}}{\modelcndfun{\Btheta_0^\cnd}{\YX = \x}}
	&=
	\vec{g}_{\Btheta_0^\cnd,\eps}(\x)
	\\
	&\quad+
	\sum_{y\in\Y}
	\big(\fhatYX(\yx) - \fepsYX(\yx)\big)
	\vec{b}_{\Btheta_0^\cnd,\eps}(\x,y)
	+ \vec{r}_{N,\x},
\end{split}
\end{equation}
where~$\vec{r}_{N,\x} = \ohp{N^{-1/2}}$ by~\eqref{eq:cnd-lemma-conditional-op} and finiteness of~$\Y$. Since~$\X$ is finite and $0\leq\fhatX(\x)\leq 1$, we thus have
\begin{equation}\label{eq:cnd-lemma-summed-remainder}
	\sum_{\x\in\X}\fhatX(\x)\vec{r}_{N,\x}
	=\ohp{N^{-1/2}}.
\end{equation}
Multiplying~\eqref{eq:cnd-lemma-stratum-expansion} by~$\fhatX(\x)$ and summing over the elements in the finite~$\X$ therefore gives
\begin{equation}\label{eq:cnd-lemma-aggregate-expansion}
\begin{split}
	-\gradient\LN^\cnd(\Btheta_0^\cnd)
	&=
	-\sum_{\x\in\X}\fhatX(\x)
	\gradient\D{\fhat^{\YX=\x}}{\modelcndfun{\Btheta_0^\cnd}{\YX = \x}}
	\\
	&=
	\sum_{\x\in\X}\fhatX(\x)\vec{g}_{\Btheta_0^\cnd,\eps}(\x)
	\\
	&\quad+
	\sum_{\x\in\X}\fhatX(\x)
	\sum_{y\in\Y}
	\big(\fhatYX(\yx)-\fepsYX(\yx)\big)
	\vec{b}_{\Btheta_0^\cnd,\eps}(\x,y)
	+\ohp{N^{-1/2}},
\end{split}
\end{equation}
where the first equality is due to the definition of the conditional loss in~\eqref{eq:loss-cnd}. 
We now rewrite both summations on the right-hand-side as averages over the~$N$ observations. First,
\begin{equation}\label{eq:cnd-lemma-g-average}
\begin{split}
	\sum_{\x\in\X}\fhatX(\x)\vec{g}_{\Btheta_0^\cnd,\eps}(\x)
	&=
	\sum_{\x\in\X}
	\left(\frac{1}{N}\sum_{i=1}^N\I{\BX_i=\x}\right)
	\vec{g}_{\Btheta_0^\cnd,\eps}(\x)
	\\
	&=
	\frac{1}{N}\sum_{i=1}^N \vec{g}_{\Btheta_0^\cnd,\eps}(\BX_i).
\end{split}
\end{equation}
Second,~\eqref{eq:cnd-lemma-conditional-rate} without the factor~$\sqrt N$ yields
\begin{equation}\label{eq:cnd-lemma-b-average}
\begin{split}
	&\sum_{\x\in\X}\fhatX(\x)
	\sum_{y\in\Y}
	\big(\fhatYX(\yx)-\fepsYX(\yx)\big)
	\vec{b}_{\Btheta_0^\cnd,\eps}(\x,y)
	\\
	&\quad=
	\frac{1}{N}\sum_{i=1}^N
	\sum_{\x\in\X}\sum_{y\in\Y}
	\left[
	\I{\BX_i=\x,Y_i=y}
	-\I{\BX_i=\x}\fepsYX(\yx)
	\right]
	\vec{b}_{\Btheta_0^\cnd,\eps}(\x,y)
	\\
	&\quad=
	\frac{1}{N}\sum_{i=1}^N
	\left[
	\vec{b}_{\Btheta_0^\cnd,\eps}(\BX_i,Y_i)
	-
	\sum_{y\in\Y}\fepsYX(y\mid\BX_i)
	\vec{b}_{\Btheta_0^\cnd,\eps}(\BX_i,y)
	\right]
	\\
	&\quad=
	\frac{1}{N}\sum_{i=1}^N
	\left\{
	\vec{b}_{\Btheta_0^\cnd,\eps}(\BX_i,Y_i)
	-\bar{\vec{b}}_{\Btheta_0^\cnd,\eps}(\BX_i)
	\right\},
\end{split}
\end{equation}
where we have used in the last equality the definition of~$\bar{\vec{b}}_{\Btheta,\eps}$.

Substituting~\eqref{eq:cnd-lemma-g-average} and~\eqref{eq:cnd-lemma-b-average} into the original expansion~\eqref{eq:cnd-lemma-aggregate-expansion} finally yields
\begin{align*}
	-\gradient\LN^\cnd(\Btheta_0^\cnd)
	&=
	\frac{1}{N}\sum_{i=1}^N
	\left(
	\vec{b}_{\Btheta_0^\cnd,\eps}(\BX_i,Y_i)
	-\bar{\vec{b}}_{\Btheta_0^\cnd,\eps}(\BX_i)
	+\vec{g}_{\Btheta_0^\cnd,\eps}(\BX_i)
	\right)
	+\ohp{N^{-1/2}}
	\\
	&=
	\vecfuncndN{\xi}{\Btheta_0^\cnd}
	+\ohp{N^{-1/2}},
\end{align*}
where the second equality follows from~\eqref{eq:xicnd}. Multiplication by~$\sqrt{N}$ proves
\[
	-\sqrt{N}\gradient\LN^\cnd(\Btheta_0^\cnd)
	=
	\sqrt{N}\vecfuncndN{\xi}{\Btheta_0^\cnd}
	+\ohp{1},
\]
as claimed.
\QED

%%%%%%%%%%%%%%%%%%%%%%%%%%%%%%%%%%%%%%%%%%

\subsubsection{Proof of Lemma \ref{lem:xilimit-cnd}}
By~\eqref{eq:xicnd},
\[
	\sqrt N\vecfuncndN{\xi}{\Btheta_0^\cnd}
	=
	\frac{1}{\sqrt{N}}\sum_{i=1}^N
	\vec{q}_{\Btheta_0^\cnd,\eps}(\BX_i,Y_i).
\]
We have that $\E{\feps^{(\BX,Y)}}{\vec{q}_{\Btheta_0^\cnd,\eps}(\BX,Y)}=\E{\fepsX}{\vec{g}_{\Btheta_0^\cnd,\eps}(\BX)}=-\gradient\Leps^\cnd(\Btheta_0^\cnd)=\vec{0}$ because~$\Btheta_0^\cnd$ is a minimizer of the population conditional loss.  
It follows that the summands in the previous display are i.i.d., have population mean zero, and have finite covariance by Condition~\ref{ass-cnd:MUfinite}. The multivariate central limit theorem therefore gives a zero-mean Gaussian limit with covariance~$\var{\feps^{\XY}}{\vec{q}_{\Btheta_0^\cnd,\eps}(\BX,Y)}$. We now prepare an application of the law of total variance on this covariance. Conditional on $\BX = \x$, the last two terms in $\vec{q}_{\Btheta_0^\cnd,\eps}(\BX,Y) = \vec{b}_{\Btheta_0^\cnd,\eps}(\BX,Y) - \bar{\vec{b}}_{\Btheta_0^\cnd,\eps}(\BX) + \vec{g}_{\Btheta_0^\cnd,\eps}(\BX)$ are nonrandom. Hence,
\begin{align*}
	\var{\feps^{(\XY)}}{\vec{q}_{\Btheta_0^\cnd}(\BX,Y)\ \middle|\ \BX=\x}
	&=
	\var{\feps^{(\XY)}}{\vec{b}_{\Btheta_0^\cnd}(\BX,Y)\ \middle|\ \BX=\x}
	\\
	&=
	\var{\feps^{\YX=\x}}{ \vec{b}_{\Btheta_0^\cnd}(\x,Y) },
\end{align*}
so the expectation of the conditional variance is
\[
	\E{\fepsX}{\var{\feps^{(\XY)}}{\vec{q}_{\Btheta_0^\cnd,\eps}(\BX,Y)\ \middle|\ \BX}}
	=
	\sum_{\x\in\X}\fepsX(\x)\var{Y|\BX=\x\sim\feps^{\YX=\x}}{ \vec{b}_{\Btheta_0^\cnd,\eps}(\x,Y) }.
\]
Furthermore,
\begin{align*}
	\E{\feps^{(\XY)}}{\vec{q}_{\Btheta_0^\cnd,\eps}(\BX,Y)\ \middle|\ \BX=\x}
	&=
	\E{\feps^{(\XY)}}{\vec{b}_{\Btheta_0^\cnd,\eps}(\x,Y)\ \middle|\ \BX=\x} 
	- \bar{\vec{b}}_{\Btheta_0^\cnd,\eps}(\BX) + \vec{g}_{\Btheta_0^\cnd,\eps}(\BX)
	\\
	&=
	\bar{\vec{b}}_{\Btheta_0^\cnd,\eps}(\BX) - \bar{\vec{b}}_{\Btheta_0^\cnd,\eps}(\BX) + \vec{g}_{\Btheta_0^\cnd,\eps}(\BX)
	\\
	&=
	\vec{g}_{\Btheta_0^\cnd,\eps}(\BX),
\end{align*}
where we have used the definition of $\bar{\vec{b}}_{\Btheta_0^\cnd,\eps}(\BX)$ in the second line. Hence, the variance of the conditional expectation is given by
\[
	\var{\fepsX}{\E{\feps^{(\XY)}}{\vec{q}_{\Btheta_0^\cnd,\eps}(\BX,Y)\ \middle|\ \BX}}
	=
	\var{\fepsX}{ \vec{g}_{\Btheta_0^\cnd,\eps}(\BX) }.
\]

Finally, apply the law of total variance to obtain
\[
	\var{\feps^{XY}}{\vec{q}_{\Btheta_0^\cnd,\eps}(\BX,Y)}
	=
	\E{\fepsX}{\var{\feps^{(\XY)}}{\vec{q}_{\Btheta_0^\cnd,\eps}(\BX,Y)\ \middle|\ \BX}}
	+
	\var{\fepsX}{\E{\feps^{(\XY)}}{\vec{q}_{\Btheta_0^\cnd,\eps}(\BX,Y)\ \middle|\ \BX}}.
\]
Substitution into this expression and the definition of $\mat{U}_\eps^\cnd(\Btheta_0^\cnd)$ yield the assertion.
\QED

%%%%%%%%%%%%%%%%%%%%%%%%%%%%%%%%%%%%%%%%%%%%%%%%%%%%%
\newpage
\section{Additional numerical results}\label{app:simulations}
This section contains numerical details in Subsection~\ref{app:numerical}, additional results of simulations in Subsections~\ref{app:simulations-additional} and~\ref{app:simulation-logistic regression}, as well as additional results of the empirical application from Section~\ref{sec:application} in Subsection~\ref{app:application-additional-results}. Regarding the simulations, Subsection~\ref{app:simulations-additional} provides additional results from the simulation in Section~\ref{sec:simulations} on the latent factor model from Example~\ref{ex:factor}, whereas Subsection~\ref{app:simulation-logistic regression} is concerned with a new simulation design on conditional $C$-estimation of a logistic regression model. 

\subsection{Numerical details}\label{app:numerical}
All estimates for $C$-estimators in this paper were obtained by minimizing the estimator's corresponding loss function. As default choice for the optimization method, we used the unconstrained BFGS algorithm. If this algorithm did not converge to a solution, encountered an error, or produced a constraint violation (e.g., non-monotonicity of the thresholds in the latent factor model in Example~\ref{ex:factor}), we instead used a constrained algorithm, namely the constrained  Nelder-Mead simplex algorithm, where constraints are explicitly imposed.

\subsection{Simulation: Composite estimation of latent factor model}\label{app:simulations-additional}

\begin{figure}[t]
    \centering
    \includegraphics[width = 0.99\textwidth]{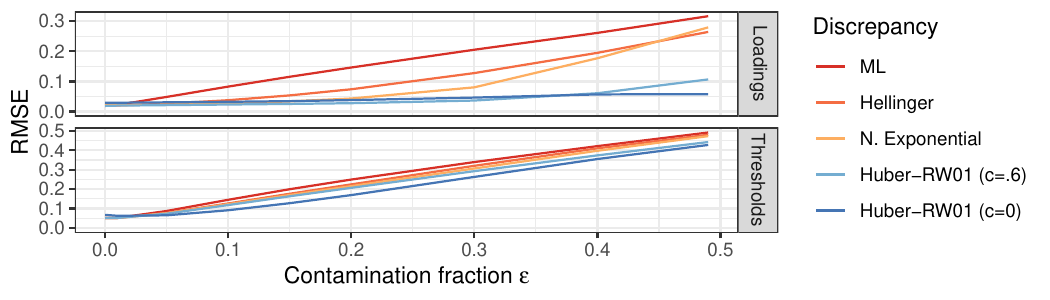}
\caption{RMSE with respect to the true values in~$\Btheta_*$ for composite estimation of the factor loadings parameters~$\Blambda$ (top) and thresholds~$\vec{\tau}$ (bottom). Results are averaged over~1,000 repetitions.}
\label{fig:simresults-cmp-rmse}
\end{figure}

Figure~\ref{fig:simresults-cmp-rmse} visualizes the root mean squared error (RMSE) of the simulation in Section~\ref{sec:simulations}. Specifically, it visualizes the RMSE with respect to the true factor loadings parameters, being 
\[
	\sqrt{ \frac{1}{p} \sum_{i=1}^p \big(\hat{\lambda}_{N,i} - \lambda_{*,i} \big)^2},
\]
and with respect to the true threshold parameters, being  
\[
	\sqrt{\frac{1}{p}\sum_{i=1}^p \frac{1}{m_i - 1} \sum_{j=1}^{m_i-1} \big(\hat{\tau}_{N,(i,j)} - \tau_{*, (i,j)} \big)^2}.
\]

\subsection{Simulation: Logistic regression with ordinal covariates}\label{app:simulation-logistic regression}
This simulation is concerned with the logistic regression model with ordinal covariates. This model is a special case of the cumulative logit model in Example~\ref{ex:cumulative-logit} where the response variable~$Y$ can only take $K=2$ distinct values. We consider a situation where there is only one single covariate~$X$ taking values in $\X = \{ -4.5, -3.5, -2.5, -1.5, -0.5,  0.5, 1.5, 2.5, 3.5,  4.5 \}$, so there are only~$|\X|=10$ unique covariate patterns. One can think of the values in~$\X$ as midpoints of ranges, like in a questionnaire where respondents are asked to choose a range from a set of ranges that a certain continuous variable of interest falls in \citep[e.g., income,][]{victoriafeser1997}. 
This setup is intentionally kept plain and simple with only $|\X \times \Y| = 20$ unique response-covariate pairs because, in such a situation, it is challenging to achieve contamination-robustness since observed data only exhibit very limited variation. As such, the goal of this simulation is to evaluate the performance of $C$-estimators in such a challenging setting.

We generate draws for the ordinal covariate~$X$ as follows. We first draw a realization~$\tilde{x}$ for a standard normal random variable~$\tilde{X}$. We then obtain an ordinal $x\in\X$ by categorizing~$\tilde{x}$ based on a binning with the thresholds~$(-\infty,-2, -1.5, -1, -0.5, 0, 0.5, 1, 1.5, 2,\infty)$. For instance, the value $\tilde{x}=-1.2$ falls into the third bin $[-1.5, -1)$, so it generates the categorical value~$x$ corresponding to the third pattern in~$\X$, being $x=-2.5$.  We then calculate the linear predictor $a = \alpha_* + x\beta_*$, where $(\alpha_*,\beta_*) = (0,1)$, and obtain realizations for the response variable~$Y$ as random draws from a Bernoulli distribution with probability parameter being set to the inverse-logit of the linear predictor~$a$, being $\exp(a)/(1+\exp(a))$. We generate $N=1,000$ observations for $(X,Y)$ according to this process.

Contamination is introduced as follows. For a given contamination fraction~$\eps$, we randomly replace~$N\eps$ observations for the covariate~$X$ by the highest possible covariate value, being~4.5. We fit the logistic model to each possibly contaminated dataset via regression $C$-estimation. We consider the same contamination fractions, the same discrepancy functions, and the same performance measures as in the simulation on composite estimation of latent factor models in Section~\ref{sec:simulations}. The confidence intervals are constructed using the estimated asymptotic covariance matrix from Theorem~\ref{thm:cnd-normality}. This procedure is repeated 1,000 times.

Unlike with the latent factor model for ordinal data from the previous simulation, there is a well-established method for robustly estimating logistic regression models, namely the $M$-estimator of \cite{cantoni2001glm}. We therefore compare this estimator (at~95\% efficiency) to the considered regression $C$-estimators.

\begin{figure}[t]
    \centering
    \includegraphics[width = 0.99\textwidth]{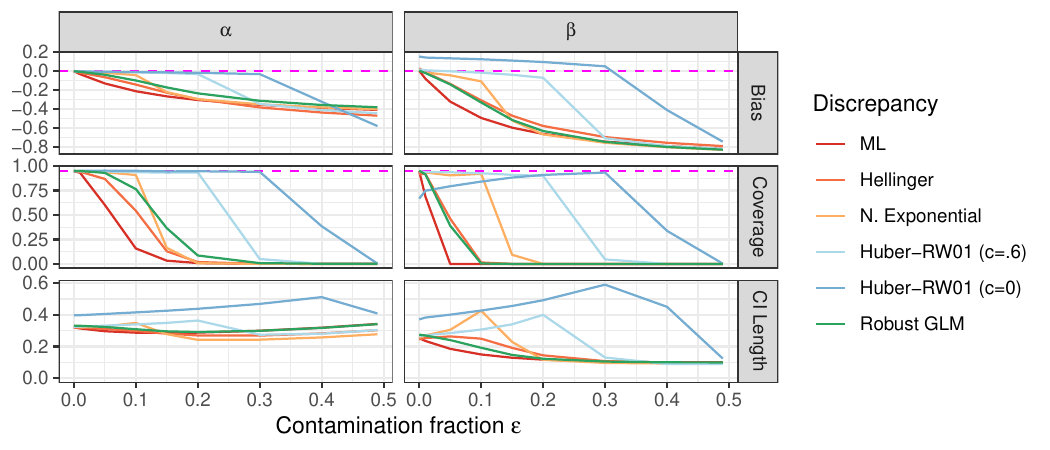}
\caption{Simulation results for estimation of the logistic regression model with an ordinal covariate using various estimators: bias with respect to the true value~$(\alpha_*, \beta_*)$, coverage of the estimated 0.95~confidence interval with respect to the true value (center; dashed line indicates nominal level), and length of that confidence interval (bottom). Results are averaged over~1,000 repetitions.}
\label{fig:simresults-cnd}
\end{figure}

Figure~\ref{fig:simresults-cnd} illustrates the simulation results. As before, ML is highly sensitive to categorical contamination, as reflected by rapidly diminishing coverage rates in the presence of contamination. The estimator of \cite{cantoni2001glm} offers only moderate robustness gains over ML, which may reflect the fact that it is primarily designed for addressing extreme values rather than categorical outliers.

In contrast, $C$-estimators can offer substantial robustness gain; particularly the estimator with Huber discrepancy function stands out with good coverage up to $\eps=0.2$ (choice $c=0.6$) and $\eps=0.3$ (choice $c=0$). However, unlike before, beyond these contamination fractions, also these estimators start to break down. This may be due to the intentionally challenging setting with only very limited variation in the data. Nevertheless, regression $C$-estimators, particularly with the Huber discrepancy function, outperformed the competitor in this design in terms of robustness even in such a challenging setting.

As in the previous simulation and for the same reasons, at $\eps=0$, the Huber discrepancy function with $c=0$ has CIs wider than all other estimators and incurs a finite-sample approximation error.

\subsection{Empirical application}\label{app:application-additional-results}

Table~\ref{tab:arias2020-thresholds} lists the factor model threshold estimates in the empirical application from Section~\ref{sec:application}. Table~\ref{tab:arias2020-PRs} lists the PR values visualized in Figure~\ref{fig:arias2020-antonym} from the same empirical application.

\begin{table}
\centering
\scriptsize

{
\renewcommand{\baselinestretch}{1}\selectfont
\renewcommand{\arraystretch}{1}
\setlength{\extrarowheight}{0pt} % if array is loaded
\setlength{\tabcolsep}{6pt}

\begin{tabular}{c c r c c r c}
& & \multicolumn{2}{c}{Robust} & & \multicolumn{2}{c}{ML}
\\
\noalign{\smallskip}\cline{3-4}\cline{6-7}\noalign{\smallskip}
Threshold & & Estimate & $\widehat{\text{SE}}$ & & Estimate & $\widehat{\text{SE}}$ 
\\
\noalign{\smallskip}\hline\noalign{\smallskip}
	1$|$1 && $-1.855$ & 0.091  && $-1.842$ & 0.090 \\ 
  1$|$2 && $-1.045$ & 0.057  && $-1.051$ & 0.057 \\ 
  1$|$3 && $-0.490$ & 0.049  && $-0.504$ & 0.049 \\\smallskip 
  1$|$4 &&    0.726 & 0.051  && 0.684    & 0.050 \\
  2$|$1 && $-1.755$ & 0.087  && $-1.609$ & 0.076 \\ 
  2$|$2 && $-0.934$ & 0.055  && $-0.886$ & 0.053 \\ 
  2$|$3 && $-0.507$ & 0.049  && $-0.478$ & 0.048 \\\smallskip
  2$|$4 &&    0.443 & 0.049  && 0.458    & 0.049 \\ 
  3$|$1 && $-1.612$ & 0.077  && $-1.603$ & 0.076 \\ 
  3$|$2 && $-0.790$ & 0.053  && $-0.799$ & 0.053 \\ 
  3$|$3 && $-0.273$ & 0.047  && $-0.288$ & 0.047 \\\smallskip 
  3$|$4 &&    0.899 & 0.054  && 0.849    & 0.053 \\ 
  4$|$1 && $-1.264$ & 0.063  && $-1.187$ & 0.061 \\ 
  4$|$2 && $-0.253$ & 0.048  && $-0.233$ & 0.047 \\ 
  4$|$3 &&    0.263 & 0.047  && 0.280    & 0.047 \\\smallskip
  4$|$4 &&    1.130 & 0.059  && 1.135    & 0.059 \\ 
  5$|$1 && $-1.528$ & 0.073  && $-1.495$ & 0.071 \\ 
  5$|$2 && $-0.667$ & 0.050  && $-0.672$ & 0.050 \\ 
  5$|$3 && $-0.094$ & 0.046  && $-0.107$ & 0.046 \\\smallskip 
  5$|$4 &&    1.083 & 0.058  && 1.045    & 0.057 \\ 
  6$|$1 && $-1.159$ & 0.061  && $-1.118$ & 0.059 \\ 
  6$|$2 && $-0.237$ & 0.047  && $-0.231$ & 0.047 \\ 
  6$|$3 &&    0.195 & 0.047  && 0.196    & 0.046 \\\smallskip 
  6$|$4 &&    0.997 & 0.056  && 0.980    & 0.056 \\ 
  7$|$1 && $-1.534$ & 0.075  && $-1.411$ & 0.068 \\ 
  7$|$2 && $-0.471$ & 0.050  && $-0.452$ & 0.048 \\ 
  7$|$3 &&    0.128 & 0.048  && 0.139    & 0.047 \\\smallskip 
  7$|$4 &&    1.077 & 0.059  && 1.054    & 0.057 \\ 
  8$|$1 && $-1.718$ & 0.083  && $-1.622$ & 0.077 \\ 
  8$|$2 && $-0.602$ & 0.051  && $-0.596$ & 0.050 \\ 
  8$|$3 &&    0.015 & 0.047  && 0.017    & 0.046 \\\smallskip 
  8$|$4 &&    0.891 & 0.055  && 0.865    & 0.053 \\ 
  9$|$1 && $-2.002$ & 0.102  && $-1.968$ & 0.099 \\ 
  9$|$2 && $-1.104$ & 0.059  && $-1.111$ & 0.059 \\ 
  9$|$3 && $-0.543$ & 0.049  && $-0.552$ & 0.049 \\\smallskip 
  9$|$4 &&    0.637 & 0.051  && 0.597    & 0.049 \\ 
  10$|$1 && $-1.943$ & 0.093 && $-1.789$ & 0.086 \\ 
  10$|$2 && $-1.015$ & 0.057 && $-0.981$ & 0.055 \\ 
  10$|$3 && $-0.606$ & 0.050 && $-0.582$ & 0.049 \\\smallskip 
  10$|$4 &&    0.210 & 0.047 && 0.218    & 0.047 \\ 
  11$|$1 && $-1.511$ & 0.074 && $-1.474$ & 0.071 \\ 
  11$|$2 && $-0.720$ & 0.052 && $-0.708$ & 0.051 \\ 
  11$|$3 && $-0.149$ & 0.047 && $-0.145$ & 0.046 \\\smallskip 
  11$|$4 &&    1.139 & 0.059 && 1.098    & 0.058 \\ 
  12$|$1 && $-1.569$ & 0.076 && $-1.438$ & 0.069 \\ 
  12$|$2 && $-0.602$ & 0.050 && $-0.563$ & 0.049 \\ 
  12$|$3 && $-0.110$ & 0.047 && $-0.084$ & 0.046 \\ 
  12$|$4 &&    0.871 & 0.054 && 0.882    & 0.054 \\ 
\end{tabular}
}
\caption{Parameter estimates and standard error estimates ($\widehat{\text{SE}}$) for the thresholds associated with the items in the \emph{neuroticism} trait in the data of \citet{arias2020}. The estimators are the composite $C$-estimator with Huber discrepancy function and tuning constant $c=0.6$,  and composite ML. A code ``$j|k$'' refers to the $k$-th threshold of the $j$-th item, $\tau^{(j)}_k$, $j=1,\dots, 12, k = 1,\dots,4$.}
\label{tab:arias2020-thresholds}
\end{table}

\begin{table}
\footnotesize
\centering
{
\renewcommand{\baselinestretch}{1}\selectfont
\renewcommand{\arraystretch}{1}
\setlength{\extrarowheight}{0pt} % if array is loaded
\setlength{\tabcolsep}{6pt}
\begin{subtable}[t]{0.46\linewidth}
\centering
\begin{tabular}{cccccc}
 & 1 & 2 & 3 & 4 & 5 \\
\midrule
1 & \makecell{1.821 \\ {\scriptsize (2)}} & \makecell{0.087 \\ {\scriptsize (8)}} & \makecell{-0.026 \\ {\scriptsize (19)}} & \makecell{0.029 \\ {\scriptsize (111)}} & \makecell{-0.088 \\ {\scriptsize (94)}} \\
2 & \makecell{0.606 \\ {\scriptsize (7)}} & \makecell{0.061 \\ {\scriptsize (28)}} & \makecell{-0.114 \\ {\scriptsize (41)}} & \makecell{-0.022 \\ {\scriptsize (133)}} & \makecell{-0.012 \\ {\scriptsize (51)}} \\
3 & \makecell{-0.234 \\ {\scriptsize (3)}} & \makecell{-0.214 \\ {\scriptsize (13)}} & \makecell{0.235 \\ {\scriptsize (27)}} & \makecell{-0.053 \\ {\scriptsize (41)}} & \makecell{0.064 \\ {\scriptsize (10)}} \\
4 & \makecell{-0.270 \\ {\scriptsize (6)}} & \makecell{0.197 \\ {\scriptsize (29)}} & \makecell{0.092 \\ {\scriptsize (27)}} & \makecell{-0.248 \\ {\scriptsize (27)}} & \makecell{0.949 \\ {\scriptsize (10)}} \\
5 & \makecell{0.025 \\ {\scriptsize (6)}} & \makecell{-0.585 \\ {\scriptsize (4)}} & \makecell{-0.849 \\ {\scriptsize (1)}} & \makecell{1.277 \\ {\scriptsize (14)}} & \makecell{27.142 \\ {\scriptsize (13)}} \\
\end{tabular}
\caption{N1}
\end{subtable}
\hfill
\begin{subtable}[t]{0.46\linewidth}
\centering
\begin{tabular}{cccccc}
 & 1 & 2 & 3 & 4 & 5 \\
\midrule
1 & \makecell{67.410 \\ {\scriptsize (2)}} & \makecell{1.413 \\ {\scriptsize (2)}} & \makecell{0.195 \\ {\scriptsize (4)}} & \makecell{0.063 \\ {\scriptsize (37)}} & \makecell{-0.122 \\ {\scriptsize (48)}} \\
2 & \makecell{-1.000 \\ {\scriptsize (0)}} & \makecell{-0.307 \\ {\scriptsize (7)}} & \makecell{-0.206 \\ {\scriptsize (19)}} & \makecell{0.111 \\ {\scriptsize (118)}} & \makecell{-0.163 \\ {\scriptsize (44)}} \\
3 & \makecell{-0.182 \\ {\scriptsize (2)}} & \makecell{-0.123 \\ {\scriptsize (17)}} & \makecell{0.387 \\ {\scriptsize (43)}} & \makecell{-0.147 \\ {\scriptsize (66)}} & \makecell{0.084 \\ {\scriptsize (19)}} \\
4 & \makecell{-0.525 \\ {\scriptsize (7)}} & \makecell{0.169 \\ {\scriptsize (67)}} & \makecell{-0.106 \\ {\scriptsize (50)}} & \makecell{-0.085 \\ {\scriptsize (72)}} & \makecell{0.994 \\ {\scriptsize (17)}} \\
5 & \makecell{0.395 \\ {\scriptsize (29)}} & \makecell{-0.319 \\ {\scriptsize (20)}} & \makecell{-0.377 \\ {\scriptsize (9)}} & \makecell{0.219 \\ {\scriptsize (12)}} & \makecell{39.368 \\ {\scriptsize (14)}} \\
\end{tabular}
\caption{N2}
\end{subtable}
%
% NEW LINE
%
\begin{subtable}[t]{0.46\linewidth}
\centering
\begin{tabular}{cccccc}
 & 1 & 2 & 3 & 4 & 5 \\
\midrule
1 & \makecell{78.399 \\ {\scriptsize (8)}} & \makecell{0.315 \\ {\scriptsize (3)}} & \makecell{0.555 \\ {\scriptsize (13)}} & \makecell{-0.155 \\ {\scriptsize (45)}} & \makecell{-0.050 \\ {\scriptsize (49)}} \\
2 & \makecell{-0.309 \\ {\scriptsize (1)}} & \makecell{-0.080 \\ {\scriptsize (15)}} & \makecell{0.018 \\ {\scriptsize (36)}} & \makecell{-0.007 \\ {\scriptsize (103)}} & \makecell{-0.029 \\ {\scriptsize (33)}} \\
3 & \makecell{-0.014 \\ {\scriptsize (3)}} & \makecell{-0.256 \\ {\scriptsize (16)}} & \makecell{-0.008 \\ {\scriptsize (32)}} & \makecell{0.076 \\ {\scriptsize (62)}} & \makecell{0.045 \\ {\scriptsize (10)}} \\
4 & \makecell{-0.628 \\ {\scriptsize (6)}} & \makecell{0.164 \\ {\scriptsize (71)}} & \makecell{-0.058 \\ {\scriptsize (55)}} & \makecell{-0.032 \\ {\scriptsize (62)}} & \makecell{0.236 \\ {\scriptsize (7)}} \\
5 & \makecell{0.231 \\ {\scriptsize (31)}} & \makecell{-0.252 \\ {\scriptsize (27)}} & \makecell{-0.268 \\ {\scriptsize (13)}} & \makecell{0.704 \\ {\scriptsize (17)}} & \makecell{20.571 \\ {\scriptsize (7)}} \\
\end{tabular}
\caption{N3}
\end{subtable}
\hfill
\begin{subtable}[t]{0.46\linewidth}
\centering
\begin{tabular}{cccccc}
 & 1 & 2 & 3 & 4 & 5 \\
\midrule
1 & \makecell{3.429 \\ {\scriptsize (14)}} & \makecell{-0.773 \\ {\scriptsize (5)}} & \makecell{-0.856 \\ {\scriptsize (4)}} & \makecell{-0.215 \\ {\scriptsize (39)}} & \makecell{1.397 \\ {\scriptsize (78)}} \\
2 & \makecell{-0.484 \\ {\scriptsize (5)}} & \makecell{-0.429 \\ {\scriptsize (29)}} & \makecell{-0.346 \\ {\scriptsize (34)}} & \makecell{0.831 \\ {\scriptsize (137)}} & \makecell{-0.634 \\ {\scriptsize (13)}} \\
3 & \makecell{-0.818 \\ {\scriptsize (2)}} & \makecell{-0.227 \\ {\scriptsize (36)}} & \makecell{1.541 \\ {\scriptsize (104)}} & \makecell{-0.585 \\ {\scriptsize (21)}} & \makecell{-0.794 \\ {\scriptsize (4)}} \\
4 & \makecell{0.217 \\ {\scriptsize (20)}} & \makecell{0.833 \\ {\scriptsize (100)}} & \makecell{-0.456 \\ {\scriptsize (22)}} & \makecell{-0.672 \\ {\scriptsize (14)}} & \makecell{-0.545 \\ {\scriptsize (6)}} \\
5 & \makecell{2.162 \\ {\scriptsize (16)}} & \makecell{-0.162 \\ {\scriptsize (10)}} & \makecell{-0.712 \\ {\scriptsize (2)}} & \makecell{-0.147 \\ {\scriptsize (5)}} & \makecell{2.733 \\ {\scriptsize (5)}} \\
\end{tabular}
\caption{N4}
\end{subtable}
%
% NEW LINE
%
\begin{subtable}[t]{0.46\linewidth}
\centering
\begin{tabular}{cccccc}
 & 1 & 2 & 3 & 4 & 5 \\
\midrule
1 & \makecell{1.369 \\ {\scriptsize (1)}} & \makecell{-0.872 \\ {\scriptsize (1)}} & \makecell{-0.479 \\ {\scriptsize (12)}} & \makecell{-0.073 \\ {\scriptsize (123)}} & \makecell{0.178 \\ {\scriptsize (163)}} \\
2 & \makecell{-0.574 \\ {\scriptsize (1)}} & \makecell{-0.434 \\ {\scriptsize (13)}} & \makecell{0.002 \\ {\scriptsize (42)}} & \makecell{0.245 \\ {\scriptsize (146)}} & \makecell{-0.533 \\ {\scriptsize (19)}} \\
3 & \makecell{-0.600 \\ {\scriptsize (1)}} & \makecell{-0.437 \\ {\scriptsize (9)}} & \makecell{1.092 \\ {\scriptsize (44)}} & \makecell{-0.346 \\ {\scriptsize (25)}} & \makecell{-0.314 \\ {\scriptsize (5)}} \\
4 & \makecell{0.011 \\ {\scriptsize (7)}} & \makecell{0.856 \\ {\scriptsize (50)}} & \makecell{-0.481 \\ {\scriptsize (13)}} & \makecell{-0.358 \\ {\scriptsize (20)}} & \makecell{0.109 \\ {\scriptsize (4)}} \\
5 & \makecell{0.900 \\ {\scriptsize (8)}} & \makecell{-0.206 \\ {\scriptsize (6)}} & \makecell{-0.754 \\ {\scriptsize (1)}} & \makecell{0.047 \\ {\scriptsize (3)}} & \makecell{54.953 \\ {\scriptsize (8)}} \\
\end{tabular}
\caption{N5}
\end{subtable}
\hfill
\begin{subtable}[t]{0.46\linewidth}
\centering
\begin{tabular}{cccccc}
 & 1 & 2 & 3 & 4 & 5 \\
\midrule
1 & \makecell{0.968 \\ {\scriptsize (2)}} & \makecell{-0.175 \\ {\scriptsize (6)}} & \makecell{-0.150 \\ {\scriptsize (14)}} & \makecell{-0.092 \\ {\scriptsize (65)}} & \makecell{0.171 \\ {\scriptsize (50)}} \\
2 & \makecell{-0.596 \\ {\scriptsize (3)}} & \makecell{-0.362 \\ {\scriptsize (21)}} & \makecell{-0.262 \\ {\scriptsize (38)}} & \makecell{0.215 \\ {\scriptsize (156)}} & \makecell{-0.114 \\ {\scriptsize (31)}} \\
3 & \makecell{-0.881 \\ {\scriptsize (1)}} & \makecell{-0.363 \\ {\scriptsize (17)}} & \makecell{0.988 \\ {\scriptsize (64)}} & \makecell{-0.178 \\ {\scriptsize (46)}} & \makecell{-0.780 \\ {\scriptsize (2)}} \\
4 & \makecell{-0.116 \\ {\scriptsize (17)}} & \makecell{0.580 \\ {\scriptsize (67)}} & \makecell{-0.212 \\ {\scriptsize (31)}} & \makecell{-0.278 \\ {\scriptsize (36)}} & \makecell{-0.235 \\ {\scriptsize (4)}} \\
5 & \makecell{1.459 \\ {\scriptsize (28)}} & \makecell{-0.227 \\ {\scriptsize (11)}} & \makecell{-1.000 \\ {\scriptsize (0)}} & \makecell{-0.453 \\ {\scriptsize (4)}} & \makecell{27.025 \\ {\scriptsize (11)}} \\
\end{tabular}
\caption{N6}
\end{subtable}
}
\caption{Each subtable holds the submodel PRs of the six pairwise submodels corresponding to the six opposite item pairs in Table~\ref{tab:neuroticism-items} in the \cite{arias2020} data, without recoding of negatively keyed items. The PRs were computed with a Huber-type composite $C$-estimator at $c = 0.6$. In each subtable, the rows correspond to the positive opposite (P) and the columns to the negative opposite (N). Each item has five response options ranging from~1 (``very inaccurate'') to~5 (``very accurate''). The numbers in parentheses denote the frequency of the pairwise response pattern within the submodel in question.}
\label{tab:arias2020-PRs}
\end{table}

\end{document}